\documentclass[numberedappendix]{emulateapj}
\usepackage{amsmath}
\usepackage{apjfonts}
\usepackage{amssymb}
\usepackage{mathrsfs}
\usepackage{color}
\usepackage{natbib}
\usepackage{ifthen}
\usepackage{comment} 

\graphicspath{{plots/}}

\def\bigplot#1{\includegraphics[width=\linewidth]{#1-color}} 

\newcommand{\msun}[0]{{\text{M}_{\sun}}}

\def\newacronym#1#2#3{\gdef#1{#3 (#2)\gdef#1{#2}}}

\newacronym{\mpm}{MPM}{moving puncture method}

\newacronym{\cra}{CRA}{Center for Relativistic Astrophysics}

\newacronym{\ghm}{GHM}{generalized harmonic method}
\newacronym{\ma}{MA}{Major Activity}
\newacronym{\hpc}{HPC}{high performance computing}
\newacronym{\cgwp}{CGWP}{Center for Gravitational Wave Physics}
\newacronym{\tat}{TAT}{Theoretische Astrophysik T\"ubingen}
\newacronym{\aei}{AEI}{Albert-Einstein-Institute (Potsdam)}
\newacronym{\jena}{Jena}{Friedrich-Schiller-University Jena}
\newacronym{\gsfc}{GSFC}{Goddard Space Flight Center}
\newacronym{\ornl}{ORNL}{Oak Ridge National Laboratory}
\newacronym{\lisa}{LISA}{Laser Interferometer Space Antenna}
\newacronym{\doe}{DOE}{Department of Energy}
\newacronym{\esa}{ESA}{European Space Agency}
\newacronym{\ligo}{LIGO}{Laser Interferometer Gravitational-wave
  Observatory} \newacronym{\lsc}{LSC}{LIGO Scientific Collaboration}
\newacronym{\Caltech}{Caltech}{California Institute of Technology}
\newacronym{\PFC}{PFC}{Physics Frontier Center}
\newacronym{\pfc}{PFC}{Physics Frontier Center}
\newacronym{\EOD}{EOD}{Education, Outreach and Diversity}
\newacronym{\MIT}{MIT}{Massachusetts Institute of Technology}
\newacronym{\sph}{SPH}{smooth particle hydrodynamics}
\newacronym{\amr}{AMR}{adaptive mesh refinements}
\newacronym{\tsi}{TSI}{Terascale Supernova Initiative}
\newacronym{\wmap}{WMAP}{the Wilkinson Microwave Anisotropy Probe}
\newacronym{\decigo}{DECIGO}{the Deci-Hertz Interferometric
  Gravitational-wave Observatory} 
  \newacronym{\cmbr}{CMBR}{cosmic
  microwave background} \newacronym{\ibbh}{IBBH}{intermediate binary
  black hole} \newacronym{\bdj}{BDJ}{Brans-Dicke-Jordan}
\newacronym{\bbo}{BBO}{Big Bang Observer}
\newacronym{\decigo}{DECIGO}{Deci-Hertz Gravitational-Wave
  Observatory} 
\newacronym{\cgwa}{CGWA}{University of Texas
  (Brownsville) Center for Gravitational Wave Astronomy}
\newacronym{\wiser}{WISER}{Women In Science and Engineering Research}
\newacronym{\mure}{MURE}{Minority Undergraduate Research Experience}
\newacronym{\igc}{IGC}{Institute for Gravitation and the Cosmos}
\newacronym{\sfb}{SFB/TR7}{Sonderforschungsbereich Transgerio 7,
  Gravitational Wave Astronomy Methods, Sources \& Observations}
\newacronym{\grb}{GRB}{Gamma-ray burst}
\newacronym{\grbs}{GRBs}{Gamma-ray bursts}
\newacronym{\csats}{CSATS}{Center for Science and The Schools}
\newacronym{\STEM}{STEM}{Science, Technology, Engineering and Mathematics}
\newacronym{\aaas}{AAAS}{American Association for the Advancement of Science}
\newacronym{\nlp}{NLP}{Natural Language Processing}
\newacronym{\heasarc}{HEASARC}{High Energy Astrophysics Science Archive Research Center}
\newacronym{\psu}{PSU}{Penn State University}
\newacronym{\gr}{GR}{general relativity}
\newacronym{\EM}{EM}{electromagnetic}
\newacronym{\gw}{GW}{gravitational wave}
\newacronym{\utb}{UTB}{University of Texas (Brownsville)}
\newacronym{\agns}{AGN}{active galactic nuclei}
\newacronym{\adaf}{ADAF}{advection-dominated accretion flow}
\newacronym{\llagns}{LLAGN}{low luminosity AGN}
\newacronym{\FOV}{FOV}{field of view}

\def\isco#1{inner-most stable orbit#1 
  (ISCO#1)\gdef\isco{ISCO}}
\def\emri#1{extreme mass-ratio inspiral#1 
  (EMRI#1)\gdef\emri{EMRI}}
\def\grb#1{Gamma-Ray Burst#1
  (GRB#1)\gdef\grb{GRB}}
\def\da#1{data analysis#1
  (DA#1)\gdef\da{DA}}
\def\nr#1{numerical relativity#1
  (NR#1)\gdef\nr{NR}}
\def\pnw#1{post-Newtonian#1
  (PN#1)\gdef\pnw{PN}}
\def\bh#1{black hole#1
  (BH#1)\gdef\bh{BH}}
\def\bbh#1{binary black hole#1
  (BBH#1)\gdef\bbh{BBH}}
\def\imbh#1{intermediate mass black hole#1
  (IMBH#1)\gdef\imbh{IMBH}}
\def\smbh#1{supermassive black hole#1
  (SMBH#1)\gdef\smbh{SMBH}}
\def\ulx#1{ultra-luminous x-ray source#1 (ULX#1)\gdef\ulx{ULX}}
\def\lmxbs{low-mass x-ray binaries
  (LMXBs)\gdef\lmxbs{LMXBs}\gdef\lmxb{LMXB}} 
\def\lmxb{low-mass x-ray binary
  (LMXB)\gdef\lmxbs{LMXBs}\gdef\lmxb{LMXB}} 
\def\emrb{extreme mass-ratio binary
  (EMRB)\gdef\emrb{(EMRB}\gdef\emrbs{EMRBs}} 
\def\emrbs{extreme mass-ratio binaries
  (EMRBs)\gdef\emrb{(EMRB}\gdef\emrbs{EMRBs}} 
\def\ns#1{neutron star#1
  (NS#1)\gdef\ns{NS}}
\def\whd#1{white dwarf#1
  (WD#1)\gdef\whd{WD}}
\def\qnm#1{quasi-normal mode#1
  (QNM#1)\gdef\qnm{QNM}}
\def\bhns#1{black hole -- neutron star#1 
  (BHNS#1)\gdef\bhns{BHNS}}
\def\nsns#1{neutron star -- neutron star#1 
  (NSNS#1)\gdef\nsns{NSNS}}
\def\ahz#1{apparent horizon#1 (AH#1)\gdef\ahz{AH}}
\def\tov#1{Tolman-Oppenheimer-Volkoff#1 (TOV#1)\gdef\tov{TOV}}

\newcommand{\siml}{\lower4pt \hbox{$\buildrel < \over \sim$}}
\newcommand{\simg}{\lower4pt \hbox{$\buildrel > \over \sim$}}

\def\simlt{\mathrel{\hbox{\rlap{\hbox{\lower4pt\hbox{$\sim$}}}\hbox{$<$}}}}
\def\simgt{\mathrel{\hbox{\rlap{\hbox{\lower4pt\hbox{$\sim$}}}\hbox{$>$}}}}

\def\ltsima{$\; \buildrel < \over \sim \;$}
\def\ltsim{\lower.5ex\hbox{\ltsima}}
\def\gtsima{$\; \buildrel > \over \sim \;$}
\def\gtsim{\lower.5ex\hbox{\gtsima}}

\def\mayakranc#1{\textsc{MayaKranc}#1}
\def\twopunctures#1{\textsc{2Punctures}#1}
\def\kranc#1{\textsc{Kranc}#1}
\def\whisky#1{\textsc{Whisky}#1}

\def\cactus#1{\textsc{Cactus}#1}
\def\carpet#1{\textsc{Carpet}#1}

\shorttitle{Binary Black Hole Mergers and their Electromagnetic Signature}
\shortauthors{Bode, Haas, Bogdanovi\'c, Laguna \& Shoemaker}
 
\begin{document}

\title{Relativistic Mergers of Supermassive Black Holes and their Electromagnetic Signatures}

\author{Tanja Bode\altaffilmark{1}, Roland Haas\altaffilmark{1},Tamara Bogdanovi\'c\altaffilmark{2,3},
Pablo Laguna\altaffilmark{1} and Deirdre Shoemaker\altaffilmark{1}}

\altaffiltext{1}{Center for Relativistic Astrophysics, School of Physics, Georgia Institute of Technology, Atlanta, GA 30332, USA}
\altaffiltext{2}{Department of Astronomy, University of Maryland, College Park, MD 207422, USA}
\altaffiltext{3}{Einstein Fellow}

\begin{abstract} 
  Coincident detections of electromagnetic (EM) and gravitational wave
  (GW) signatures from coalescence events of supermassive black holes
  are the next observational grand challenge. Such detections will
  provide the means to study cosmological evolution and accretion
  processes associated with these gargantuan compact objects. More
  generally, the observations will enable testing general relativity
  in the strong, nonlinear regime and will provide independent
  cosmological measurements to high precision. Understanding the
  conditions under which coincidences of EM and GW signatures arise
  during supermassive black hole mergers is therefore of paramount
  importance.  As an essential step towards this goal, we present
  results from the first fully general relativistic, hydrodynamical
  study of the late inspiral and merger of equal-mass, spinning
  supermassive black hole binaries in a gas cloud.  We find that
  variable EM signatures correlated with GWs can arise in merging
  systems as a consequence of shocks and accretion combined with the
  effect of relativistic beaming.  The most striking EM variability is
  observed for systems where spins are aligned with the orbital axis
  and where orbiting black holes form a stable set of density wakes,
  but all systems exhibit some characteristic signatures that can be
  utilized in searches for EM counterparts.  In the case of the most
  massive binaries observable by the Laser Interferometer Space Antenna,
  calculated luminosities imply that they may be identified by EM
  searches to $z\approx1$, while lower mass systems and binaries
  immersed in low density ambient gas can only be detected in the
  local universe.

\end{abstract}


\section{Introduction}\label{sec:introduction}

Galactic mergers are expected to be the major route for the formation
of \bbh{s}, in agreement with predictions of hierarchical models of
structure formation \citep*{hk02, volonteri03} and with observations
showing that the majority of galaxies harbor \smbh{s} in their centers
\citep[e.g.,][]{kr95, richstone98, pw00,ff05}.  Some fraction of newly
formed galaxies are thus expected to host binary systems of \smbh{s},
whose inspiral and merger lie in the ``sweet spot'' of the future
\lisa{}. These cataclysmic cosmic events produce copious amounts of
gravitational radiation ($\sim 10^{57}\,{\rm erg\,s^{-1}}$) during the
final stages of merger, radiation which is imprinted with detailed
information about the inspiral and merger of the binary as well as the
state (mass, spin and kick) of the resulting \bh{}. Detecting and
characterizing these binary mergers will thus be an important resource
for understanding formation mechanisms and the cosmological evolution
of \smbh{s} \citep{volonteri2003, volonteri2004, micic2007, sesana07}.
A synergistic detection of \EM{} and \gw{} radiation from merging
\smbh{s} has been defined as the next observational grand challenge
\citep[][for example]{kocsis06}. A combination of \EM{} and \gw{}
signatures will provide independent measurements of redshift and
luminosity distance, allowing cosmological measurements at a precision
of $\sim 10\%$, limited mainly by uncertainties due to weak
gravitational lensing \citep{hh03}. Similarly, \EM{+}\gw{}
observations can be used to test whether gravitons travel at the speed
of light, as required by general relativity, and in such a way test one
of the fundamental principles of the theory. Furthermore, these
multi-messenger observations will provide new insight on accretion
processes associated with \smbh{} binary systems in the final stages
of their evolution.

One of the outstanding astrophysical questions with direct bearing on
the feasibility of \EM{+}\gw{} detections is regarding the physical
properties of the gaseous environment surrounding a binary before and
during coalescence. Observationally, a significant sample of these
objects is yet to be attained, and finding them in EM searches is a
challenging task \citep{bogdanovic09}. Consequently, most of the
information about these systems has to be drawn from theoretical
perspectives. Non-relativistic hydrodynamical simulations have, over
the past several years, significantly contributed to our understanding
of the evolution of \bh{} pairs and interstellar gas, both during and
after the galactic mergers \citep{kazantzidis05, an02, escala04,
escala05, dotti07, mayer07, colpi07, mm08, hayasaki08, cuadra09}.
However, simulations spanning the entire dynamical range, from
galactic merger scales ($\sim 10^2 \,\hbox{kpc}$) to binary
coalescences ($<10^{-2}\,\hbox{pc}$), are still prohibitively
computationally expensive. As a consequence, non-relativistic 
simulations stop at binary separations of order $1\,\mathrm{pc}$
while fully general relativistic simulations are possible only
at separations of order $10^{-5}\,\mathrm{pc}$.  Thus the properties and
structure of accretion flows around binaries are uncertain. For
instance, the presence of gas on larger scales in the aftermath of a
gas rich galactic merger may not guarantee an abundant supply of gas
for accretion once a gravitationally bound binary is formed. This is
because the binary torques can evacuate most of the surrounding gas,
thus preventing any significant accretion on either member of the
binary \citep{mp05}. The scenario in which binary torques clear a
central low density region is commonly described in the literature as
the {\it circumbinary disk}. Alternatively, it is also plausible that
if the surrounding gas is sufficiently hot and tenuous, the binary may
find itself engulfed in a radiatively inefficient, turbulent flow all
the way through coalescence. Such conditions could arise in gas
deficient mergers and are indeed expected to exist in nuclear regions
of some low luminosity active galactic nuclei \citep[AGNs;][for
example]{quataert99, ptak04, nemmen06, eh09}. We refer to this
scenario as the {\it gas cloud} and note that if binaries indeed do
exist in radiatively inefficient flows, then the \emph{circumbinary
disk} and \emph{gas cloud} scenarios effectively bracket the range of
physical situations in which pre-coalescence binaries may be found in
centers of galaxies.

In addition to \EM{} signatures and accretion, a gaseous environment
could potentially have a profound effect on the dynamics of the
\bh{s}. For instance, accretion torques in gas rich mergers could
force the \bh{s} to have a ``preferential'' spin orientation, aligning
the spins of the \bh{s} with the angular momentum of the large-scale
gas disk \citep{bogdanovic07}. Since \nr{} simulations have
established that the magnitude of the gravitational recoil on the
final \bh{} depends on the spin orientation of the merging
\bh{s}~\citep{2007PhRvD..76h4032H,2007PhRvL..99d1102K,
2007ApJ...659L...5C, 2007PhRvL..98w1102C, 2007PhRvL..98w1101G,
2007ApJ...668.1140B, 2007ApJ...662L..63S}, the presence or absence of
gas in the vicinity of the \bbh{} can have direct implications for the magnitude of the
kick inflicted on the final \bh{.}

In this work, we build upon the framework of larger-scale simulations
as well as the initial relativistic calculations that investigated
both the dynamics of test particles \citep{vanmeter09} and the
evolution of \EM{} fields \citep{palenzuela09,ply09} in the
gravitational potential of a coalescing binary. We use fully general
relativistic numerical hydrodynamics simulations to follow the
interaction of a \bh{} binary in a gaseous environment through
coalescence. We focus on the final stages of the binary evolution
(last few orbits and merger) and consider only equal-mass \smbh{}
binaries surrounded by a hot and tenuous gas cloud. The main objective
of this work is to characterize the \EM{} and \gw{} signatures that
arise during coalescence\footnote{Animations of the
  simulations discussed in this paper can be found at 
  \url{http://www.cra.gatech.edu/numrel/papers/BBH\_GasCloud.shtml}}.

The remainder of this paper is organized as follows: The computational
methodology used is described in \S\ref{sec:methods}, followed by the
initial conditions used in the simulations in \S\ref{sec:ID}. We
describe the general gas dynamics in \S\ref{sec:sysdyn}.  In
\S\ref{sec:emo}, we present a discussion of the \EM{} and \gw{}
signatures, followed by our conclusions in \S\ref{sec:conclusions}.


\section{Computational Methodology}\label{sec:methods}

The results in this paper were obtained with the new version of the
\mayakranc{} code of the \nr{} group at Georgia Tech. The new code is
an enhanced version of the code that was used primarily for studies
involving vacuum spacetimes containing \bh{}
singularities~\citep{2008PhRvL.101f1102W,2008PhRvD..77h1502H,
  2007ApJ...661..430H,2007CQGra..24...33H,2008PhRvD..77d4027B,Bode:2009fq}.
As with the previous code, the new \mayakranc{} code is based on the
BSSN formulation and the moving puncture
method~\citep{2006PhRvL..96k1101C,2006PhRvL..96k1102B}.  The code
source is generated by the package \kranc{}~\citep{Husa:2004ip}, which
produces a set of \emph{thorns} that work under the \cactus{}
infrastructure~\citep{Allen99a} and \carpet{} mesh
refinement~\citep{Schnetter-etal-03b}.  The new main feature in the
\mayakranc{} code is the inclusion of general relativistic
hydrodynamics.  The hydrodynamics code in \mayakranc{} is a modified
version of the public version of the \whisky{} hydrodynamics code
developed by the European Union Network on Sources of Gravitational
Radiation~\citep{whisky-web,Baiotti03a}.  Unless otherwise specified, we use 
geometrized units where $G=c=1$ and normalized to the mass of the 
system $M$, using the metric signature $(-,+,+,+)$.

\mayakranc{} assumes a perfect fluid with energy momentum tensor
\begin{equation}
T^{\mu\nu} = \rho\,h\, u^\mu u^\nu +P g^{\mu\nu}
\end{equation}
with $\rho$ the rest mass density, $h = 1+\epsilon + P/\rho$ the
enthalpy, $P$ the pressure, $\epsilon$ the internal energy per unit
mass, and $u^\mu$ the 4-velocity of the fluid. The 3-velocity of the
fluid is given by $v^i = (u^i/u^0 + \beta^i)/\alpha$, with $\alpha$
and $\beta^i$ the lapse function and shift vector, respectively. The
quantities $\rho$, $v^i$ and $\epsilon$ are considered
\emph{primitive} variables.  The pressure $P$ and the Lorentz factor
$W = \alpha\,u^0$ are viewed as \emph{auxiliary} variables. The
pressure $P$ for all the simulations in the present work is calculated
from the $\Gamma$-law equation of state $P = \rho \epsilon (\Gamma-1)$
with $\Gamma = 5/3$. The Lorentz factor is obtained from $u^\mu
\,u_\mu = -1$.

\whisky{} uses the \emph{Valencia
  formulation}~\citep{Marti91,Banyuls97,Ibanez01} of numerical 
hydrodynamics. That is, the
evolution equations are cast into a set of conservation equations of the 
form
\begin{eqnarray}
   \partial_0 \vec{F^0}(\vec{w}) + \partial_i\vec{F^i}(\vec{w}) = \vec{S}(\vec{w})\,,
\end{eqnarray}
where $\vec{w} = (\rho,v^i,\epsilon)$ is the vector of primitive
variables and $\vec{F^0} = (D,S^i,\tau)$ the corresponding vector of
\emph{conservative} variables defined as:
\begin{subequations}
\begin{eqnarray}
  D &=& \sqrt{\gamma} \rho W \label{eqn:valencia-conservatives-begin} \\
  S^i &=& \sqrt{\gamma} \rho h W^2 v^i \\
  \tau &=& \sqrt{\gamma} \left(\rho h W^2 - P\right) - D \label{eqn:valencia-conservatives-end}
\end{eqnarray}
\end{subequations}
with $\gamma$ the determinant of the spatial metric. The vectors
$\vec{F}^i$ and $\vec{S}$ are the fluxes and source terms,
respectively. For details see~\cite{Banyuls97,Ibanez01}.

Our code, \mayakranc{}, modifies the \whisky{} code in several important 
ways.  In order to improve efficiency as well as simplify the interface
between the hydrodynamics and geometry sectors, we implemented a new construction 
of the stress-energy tensor $T_{\mu\nu}$ and the corresponding sources
appearing in the BSSN equations.  Furthermore, in
the regions within the \ahz{s}, we impose, as suggested
by~\cite{2007PhRvD..76j4021F}, the dust limit of the hydrodynamics
equations when unphysical data appears during the evolution.  We also
modified the atmosphere treatment used to model vacuum regions. A
filter is in place in the atmosphere domains to avoid spurious
increases of fluid densities above the atmospheric threshold limit
during temporal updates.  The overall structure of the \whisky{} code
has also slightly changed to be able to work seamlessly with the rest
of the \mayakranc{} code.

For \bbh{} initial data, the \mayakranc{} code uses the
\twopunctures{} spectral code developed by~\cite{Ansorg:2004ds}.  We
modified this solver to include matter fields. Since the present work
focuses on matter fields initially at rest, the modifications to the
\twopunctures{} code involved only the Hamiltonian constraint. The
solutions to the momentum constraint remained unchanged; specifically
the Bowen-York type solutions continue to be
applicable~\citep{1980PhRvD..21.2047B}.  As with the vacuum case, we
construct initial data for a \bbh{} in a gaseous
environment assuming both conformal flatness and a vanishing trace of
the extrinsic curvature. The Hamiltonian constraint thus reads
\begin{equation}
\label{eq:ham}
\tilde{\nabla}^2 \psi + \frac{1}{8} \psi^{-7} \tilde A_{ij} \tilde A^{ij} 
   = -2\pi\psi^5 \rho_\circ = -2\pi\psi^{-3} \tilde{\rho}_\circ\,,
\end{equation}
where $\rho_\circ= n^\mu n^\nu T_{\mu\nu} = \rho\,h\,W^2-P$ with $n^\mu$
the unit normal to the constant time hypersurfaces.  In
Eq.~\ref{eq:ham}, conformal quantities have tildes, $\tilde A_{ij}$ is
the traceless part of the extrinsic curvature, and $\psi$ is the conformal
factor such that the spatial metric transforms like $\gamma_{ij} =
\psi^4\,\eta_{ij}$ with $\eta_{ij}$ the flat metric.  The conformal
rescaling $\tilde \rho_\circ = \psi^8 \rho_\circ$ is necessary in order to
guarantee the existence of a solution to the constraint
equation~\citep{York79}.  It is important to emphasize that the freely
specifiable matter field is $\tilde \rho_\circ$. We construct $\tilde
\rho_\circ$ from $\tilde \rho_\circ = \psi_{vac}^8 \rho_\circ$ in which
$\psi_{vac}$ is the solution to Eq.~\ref{eq:ham} in the absence of
matter.  Given $\tilde\rho_\circ$ constructed in this fashion, the
solution to the Hamiltonian constraint yields a different $\psi$, and 
thus the new $\rho_\circ$ does not exactly correspond to the primitive 
matter fields, $\tilde\rho_\circ$, used in setting up the conformal factor.

The \mayakranc{} code also includes infrastructure to analyze the data
produced by the simulations.  This infrastructure consists of a set of
analysis tools or modules to construct gravitational waveforms,
estimating \bh{} kicks and spins, and tracking \ahz{s}. With the
inclusion of hydrodynamics, more modules were needed.  Of particular
relevance to the present work, we developed tools to monitor
conservation of total rest mass in the system, to compute fluxes of
matter both through \ahz{s} (accretion) and in the form of outflows from
the immediate vicinity of the \bh{s}, and to calculate the integrated
luminosity of the emitted light.

We tested the implementation of hydrodynamics in \mayakranc{} through
a suite of standard tests. One of the important tests includes
verifying the conservation of baryonic mass.  While baryonic mass 
should be conserved everywhere in the spacetime, we consider here the 
conservation within sub-volumes of the computational domain, which
also aids us in testing in which sub-volumes conservation may be 
problematic.  That is, integrating \citep[see][]{Banyuls97}
\begin{equation}
\partial_t D + \partial_i \left[ D\,(\alpha v^i - \beta^i) \right]= 0
\end{equation}
over a constant volume ${\cal V}$ with boundary $\partial{\cal V}$ yields
\begin{equation} \label{eq:mdot}
\dot {\cal M}_{{\cal V}} + \int_{\partial{\cal V}} D\,(\alpha v^i -\beta^i) \,\hat r_i\, d{\cal S} = 0\,,
\end{equation}
with $\hat r^i$ the outgoing unit vector to ${\cal V}$ and 
\begin{equation} \label{eq:volmass}
\mathcal{M}_{\cal V} =\int_{\cal V} D\,d^3x\,.
\end{equation}
Time integration of Eq.~\ref{eq:mdot} yields
\begin{equation} \label{eq:totmass}
 \mathcal{M}(t) = \mathcal{M}_{\cal V}(t) + \mathcal{M}_{\partial{\cal V}}(t) = \hbox{constant.}
\end{equation}
where
\begin{equation} \label{eq:massflux}
\mathcal{M}_{\partial{\cal V}}(t) = \int_0^t dt' \int_{\mathcal{S}_h}
D\,(\alpha v^i - \beta^i)\,\hat r_i\, d\mathcal{S}\,.
\end{equation}
On a given step, we compute ${\cal M}$ from Eq.~\ref{eq:totmass} to check
how well it remains constant.

\begin{figure}[h]
\includegraphics[width=0.96\linewidth]{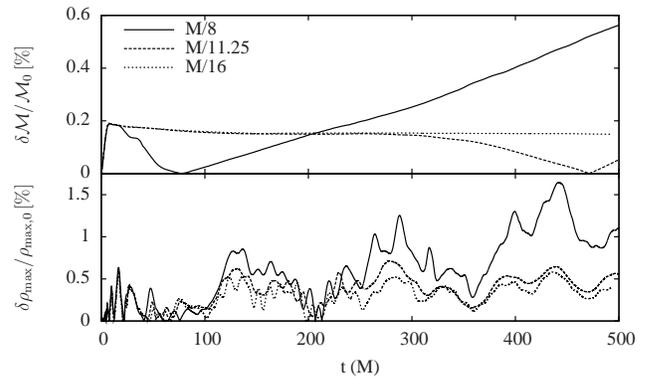}
\caption{ Top panel shows the errors in baryonic mass conservation as
  a function of time from the evolution of a TOV at different
  numerical resolutions. Bottom panel shows the corresponding errors
  in the central density.}
\label{fig:tovs}
\end{figure}

As a test of the code, we present here the evolution of a \tov{}
star~\citep{Tolman39,Oppenheimer39b,Misner73} of mass $1.4\,M$, where
$M$ is the arbitrary mass scale in the simulation. A \tov{} star is 
a static, spherically symmetric solution to the Einstein equations
for a spacetime with a perfect fluid.  As such, the stability of an
evolved \tov{} is a standard test in numerical relativity with 
hydrodynamics.  The initial data
is created by solving the \tov{} equations, assuming a polytropic
equation of state $P = \kappa\,\rho^\Gamma$ with $\kappa = 100$ and
$\Gamma = 5/3$.  The computational grid for these test runs consisted
of 5 refinement levels extending to an outer boundary of $512\,M$, the
finest of which lay within the star.  We evolved this system for three
resolutions where we chose the resolutions on the finest refinement
level as $M/8$, $M/11.25$, and $M/16$.  In the top panel of
Fig.~\ref{fig:tovs} we plot the errors in baryonic mass conservation,
$\delta{\cal M}/{\cal M}_0 = ({\cal M}-{\cal M}_0)/{\cal M}_0$, as a
function of time with ${\cal M} $ computed from Eq.~\ref{eq:totmass}
and ${\cal M}_0 = {\cal M}(t=0)$. In the bottom panel of
Fig.~\ref{fig:tovs}, we show the corresponding errors in the central
density, $\rho_{\text{max}}$, over $500\,M$ of evolution time.  The
central density stays within an acceptable $1.5\%$.  The baryonic mass
conservation on the other hand shows a linear increase of errors for
the coarsest level of resolution of $M/8$. This error growth arises from
the refinement boundaries in the vacuum regions, which are modeled as
a low density atmosphere.  As each boundary comes into causal contact 
with the surface of the star, the density rises above atmospheric levels
despite our atmospheric filters, adding a mass violation.  While this 
effect is significant in the coarsest run, the errors remain below $0.2\,\%$
in the other, higher resolution runs.  Improvements in
atmosphere handling are thus needed in our code for systems involving
vacuum regions (e.g. \bh{} + neutron star mergers). Later on, in
Section~\ref{sec:sysdyn}, we show that errors in mass conservation do
not affect the results presented in this paper since the relevant dynamics
remain completely contained within the gas cloud.

\section{Initial Data}\label{sec:ID}

With the uncertainties in our understanding of accretion flow properties
around binaries, it is not possible to uniquely define
initial conditions for simulations of \bbh{s} and gas on very small
scales ($\sim 10^{-5}\mathrm{pc}$), such as those considered in the
present work. Nevertheless, the physical properties of the gas in the
vicinity of the \bh{s} are likely to be bracketed by two
characteristic scenarios depending on the balance of heating and
cooling mechanisms in the accretion flow. In the first scenario, a hot
and turbulent accretion flow forms as a consequence of relatively
inefficient cooling processes; the \bbh{} is then immersed in a
pressure-supported, geometrically thick torus with a scale height $H$
comparable to its radius $R$. Hereafter, we refer to this scenario as
the \emph{gas cloud} model. In the second scenario, relatively
efficient cooling processes result in the gas settling into a
rotationally-supported, geometrically thin ($H<R$) accretion disk
around the binary. This scenario is commonly referred to as the
\emph{circumbinary disk} model and has been previously studied by
means of non-relativistic hydrodynamic simulations
\citep{an02,mp05,hayasaki07, hayasaki08, mm08,cuadra09}. In this
paper, we consider the gas cloud scenario and defer the numerical
investigation of the circumbinary disk with a fully relativistic
treatment to a future study.

Though the majority of \bbh{s} in the universe are expected to be
unequal-mass binaries, we will consider for computational
accessibility and as a first step only equal-mass binaries where 
we can exploit symmetries of the system. Our
simulations thus consist of equal-mass binaries with total mass $M$
and initial separation of $R = 8\,M\approx 10^{-5}\,M_7\,\mathrm{pc}$,
where $M_7 = M/10^7 \msun$.  This choice of mass scaling yields
\bbh{s} which can be detected in the \lisa{} band during the plunge
and coalescence and therefore in a regime where modeling is only
accessible via \nr{.}  As these systems are among the higher mass
\bbh{} systems in \lisa{'s} sensitivity window, they are relatively
luminous in gravitational radiation and, if associated with an \EM{}
signature, likely to be among the most \EM{} luminous binaries that
\lisa{} can detect.

The initial data for our simulations consist of a \bbh{} immersed
in a gas cloud with a radius of $60\,M$. The radius of the cloud is
selected arbitrarily and in such a way that it entirely encloses the
binary orbit.  The gas cloud is initially static with a Gaussian
density profile
\begin{equation}
\rho = \rho_\circ e^{-\frac{r^2}{2\sigma^2}}
\label{eq:denfun}
\end{equation} 
where $\sigma = 10.83\,M$ and $\rho_\circ$ is the central density in
the cloud.  The $\sigma$ was chosen to avoid non-atmosphere gas 
densities from reaching the outer boundary initially, though subsequent
simulations show larger $\sigma$ do not change the results qualitatively.
The value of $\rho_\circ$ adopted in our simulations is
obtained by extrapolating the results of larger scale simulations
that follow the evolution of \bbh{s} in gaseous environment in the
aftermath of galactic mergers. These results suggest that the amount
of gas which remains strongly bound to the binary on scales of
$\lesssim 10^{-2} \mathrm{pc}$ can be of the order of $1\%$ of the
total mass $M$ of the binary~\citep{colpi07}.  This implies a mean gas
density of
\begin{equation}
\rho\approx \frac{0.01\,M}{(0.01\,\mathrm{pc})^3} \approx 10^{-11}\,M_7 \,{\rm g\,cm^{-3}}\,.
\label{eq:den}
\end{equation}

Note that the internal units in the \mayakranc{} code are given in
terms of the total mass of the binary, $M$. For vacuum simulations,
this implies that the results from a \bbh{} simulation with mass $M$
can be scaled to arbitrary \bh{} physical masses. For non-vacuum
simulations such as those in the present work, the \bh{} mass also
determines the scaling of hydrodynamical variables (e.g,. density,
pressure, internal energy, etc.). In particular, density scales as
$M^{-2}$. Thus, in non-vacuum simulations scaling with $M$ is not
arbitrary and the mass parameter should be chosen in such a way as to
reflect a plausible range of densities. We set the initial central
density of the gas cloud to $\rho_\circ = 7 \times 10^{-12}
M_7^{-2}\,\mathrm{g\,cm^{-3}}$, a value consistent with the mean
density estimate in Eq.~\ref{eq:den}. The total rest mass of the gas
cloud in the computational domain is, initially, $\sim
10^{-4}\,M_7\,\msun{}$, about 11 orders of magnitude lower than the
\bh{} masses. For computational facility, we surround the gas cloud
with a uniform, low density atmosphere of density $10^{-5}\rho_\circ$.

We use a polytropic equation of state $P = \kappa\,\rho^\Gamma$ to
construct the initial data; the temperature of the gas is then given
by
\begin{equation}
T = \frac{m_p c^2}{k_B}\frac{P}{\rho} = \frac{m_p c^2} {k_B} \kappa\,\rho^{\Gamma-1}\,.
\label{eq:tempfun}
\end{equation}
In this gas cloud model, the thermal energy of the gas is comparable to
its gravitational potential energy. Thus, the velocity of the fluid in
the vicinity of the \bh{s} is
\begin{equation}
 v_\mathrm{th}=\sqrt{\frac{3 k_B T}{m_p}}\simeq
0.35\mathrm{c} \sqrt{\frac{8\,M}{R}}\,,
\end{equation}
where $T$ is the temperature of the gas.  It follows that the gas in
the vicinity of the binary is a high temperature plasma, $T \sim
10^{12}$~K, with thermal velocities comparable to the binary orbital
speed. We then estimate the adiabatic constant $\kappa$ in the
polytropic equation of state by evaluating Eq.~\ref{eq:tempfun} at the
center of the cloud, where $T_\circ = 10^{12}$~K and $\rho_\circ=7
\times 10^{-12} \mathrm{g\,cm^{-3}}\,M_7^{-2}$, and obtain
$\kappa=2.51\times 10^8\,\mathrm{(cm^3/g)^{\Gamma-1}}$, where $\Gamma
= 5/3$. For these values and an ideal fluid equation of state, we find
the speed of sound at the center of the cloud,
\begin{equation}
 c_\mathrm{s}=\sqrt{\frac{1}{h} \left(\frac{\partial P}{\partial \rho}
  +\frac{P}{\rho^2}\frac{\partial P}{\partial \epsilon}\right)}=
  \sqrt{ \frac{(\Gamma-1) (\epsilon + P/\rho)}{h}} \simeq
  0.28\,\mathrm{c}\,.
\end{equation}
Because the speed of sound is comparable though smaller than the
thermal velocity and the orbital velocity of the binary, it follows
that shocks can be expected to develop in vicinity of the \bh{s}.

The gas cloud surrounding the \bbh{} is not in hydrostatic
equilibrium. Namely, if the cloud were placed in a potential well of a
single \bh{}, the top layers of the cloud would be gravitationally
unbound and the gas cloud would expand, doubling its volume over
approximately $130\,M$ of evolution.  However, a condition for
hydrostatic equilibrium in the case of an isolated \bh{} is of limited
utility for the binary scenario considered here.  In a realistic case,
the binary torques will act to unbind some fraction of gas and, as a
result, the outer layers of the cloud in the binary scenario can be
expected to be hotter (i.e., less gravitationally bound) compared to
the cloud in equilibrium around a single \bh{.}  The simple initial
conditions we choose emulate this behavior. Note however that in order
to achieve more realistic initial conditions one should, in principle,
evolve the cloud in the potential of a rotating \bbh{} long before the
inspiral and merger until quasi-steady state is achieved. Such
relaxation is nontrivial and computationally expensive in the case of
a general relativistic system. Instead, we start the evolution of the
gas cloud in the initially \emph{frozen}, non-rotating gravitational
potential of the \bbh{} for $32\,M$ of evolution time, or four times
the crossing-time of the system.  During this phase, the cloud
passively evolves to form a gravitationally bound core with two
density peaks, each associated with one of the \bh{s}. After this
period, we ``unfreeze'' the binary and evolve it on its orbit for an
additional $\sim 60\,M$ (half orbit), during which the gas adjusts to
the dynamics of the binary. Any transients that develop as a
consequence of this initial relaxation are the artifact of imperfect
initial conditions and are not considered as a true signature of the
system.

The initial distribution of velocity vectors of fluid elements is
isotropic about the center of mass of the system. Given the
high thermal velocities, it follows that in such a flow, the radial
inflow speed of the gas will always be sufficiently high that the
binary torques will be incapable of evacuating all of the gas and
creating a low density ``hole'' in the center of the cloud, a property
that is a distinguishing feature from the circumbinary disk model.
The only mechanisms for heating and cooling of gas in our simulations
are adiabatic compression and expansion in the potential of the
binary. We do not account for heating of the gas by the \agns{} nor for
cooling by emission of radiation.

The initial orbital parameters for the \bbh{s} are calculated as
though the binary were in vacuum. This is justified since the total
mass in the gas cloud is negligibly small compared to the mass of the
\bh{s}. We calculate the initial momentum required for quasi-circular
orbits from \pnw{} equations.  For the non-spinning \bh{s}, we evolve
the 3rd order \pnw{} equations of motion~\citep{Buonanno:104005} from
a separation of $40\,M$ to the desired separation.  We also consider
binaries with spinning \bh{s} where spins are parallel and either
aligned or anti-aligned with the orbital angular momentum. In this
case we calculate the initial momentum of the \bh{s} from the
3rd-order \pnw{} angular momentum found in Eq. (4.7)
of~\citep{PhysRevD.52.821}, assuming purely tangential momenta.
Table~\ref{tab:bbh_id} lists parameters for the \bbh{} systems
discussed in this paper.  For each \bh{,} labeled 1 and 2, $a_{1,2}/m$
is the dimensionless spin parameter, where $m$ is the mass of each
\bh{} set in such a way that $m/M = 1/2$, with $M$ the total mass of
the binary.  As mentioned before, the \bh{s} are placed at a
coordinate distance $8\,M$ along the $x$-axis, with \bh{$_1$} located
at $x=4\,M$ and \bh{$_2$} at $x = -4\,M$. In Table~\ref{tab:bbh_id},
$P^x$ and $P^y$ are the components of linear momentum for
\bh{$_1$}. Finally, $m_i$ is the irreducible mass for the individual
\bh{s} computed from the area of their \ahz{,} and $M_\mathrm{ADM}$ is
the total initial energy of the system.

The simulations are run on a grid whose outer boundary is located at
$317.44\,M$, with 9 levels of mesh refinement. The 4 coarsest levels
are fixed about the center of mass while the 5 finest levels are
constructed around and free to follow the \bh{s}.  For runs G0, G1,
and G3, the finest resolution is $M/51.6$ and the refinement
boundaries are chosen such that the 5 finest have radii of 32
grid points while the 4 coarsest have radii of 64 grid points.  The mesh
setup for G2 was altered somewhat as higher spins require a higher
resolution at the puncture. The physical refinement boundaries of G2
remain stationary while the resolution is increased for a finest resolution
of $M/67.7$.  For all runs, we halve the evolved computational domain
by using the reflection symmetry about the $xy$-plane.  For runs G0,
G1, and G2, we halve the domain again using rotational symmetry.

\begingroup
\begin{center}
\begin{table}[ht]
\begin{ruledtabular}
\begin{tabular}{c|cccccc}
  Run & $a_1/m$ & $a_2/m$ & $P^x/M$ &  $P^y/M$ & $m_i/M$ & $M_\mathrm{ADM}/M$\\
\hline
  G0  &    0 &    0 & $-2.0902\times10^{-3}$ & 0.11237 & 0.5000 & 0.9878 \\ 
  G1  & +0.4 & +0.4 &                      0 & 0.10862 & 0.4893 & 0.9875 \\ 
  G2  & +0.6 & +0.6 &                      0 & 0.10677 & 0.4736 & 0.9874 \\ 
  G3  & +0.4 & -0.4 &                      0 & 0.11237 & 0.4893 & 0.9878 \\ 
\end{tabular}
\end{ruledtabular}
\caption{\bbh{} initial parameters.} 
\label{tab:bbh_id}
\end{table}
\end{center}
\endgroup
%


\section{Dynamics of the Binary and Gas}\label{sec:sysdyn}

We now describe the general features and dynamics of the binaries and gas
in our modeled systems. \bbh{s} with an initial separation of $8\,M$ evolve
for approximately 3 orbits before they plunge and finally merge.
Since the mass of the gas cloud is small compared to the \bbh{}, the
binary dynamics are practically indistinguishable from the equivalent
situation in vacuum. The only differences among the cases we
considered are due to the ``hang-up'' or delay induced by the \bh{}
spins~\citep{2006PhRvD..74d1501C}. The excess of angular momentum 
from the \bh{'s} spins result in runs G1 and G2 taking longer
to merge compared to runs G0 and G3. In run G2 for example, the
hang-up extends the inspiral to about 5 orbits before merger.

\begin{figure*}
\bigplot{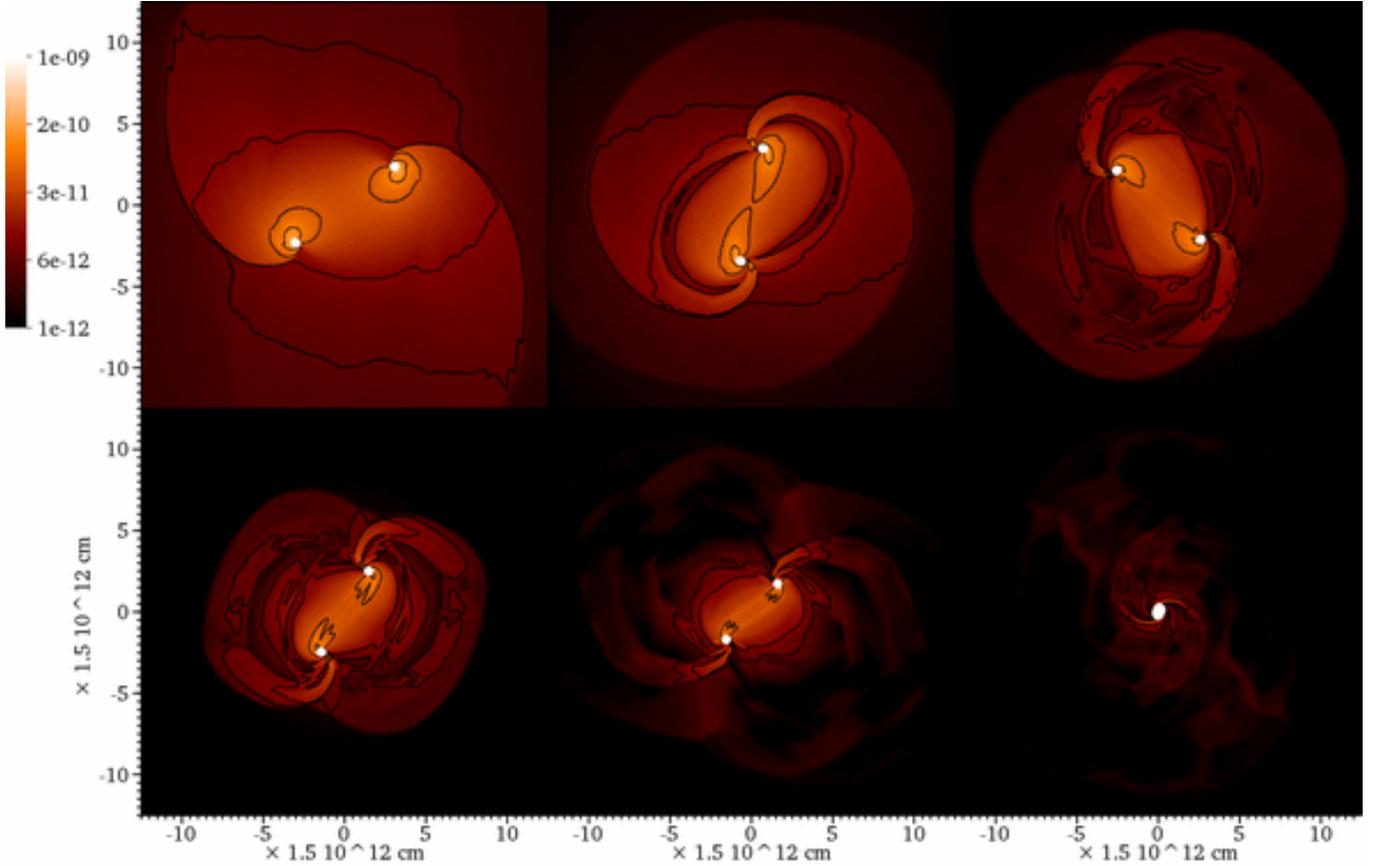}
  \caption{Snapshots of the baryonic rest mass density, $\rho$ $[{\rm
    g\,cm^{-3}}]$, in the orbital plane of the binary shown for run
    G2. From upper left to lower right the snapshots correspond to 
    $t = 7.6\times10^2\,M_7\,{\rm s}$, $1.3\times10^3\,M_7\,{\rm s}$, 
    $1.9\times10^3\,M_7\,{\rm s}$, $2.5\times10^3\,M_7\,{\rm s}$, 
    $3.1\times10^3\,M_7\,{\rm s}$, and $3.7\times10^3\,M_7\,{\rm s}$
    respectively.
    The color scale
    is logarithmic, with contour lines plotted at half
    order-of-magnitude intervals.}
\label{fig:2DrhoG2}
\end{figure*}

\begin{figure*}[tbp]
\bigplot{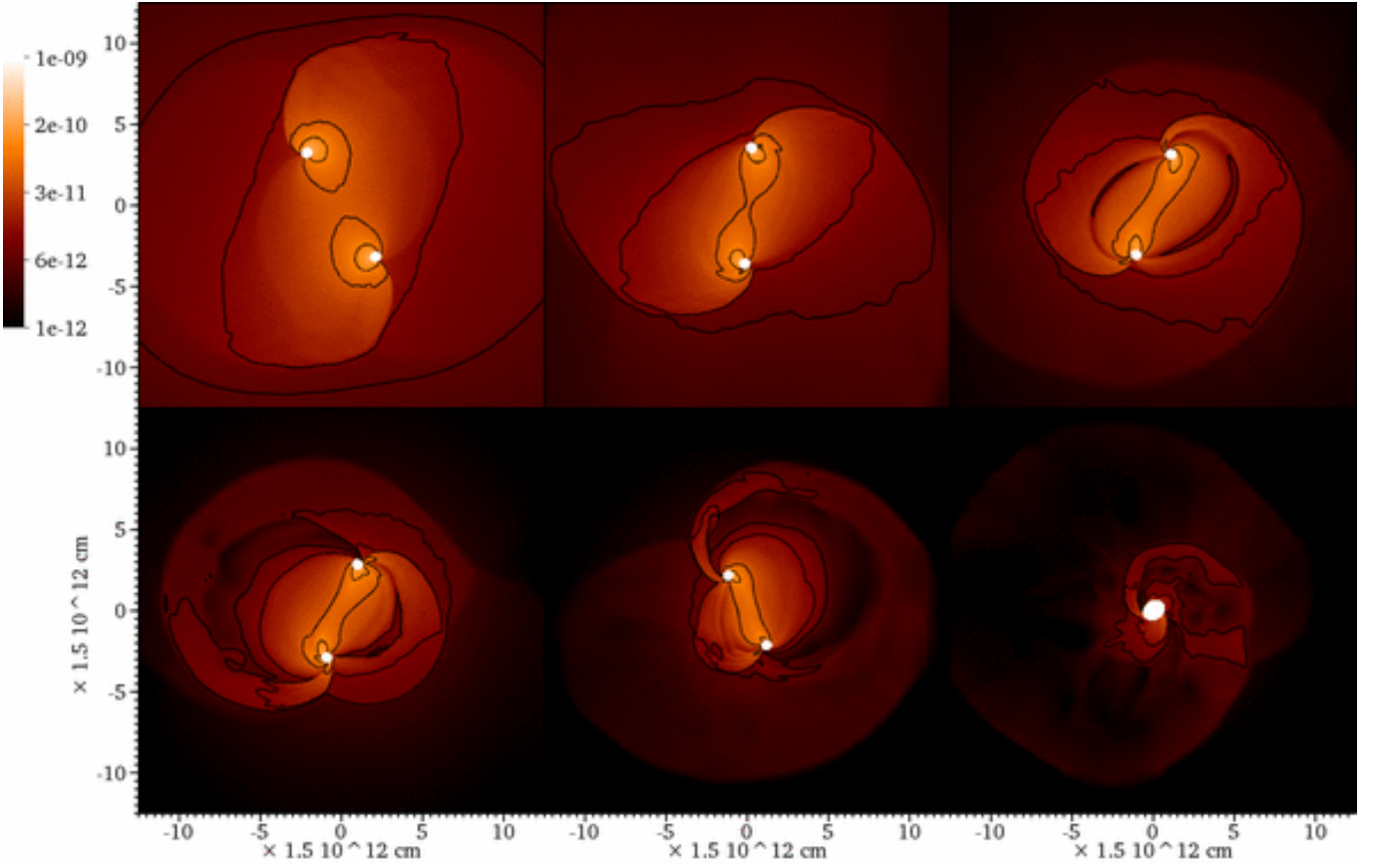}
\caption{Same as in Fig.~\ref{fig:2DrhoG2} but for the run G3. 
From upper left to lower right the snapshots correspond to 
$t = 3.1\times10^3\,M_7\,{\rm s}$, $6.2\times10^3\,M_7\,{\rm s}$, 
$9.2\times10^3\,M_7\,{\rm s}$, $1.2\times10^4\,M_7\,{\rm s}$, 
$1.5\times10^4\,M_7\,{\rm s}$, and $1.8\times10^4\,M_7\,{\rm s}$
respectively.}
\label{fig:2DrhoG3}
\end{figure*}

\begin{figure*}[tbp]
\bigplot{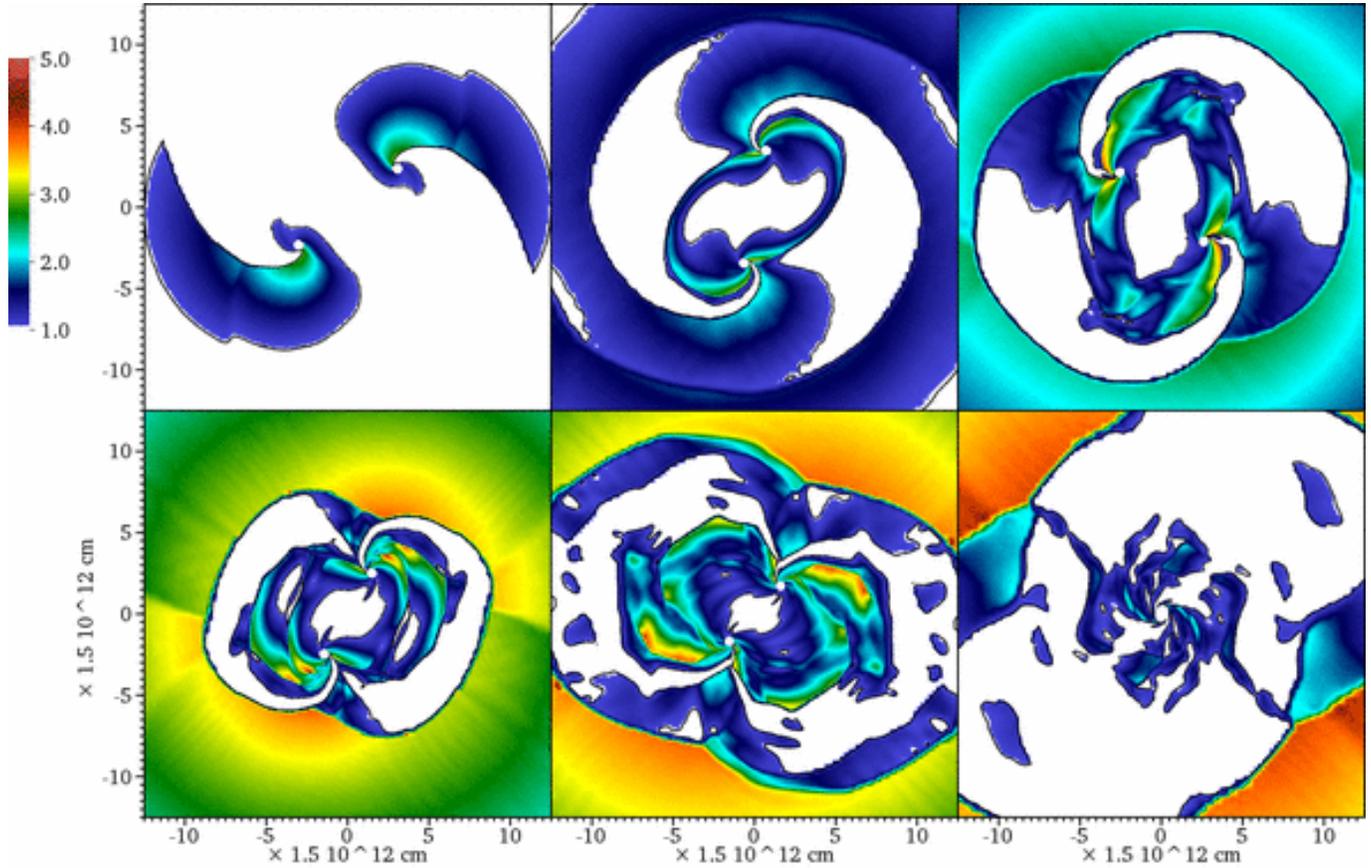}
  \caption{Snapshots of the Mach number, $\beta_s = v/c_s$, measured
    in the orbital plane of the binary in run G2, shown at the same
    times as in Fig.~\ref{fig:2DrhoG2}.  Only values $\beta_s \geq 1$
    are shown. The color scale is linear ranging from light (blue) for
    $\beta_s = 1$ to dark (red) for $\beta = 5$, with the black 
    contour line
    marking the location of $\beta_s = 1$.}
\label{fig:2DMachG2}
\end{figure*}

\begin{figure*}[tbp]
\bigplot{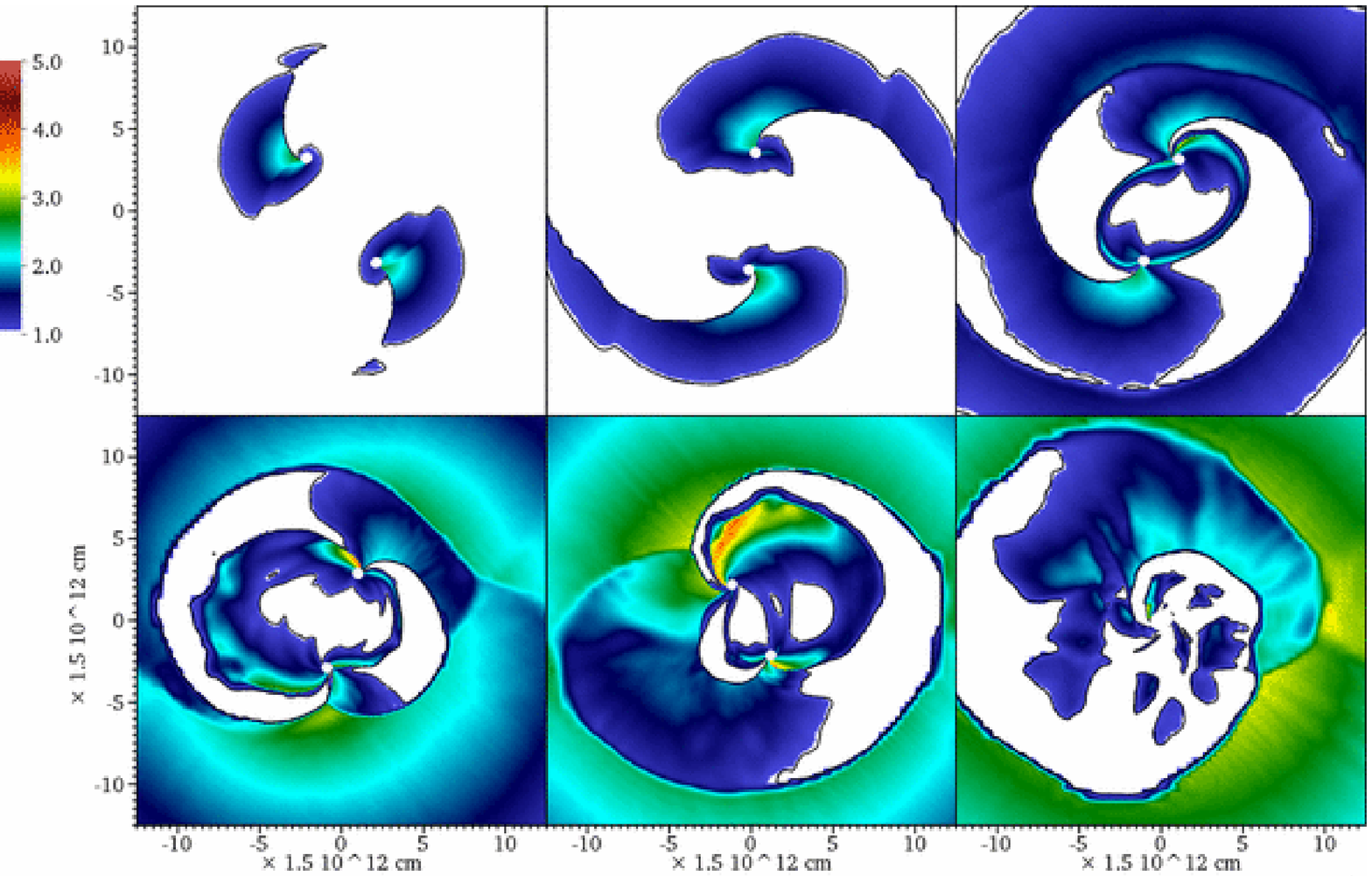}
\caption{Same as in Fig.~\ref{fig:2DMachG2}  but for the run G3.}
\label{fig:2DMachG3}
\end{figure*}

In Figures~\ref{fig:2DrhoG2} and \ref{fig:2DrhoG3} we show snapshots
of the baryonic rest mass density for the cases G2 and G3.  The
distinct features that arise in the gas during the binary's evolution
are the density waves that develop behind the inspiraling \bh{s} and a
bar of high density gas connecting the two \bh{s}. This pattern of
density distribution of gas is not unique to relativistic systems and
was previously noticed in simulations of binary systems interacting
with gas on a variety of scales, including binary galactic nuclei and
TTauri stars. In order to examine whether the gas dynamics driven by the
binary give rise to shocks, we calculate the Mach number, $\beta_s =
v/c_s$, by comparing the velocity of the fluid element to the speed of
sound measured locally in the simulation\footnote{Though $\beta_s>1$ is
necessary, it is not a sufficient condition to locate shocks (e.g., cold 
fronts).  Given the long cooling timescale of the hot, diffuse gas, 
however, discontinuities in the Mach number within our simulations do 
select regions where shocks are likely to form.}. Figures~\ref{fig:2DMachG2}
and \ref{fig:2DMachG3} show sequences of snapshots with the
distribution of the Mach number value for the gas in the plane of the
binary of runs G2 and G3. In all of the panels, only transonic and
supersonic parts of the flow were plotted, where $\beta_s \geq
1$. Early in the inspiral, the shocks are confined to the density
wakes expanding externally to the binary orbit. Later in the inspiral,
the bar of high density gas forms between the \bh{s}, and its expansion
gives rise to another shock region emanating from the central part of
the binary orbit. As the \bbh{} evolves closer to merger, the central
high density region decreases in size, being replaced after the merger
by a steadier inflow in the immediate vicinity of the final \bh{.}

Here we briefly make a note of the degree to which baryonic mass is
conserved in our simulated binary systems, as discussed in the context
of a TOV star at the end of \S\ref{sec:methods}. We monitor the
conservation of mass in two separate volumes: (1) the volume $r \le
8\,M$ excluding the regions inside the \ahz{s} and (2) the volume
$8\,M \le r \le 20\,M$, external to the binary orbit. This allows us
to calculate the total error in mass conservation over the whole
computational domain but also to more closely monitor the inner region
containing the binary. From run G0 we measure that the baryonic mass is
conserved in both volumes at the level of $< 1\%$ over the duration of
the simulation. Due to similarities in computational setup and the 
physical systems, similar mass conservation is expected in the other runs.

\section{Electromagnetic and Gravitational Wave Signatures}\label{sec:emo}

In this section, we discuss the \EM{} and \gw{} signatures expected to
arise from the binary dynamics and accretion flows during the late inspiral,
plunge, and merger of the \bh{s}.  We analytically estimate the
characteristic timescales and emission properties of the hot gas in
Section~\ref{S_properties}. In Section~\ref{S_brems}, we evaluate the
bremsstrahlung emission from our simulations and compare it to the
emission powered by accretion onto the two \bh{s}
(Section~\ref{S_accretion}). In Section~\ref{S_gw}, we characterize the
\gw{} signatures and highlight combined EM+GW signatures that may
enable us to distinguish binary systems from isolated \bh{s}. Finally, in
Section~\ref{S_observability} we discuss the observability of coalescences
in the gas cloud scenario.

\subsection{Characteristic timescales and properties of the gas}\label{S_properties}

We compare the physical properties of the hot plasma in our
simulations with the hot flows that are expected to arise in a
fraction of \agns{s} as radiatively inefficient accretion flows
(RIAFs). From these characteristic timescales, we find that the gas
flow in our simulations in some respects resembles RIAFs. We do not,
however, carry out a self-consistent simulation of the multi-component
plasma or radiative transfer in such flows as such a detailed
treatment would be beyond the scope of this study.  The main property
of radiatively inefficient flows is that very little energy generated
by accretion and turbulent stresses is radiated away, being instead
stored as thermal energy in the gas; this gives rise to a very hot
flow \citep{ichimaru77,rees82,ny94}. Since thermal pressure forces
within the gas are significant, the hot accretion flows are expected
to be geometrically thick. Given high thermal velocities, the inflow
speeds in the flow are comparable to the speed of sound and the
orbital velocity that a test particle would have at a given radius.

Below some critical gas density, the Coulomb collision time in
RIAF-type flows becomes longer than the inflow time of the gas,
resulting in a two-temperature flow in which the ion plasma remains at
$T_p\sim 10^{12}\,{\rm K}$ and the electron plasma cools to
temperatures in the range $T_e\sim10^{10}-10^{12}\,{\rm K}$. Since
electrons are more efficient radiators, the observed radiation will
thus be determined by the properties of the electron population. In
order to determine whether the hot plasma in the cloud developed a
two-temperature flow, we evaluate the characteristic timescales for the
inflow of gas $t_{\rm inflow}$ and Coulomb collisions $t_{\rm
Coulomb}$ in the vicinity of the binary. We estimate that for the gas
cloud under consideration
\begin{eqnarray}
t_{\rm inflow} & \approx & 0.4\,{\rm hr} \left(\frac{R}{10M} \right) 
\left(\frac{c_s}{0.3c}\right)^{-1} , \\
t_{\rm Coulomb} & = & \frac{1}{n\,\sigma c_s} \nonumber \\
   & \approx & 0.5\,{\rm hr} \left(\frac{\rho}{10^{-11} {\rm g\, cm^{-3}}}\right)^{-1} 
\left(\frac{c_s}{0.3c}\right)^{-1} \left(\frac{T_e}{10^{10}\,{\rm K}}\right)^2 \,.
\end{eqnarray}
The size of the region under consideration here, $R$, is comparable 
to the Bondi radius of gravitational influence of the \bh{}
binary. $n \approx n_p \approx n_e \approx \rho/m_p$ is the number
density of the gas, $c_s$ is the speed of sound, $\rho$ is the
characteristic density of the cloud, and $\sigma\approx 0.3 \,Z^2 e^4
/ (kT_e)^2$ is the cross-section for Coulomb scattering of an electron
with kinetic energy $\sim kT_e$ on a more massive ion, evaluated for a
population of electrons with $T_e\sim10^{10}\,{\rm K}$.
  
Because $t_{\rm Coulomb} \gtrsim t_{\rm inflow}$ for the gas
considered here, we conclude that the gas cloud region can be
described as a two temperature flow with $T_e < T_p$. However, given
that the two timescales are not far apart in the center of our cloud,
the electrons and ion plasma may remain weakly coupled and, via
occasional scattering with ions, electrons near the \bh{s} may largely
preserve the thermal energy distribution.  Thus, we estimate the
bremsstrahlung luminosity from a thermal distribution of electrons
\citep{rl86} 
\begin{eqnarray}
L_{\rm brem} & \approx & 4\times 10^{44}\,{\rm erg\,s^{-1}} \left(\frac{\rho}{10^{-11} 
{\rm g\, cm^{-3}}}\right)^2 \left(\frac{R}{10M}\right)^3 \nonumber \\
 & &  M_7^3 \left(\frac{T_e}{10^{10}\,{\rm K}}\right)^{1/2} 
     \left[ 1 + 4.4\times\left(\frac{T_e}{10^{10}\,{\rm K}}\right) \right]_{5.4}\,. 
\label{eq_Lbrem}
\end{eqnarray}
where subscript ``5.4'' indicates that a numerical factor in the
square brackets is normalized to 5.4.  For comparison, the Eddington
luminosity of the system is $L_{Edd}\approx 1.3\times10^{45}\,{\rm
erg\,s^{-1}}$. We therefore estimate that thermal bremsstrahlung is
the dominant emission mechanism of the hot flow, as the synchrotron
radiation and inverse Compton scattering are comparably
smaller\footnote{Synchrotron and inverse Compton scattering
luminosities were derived from the standard expressions; see
\citep{rl86}, for example.}:
\begin{eqnarray}
L_{\rm synchro} & \approx & 8\times10^{36}\,{\rm erg\,s^{-1}} 
\left(\frac{\rho}{10^{-11} {\rm g\, cm^{-3}}}\right)
\left(\frac{R}{10M}\right)^3 \nonumber \\
& & \left(\frac{B}{1G}\right)^2  M_7^3 \label{eq_Ls} \\
L_{\rm IC} & \approx & 3\times10^{-8}\, L_{\rm soft} 
\left(\frac{\rho}{10^{-11} {\rm g\, cm^{-3}}}\right)
\left(\frac{R}{10M}\right)^3  \nonumber \\
& & \left(\frac{R_{\rm tran}}{10^5 M}\right)^{-2} 
M_7\label{eq_Lic}
\end{eqnarray}
where the relativistic factor $\beta=v/c$ and Lorentz factor $W$ have
been evaluated for $v/c \approx 0.3$.  Here $L_\mathrm{soft}$ is 
a supply of low energy photons transported from the edge of the RIAF,
a distance of $R_\mathrm{tran}$ away.

Note that the luminosity of the synchrotron emission remains below
that of the bremsstrahlung radiation unless the magnetic field
strength is close to the equipartition value, which in our case is
$B_{\rm equip}\sim 10^5\,G$. Because our simulations are purely
hydrodynamical, we have no means of constraining the magnetic field
strength and its dynamics around the coalescing binary.  Thus in
estimating $L_{\rm synchro}$, we have assumed that magnetic fields are
sufficiently weak and that the synchrotron component is not the
dominant one. 

Similarly, in order for the inverse Compton luminosity to be
significant, a supply of soft, lower energy photons, presumably
produced externally to the hot binary accretion flow, is required. In
estimating $L_{\rm IC} $, we assumed that this soft photon component
is produced at large radii, where radiative cooling is efficient and
the geometrically thick flow described here transitions into a
geometrically thin accretion disk or a lower temperature ambient
medium. In the case of Sgr~${\rm A}^\star$ for example, observations
indicate that a radiatively inefficient accretion flow extends into
the ambient medium out to distances $R_{\rm tran} \sim 10^5\,M$ away
from the center~\citep{quataert03}. The estimate for $L_{\rm IC}$
obtained with this value of the transition radius implies that inverse
Compton scattering is a very inefficient process even if a generous
supply of low energy photons is available from the ambient medium,
parametrized here in terms of the luminosity $L_{\rm soft}$.

Note that a subsequent paper to the current work \citep{Farris:2009mt}
suggests that the spinning \bh{s} could amplify the magnetic fields in
their vicinity by several orders of magnitude to nearly the
equipartition value.  In the context of a scenario considered by
\citet{Farris:2009mt}, where the magnetic field strength is about an
order of magnitude less than at equipartition (i.e., $B\sim 10^4\,G$
given the properties of our gas cloud), the estimate for luminosity of
synchrotron radiation becomes $L_{\rm synchro}\sim 10^{45}\,{\rm erg\,s^{-1}}$ 
(see equation~\ref{eq_Ls}).  In this scenario, not only is $L_\mathrm{synchro}$ 
comparable to $L_\mathrm{brem}$, but the synchrotron radiation could feed the
inverse Compton luminosity of comparable magnitude (equation~\ref{eq_Lic}) by 
providing an immediate source of soft photons.  In the present work, however, 
we have no means of judging the strength of the magnetic field in the immediate vicinity 
of the binary and we focus most of our discussion on bremsstrahlung emission, 
keeping in mind the potential importance of the other two emission mechanisms. 

The corresponding cooling timescale of the plasma at  $T_p\sim10^{12}$~K due to
bremsstrahlung radiation from $T_e\sim10^{10}$~K electrons is
\begin{eqnarray}
t_{\rm cool} & \sim & 8\,{\rm hr} 
\left( \frac{\rho}{10^{-11} {\rm g\, cm^{-3}}}\right)^{-1} 
\left(\frac{T_p}{10^{12}\,{\rm K}} \right)
\left(\frac{T_e}{10^{10}\,{\rm K}} \right)^{-1/2} \, .
\end{eqnarray}
Note that $t_{\rm cool} > t_{\rm Coulomb} > t_{\rm inflow}$
implies that the hot gas plunges into the \bh{s} before it had a
chance to radiatively cool and settle into an accretion disk, thus
justifying the initial assumption of a hot, geometrically thick gas
cloud and implying that radiative cooling of the cloud can indeed be
neglected in this case. It is also worth noting that the expressions
presented in this section scale with density and thus can be used to
estimate luminosity components from a lower density plasma than that 
considered here. The scaling relations however break down for plasma 
densities higher than $10^{-11} {\rm g\, cm^{-3}}$ because in this regime 
Coulomb collisions become sufficiently frequent as to
thermalize the electrons and produce a single temperature plasma
flow. Once the plasma flow of electrons and ions is fully thermally
coupled, it can cool efficiently via electron-emitted radiation,
yielding an evolution more similar to the accretion disk scenario.

\subsection{Bremsstrahlung emission from the gas}\label{S_brems}

We now evaluate the characteristic luminosity arising from the gas
near the \bbh{} in our simulations. At a given time step during
the simulations, we calculate the local thermal bremsstrahlung
emissivity ${\cal E}_{\rm brem}$ (i.e., luminosity per unit volume)
from~\cite{rl86} as
\begin{eqnarray}
{\cal E}_{\rm brem} & = & 2.8\times 10^4\,{\rm erg\,s^{-1}\,cm^{-3}} \left(\frac{\rho}{10^{-11} 
{\rm g\, cm^{-3}}}\right)^2 \nonumber \\
 & & \left(\frac{T_e}{10^{10}\,{\rm K}}\right)^{1/2} 
     \left[ 1 + 4.4\times\left(\frac{T_e}{10^{10}\,{\rm K}}\right) \right]_{5.4}\,, 
\label{eq_Ebrem}
\end{eqnarray} 
where we assumed that $T_e = 10^{-2}\,T_p$, with $T_p$ computed from the
internal energy $\epsilon$ in our simulations as
\begin{equation}
\frac{k\,T_p}{m_p c^2} = (\Gamma-1)\,\epsilon\,.
\end{equation}

\begin{figure*}[tbp]
\bigplot{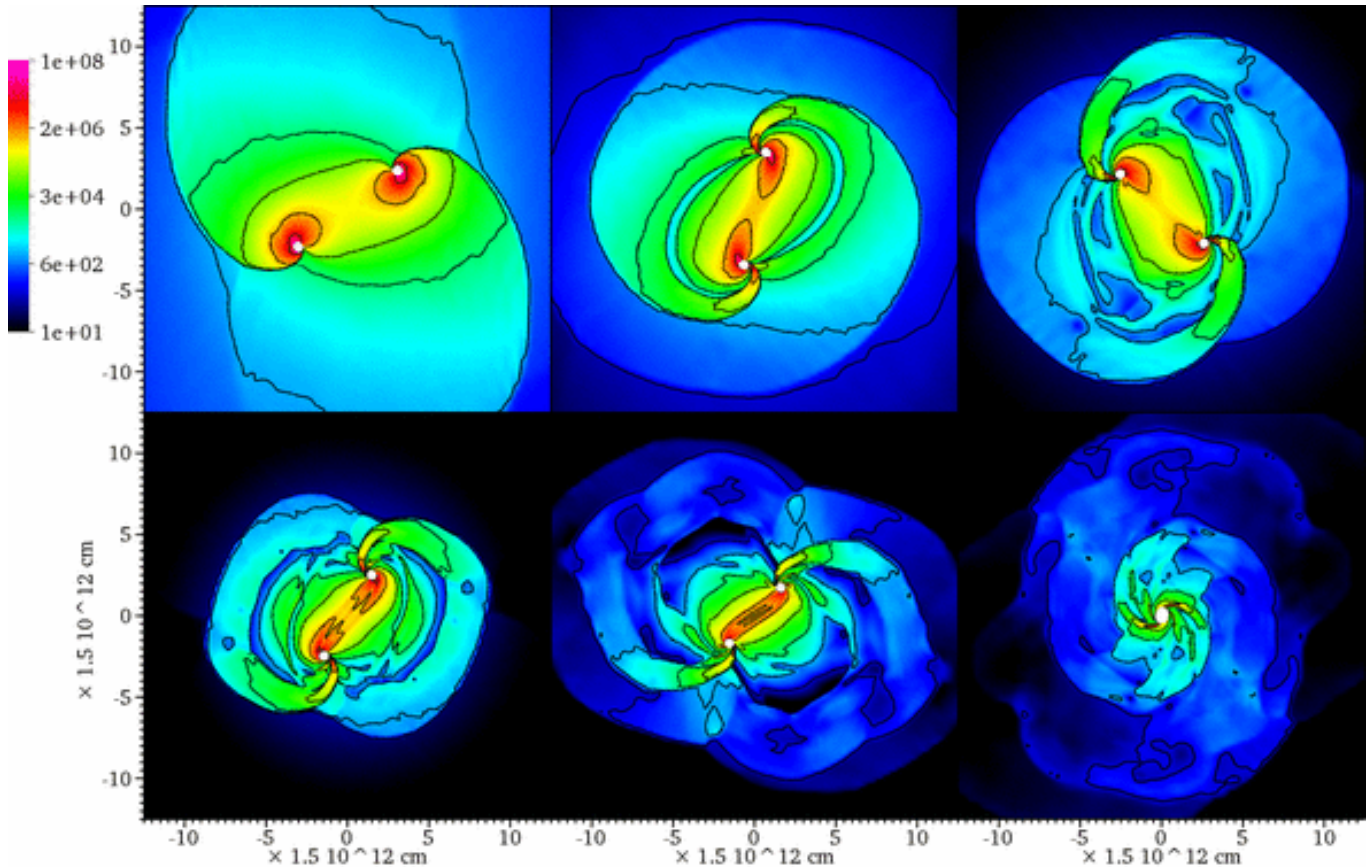}
\caption{Snapshots of the bremsstrahlung emissivity, in units of
$[{\rm erg\,s^{-1}\,cm^{-3}}]$, shown in the orbital plane of the
binary for run G2. The time sequence of snapshots corresponds to
Figures~\ref{fig:2DrhoG2} and \ref{fig:2DMachG2}. The color scheme is
logarithmic and contours are plotted at half order-of-magnitude
intervals.}
\label{fig:BremsG2}
\end{figure*}

\begin{figure*}[tbp]
\bigplot{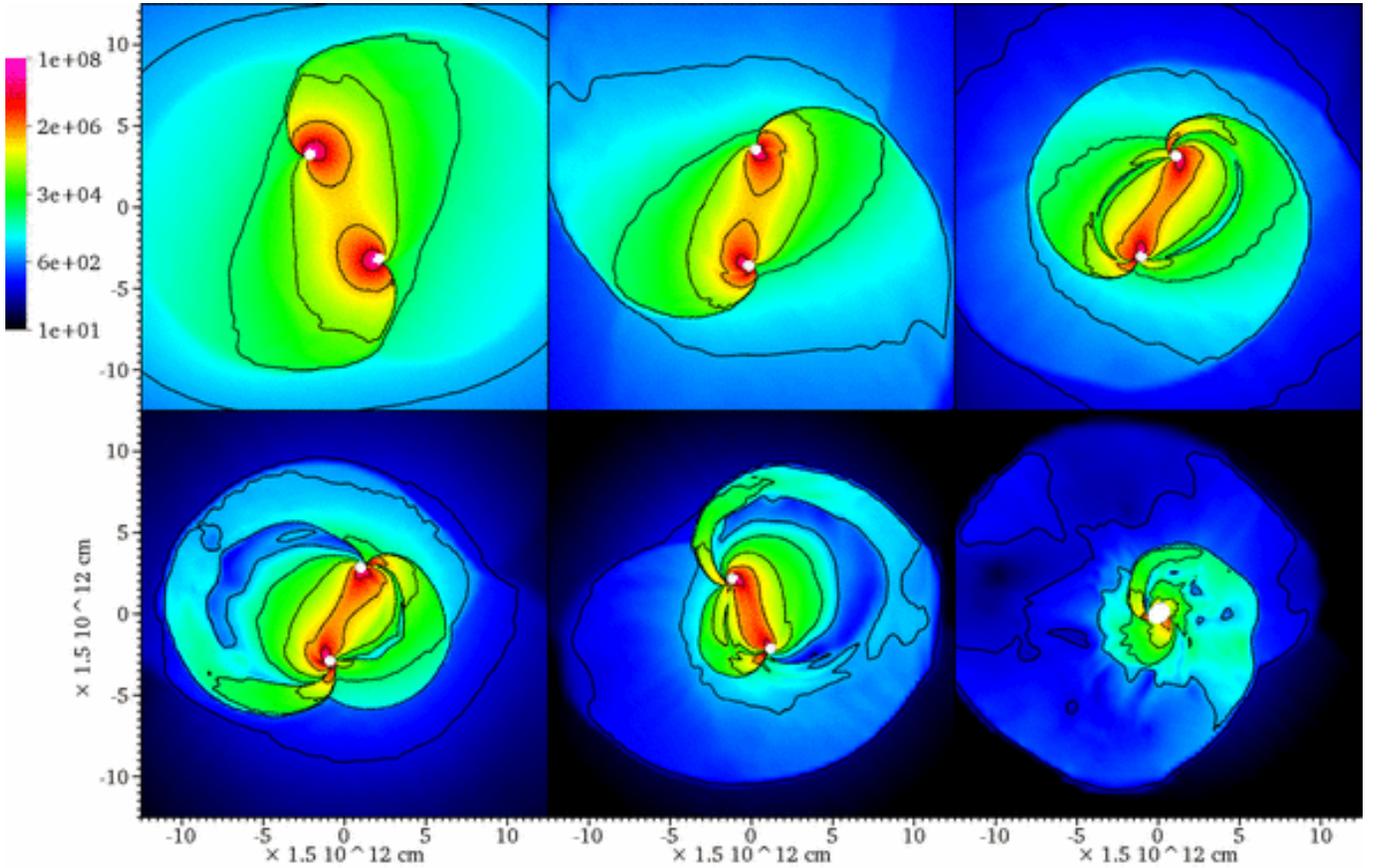}
\caption{Same as in Fig.~\ref{fig:BremsG2} but for the run G3. The
time sequence of snapshots corresponds to Figures~\ref{fig:2DrhoG3}
and \ref{fig:2DMachG3}.}
\label{fig:BremsG3}
\end{figure*}

Figures~\ref{fig:BremsG2} and \ref{fig:BremsG3} show snapshots of the
bremsstrahlung emissivity ${\cal E}_{\rm brem}$ for the gas located in
the plane of a binary in runs G2 and G3. The time sequences of the
snapshots correspond to those in Figs.~\ref{fig:2DrhoG2} --
\ref{fig:2DMachG3} for the density and Mach number.  Not surprisingly,
the highest bremsstrahlung emissivity is found in the regions of high
density, namely at the wakes trailing behind the \bh{s} and, at later
times, around the central bar of gas between the \bh{s}. By comparing
Figure~\ref{fig:BremsG2} (emissivity) with Figure~\ref{fig:2DMachG2}
(Mach number) in runs G2 and G3, it is evident that regions of high
emissivity are closely associated with regions where shocks arise and
expand into the surrounding cloud. In the late stages of the merger,
most of the emission is produced by the bridge of shocked gas between
the \bh{s}. This region decreases in size as the \bh{s} approach
coalescence and is rapidly swallowed by the \bh{s} when they merge to
form a single, final \bh{}. As we will discuss next, this gives rise
to a characteristic variability in the bremsstrahlung light emitted by
the gas.


\begin{figure}[ht]
\includegraphics[width=0.96\linewidth]{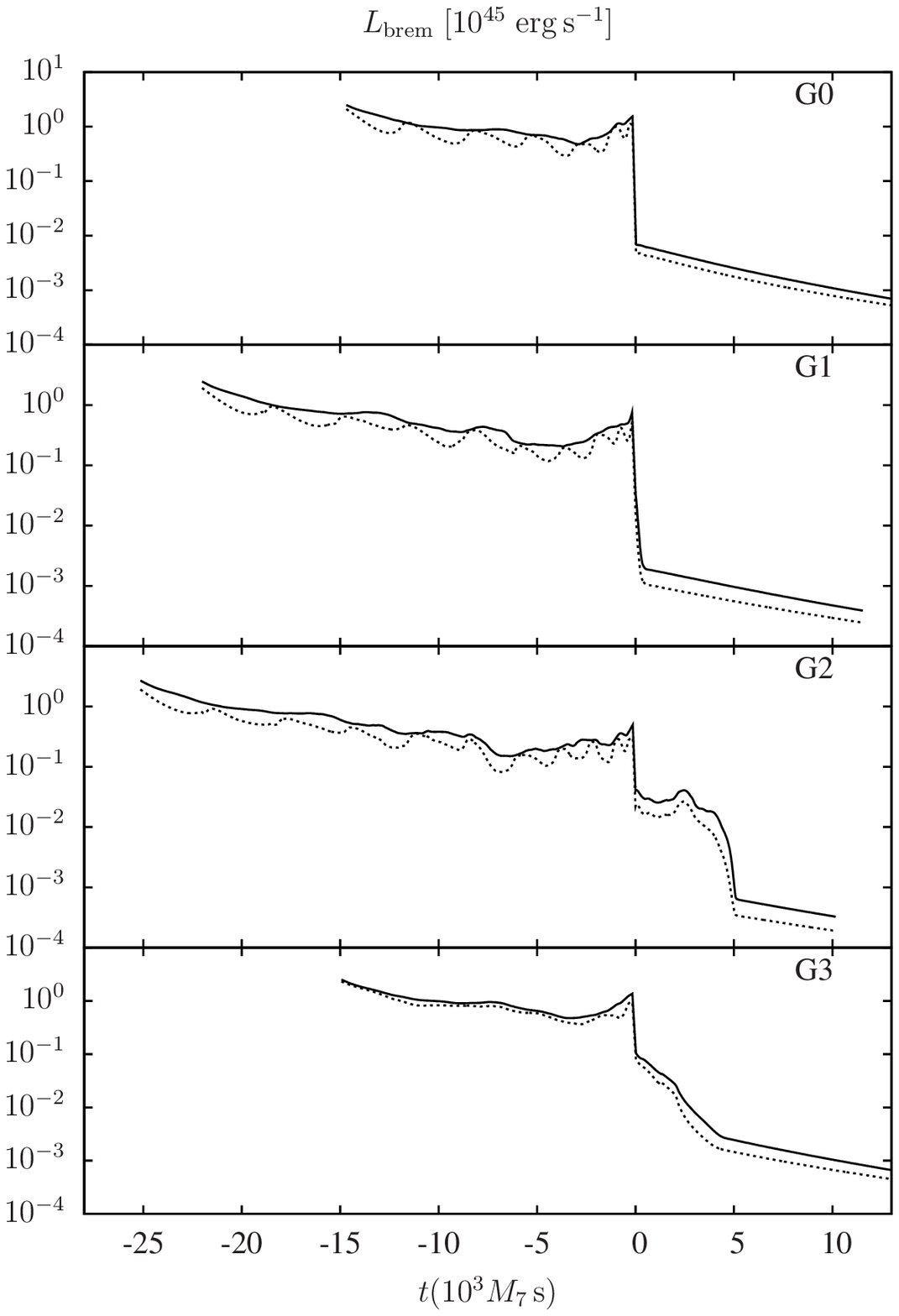}
\caption{Thermal bremsstrahlung luminosity as a function of time
without (solid line) and with relativistic beaming (dashed line),
calculated within a sphere of radius $10\,M$ about the center of mass
of the binary but excluding the region within the \ahz{s}.  Effect of 
beaming is calculated for an observer placed
at infinity, in the orbital plane of the binary.  Merger occurs at $t\sim0$.}
\label{fig:brems1d}
\end{figure}

To obtain the bremsstrahlung luminosity $L_\mathrm{brem}$, we
integrate ${\cal E}_{\rm brem}$ over a spherical volume of radius
$10\,M$, enclosing the binary orbit but excluding the volumes inside
the \ahz{s}.  By repeating this procedure at regular time intervals
during the simulations, we construct a light curve for thermal
bremsstrahlung.  Figure~\ref{fig:brems1d} shows the light curves for
all four runs. In this and subsequent figures, $t = 0$ marks the time
when we first find the \ahz{} of the final \bh{.} We estimate
that the actual merger, i.e. first appearance of a common \ahz{,}
takes place within $5\times10^2\,M_7\,{\rm s}$ before this. 

We first discuss the light curves in Figure~\ref{fig:brems1d}
represented by solid lines. During the early part of the inspiral, $t
\ltsim -5\times10^3\,M_7\,{\rm s}$, the luminosities in all systems 
decay from the 
Eddington luminosity level to ${\rm few} \times 10^{44} \;
\mathrm{erg\,s^{-1}}$. At about $t \sim 5\times10^3\,M_7\,{\rm s}$ 
before the merger,
when the binary enters the final plunge, the luminosity starts
increasing, leading to a broad flare lasting until the binary merges.
The flare reaches its maximum during 
$-5\times10^2\,M_7\,{\rm s} \ltsim t \ltsim 0\,{\rm s}$, a
period where we estimate the \ahz{s} of the two \bh{s} merge. We
measure the maximum luminosity of the flare in G0, G1, G2 and G3 run
to be $L_\mathrm{brem} = \lbrace 15.5, 7.56, 4.96, 13.7 \rbrace \times
10^{44} \; \mathrm{erg\,s^{-1}}$, respectively, and estimate the error
associated with these measurements to be between $5$ and $10\,\%$. Our 
measurements give a lower limit with errors arising because the volume 
integrated luminosity depends on the formation and geometry of the ``common,'' final \ahz{}, whose early
distorted (peanut-like) shape is difficult to localize immediately
after its formation.  As mentioned before, we estimate that the
common, final \ahz{} forms during 
$-5\times10^2\,M_7\,{\rm s} \ltsim t \ltsim 0\,{\rm s}$.  For this
reason, we do not include in Fig.~\ref{fig:brems1d} luminosities
during this time interval.

Another characteristic feature in the luminosity curves depicted in
Figure~\ref{fig:brems1d} is a sudden drop that occurs soon after
the \bh{s} have merged. This feature is most dramatic in run G1 where the
luminosity decreases by nearly 3 orders of magnitude. A sudden
decrease can be attributed to the disappearance of the dynamic region 
of high emissivity between the two \bh{s}, which is rapidly swallowed by 
the \bh{s} in the process of coalescence.

In all cases, the luminosity eventually decays exponentially. The
luminosity in runs G0 and G1 decay in this manner immediately after
merger, but runs G2 and G3 exhibit an additional variability before
the exponential decay. This exponential decay is due to the steady
state accretion of the left-over gas surrounding the final
\bh{}. 

Since $L_{\rm brem}\propto\rho^2$, any density variations within the
emitting volume are mirrored in the calculated luminosity curve. The
variability following the luminosity drop in runs G2 and G3 is
especially interesting since it results in a more gradual drop and
also highlights some differences in the evolution of the gas in these
two runs with respect to runs G0 and G1. Specifically, in run G2,
which is the longest of the four runs, the spiral wakes trailing
behind the \bh{s} have sufficient time to interact with each other,
and, as a result, the final \bh{} finds itself embedded in a highly
turbulent medium. Because turbulence heats the gas and creates density
inhomogeneities in the cloud, this gives rise to the variability
observed in the aftermath of coalescence. Similarly, in run G3, the
asymmetry of the system seeded by the prograde and retrograde spinning
\bh{s}, leads to interactions of the spiral wakes early in the
simulation and gives rise to turbulence and excess luminosity after
the coalescence. Since the medium surrounding the binary can be
expected to be turbulent in realistic cases, we argue that realistic
light curves may resemble runs G2 and G3 more than cases G0 and G1.
Namely, we expect that turbulence would arise naturally in simulated
systems that have been evolved for a longer period of time before the
coalescence.

We now highlight additional features of bremsstrahlung light curves
that arise when relativistic beaming and Doppler boosting are included
in the calculation of luminosity (see the curves plotted with dashed
lines in Fig.~\ref{fig:brems1d}). For simplicity, we neglect
relativistic bending of photon trajectories and gravitational redshift
of photons in the potential well of the binary. We include the special
relativistic Doppler effect by multiplying the broadband
bremsstrahlung emissivity ${\cal E}_{\rm brem}$ with the factor
$D^4=\left(W (1-\beta \cos(\theta))\right)^{-4}$ where $\theta$ is the
angle between the line-of-sight to the observer and the velocity
vector of the gas.  It follows that, depending on the position of an
observer relative to the orientation of the binary, the changing
beaming pattern of the orbiting binary surrounded by emitting gas can
potentially give rise to modulations in the observed luminosity of the
system. To judge the importance of this effect, we evaluate the
bremsstrahlung luminosity for a configuration in which the modulations
are expected to be largest by placing a fiducial observer in the plane
of the binary at infinity. In this way, the observer is placed
directly in the path of the sweeping emission beams associated with
the two \bh{s} and can sample both the minimum and maximum luminosity
of the system, depending on its orbital phase. As shown in
Fig.~\ref{fig:brems1d}, see dashed lines, quasi-periodic oscillations
in luminosity indeed arise in runs G0, G1, and G2 prior to
coalescence. The amplitude of the variations in luminosity is
approximately a factor of $2$ between subsequent peaks and
troughs, oscillating about a value lower than the unboosted luminosity
as boosting into the gas frame reduces the energy of the emitted photon
by a factor $W^{-1}$. Furthermore, it is the beamed emission from the shocks,
launched by the \bh{s} into the surrounding layers of the cloud, that
contribute the most to the modulation of luminosity. This argument is
strengthened when oscillations in runs G0, G1 and G2 are compared with
those in the asymmetric case, G3, where the binary does not form a
stable set of density wakes. As a result, the luminosity oscillations
in run G3 are much weaker (not easily discernible in
Fig.~\ref{fig:brems1d}) and appear at roughly half the frequency seen
in the other cases.

The discussed variability stands a chance of being seen by a distant
observer, as long as photons emitted in the vicinity of the \bh{s} are
not absorbed within the cloud and the surrounding medium or
reprocessed in such a way that the variability signature is lost. To
estimate these effects, we consider separately the optical depth in the 
portion of the gas which serves as the source of bremsstrahlung photons 
and the remainder of the gas cloud.  Assuming that
the electron population at a temperature $T_e\sim 10^{10}\,\mathrm{K}$ will
give rise to a range of photon energies up to $kT_e\sim 1\,\mathrm{MeV}$, 
we estimate the optical depth for Compton scattering as 
$\tau_\mathrm{Compton} \approx n \sigma_\mathrm{T} R$.  For simplicity we 
replaced the cross-section for Compton scattering by that for Thomson 
scattering and neglect a factor of few discrepancy between the two that 
arises for the highest energy photons.  Within the emitting portion of 
the cloud, approximately the central $10\,M$ or $10^{-6}\,\mathrm{pc}$ 
of the system, the number density of the gas cloud ranges from 
$10^{12}\,-\,10^{15}\,\mathrm{cm^{-3}}$ resulting in an optical depth of 
the order $10\,-\,10^4$. By contrast, the number
densities outside this region drop to order $10^{11}\,\mathrm{cm^{-3}}$ 
and below, implying an optical depth of order unity and smaller.
Therefore, if the photon escapes the region of emission it is likely that
the variability in luminosity can be seen by an observer.

\subsection{Accretion onto the black holes}\label{S_accretion}

\begin{figure}[ht]
\includegraphics[width=0.96\linewidth]{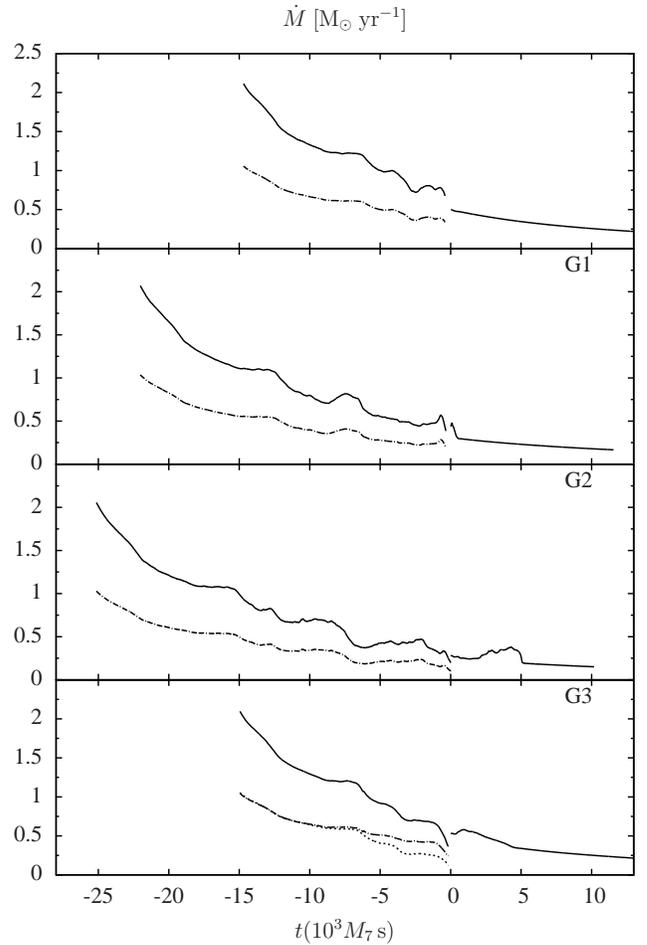}
\caption{Accretion rates across the \ahz{s}.  Dash-dot lines mark the
  accretion rates onto individual \bh{s} before the merger, $t <
  0\,{\rm s}$. Accretion curves of pre-merger \bh{s} in runs G0, G1, 
  and G2
  overlap due to identical accretion rates. In G3, the dash-dot line
  marks the accretion rate onto the prograde-spinning \bh{} and the
  dotted line marks the accretion onto the retrograde-spinning \bh{}.
  In all cases solid line shows the total accretion rate of the system
  onto the \ahz{s}. Merger occurs at $t\sim0$.}
\label{fig:Accretion}
\end{figure}
\begin{figure}[ht]
\begin{center}
\includegraphics[width=0.96\linewidth]{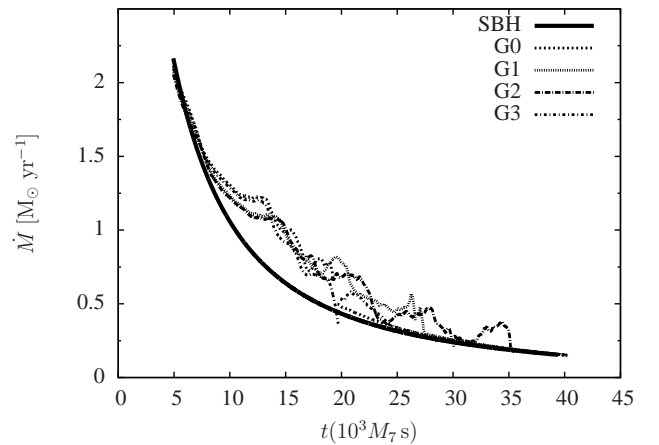}
\caption{Accretion rate of gas onto the single spinning \bh{} 
(solid line) and the total (sum of individual \bh{}) accretion 
rates for the binary runs G0-G3 (dashed lines), again omitting
the initial transients.}
\label{fig:sbh_accretion}
\end{center}
\end{figure}
In this section we discuss the properties and structure of 
the accretion flows modeled in our simulations.
Fig.~\ref{fig:Accretion} shows the accretion rates measured
across the apparent horizons of the \bh{s}.  The dash-dot lines at $t
< 0\,{\rm s}$ mark the accretion rates onto the individual \bh{s} 
before the
merger. Since the accretion rates of the inspiraling \bh{s} in runs
G0, G1 and G2 are identical, only one curve is visible.  In run G3,
the dash-dot line at $t < 0\,{\rm s}$ marks the accretion rate for the
prograde spinning \bh{} and the dotted line that of the retrograde
spinning \bh{} (prograde direction is defined as a spin parallel to
the orbital angular momentum). In all runs, the solid line represents
the sum of the accretion rates of the two \bh{s} before the merger and
the accretion rate onto the final \bh{} after the merger.

It is prudent here to separate any effects on the accretion rate that 
stem from our simple assumptions about the initial structure of the
gas cloud from physical effects that arise as a consequence of the 
binary dynamics.  To do so we consider the evolution of the same 
gas cloud, replacing the \bbh{} by a single stationary \bh{} with a mass of 
$1\,M\simeq 10^7\,\msun{}$ and spin $a/M=0.62$. This choice of mass
and spin approximates the final \bh{} in the G0 case.  We overlay
the accretion rate for this single \bh{} case and the binary runs in
Fig.~\ref{fig:sbh_accretion} leaving the time axes unshifted such that 
the starting moment of each simulation coincides (not the moment of 
coalescence as in previous figures), again
omitting the initial transients.
In all cases, the average gas density in the center of the gas 
cloud decreases exponentially as the gas cloud of finite size is swallowed 
by the BHs. This decline is a consequence of our choice of initial 
conditions and should not be regarded as a prediction of the simulations,
as in reality gas may be continuously supplied to the \bh{} and a 
leveled accretion rate maintained over longer periods of time.
On the other hand, the excess variability in accretion rate noticeable
in the four binary scenarios is a consequence of the binary motion
which stirs the gas, causing turbulence.  It is this variability 
that we regard as a true signature of the binary.

The two orbiting \bh{s} accrete from a hot, turbulent flow in a
Bondi-like fashion. We measure the total accretion rate of the \bh{s}
in the simulations to be in the range $0.2-2\,\msun\,{\rm yr^{-1}}$,
in good agreement with the analytic expectation for Bondi accretion
rate of this system
\begin{equation}
\dot{M}_{\rm B} \approx 0.84\,M_{\odot}\,{\rm yr^{-1}} 
\left(\frac{c_s}{0.3c}\right)^{-3} 
\left(\frac{\rho}{10^{-11} {\rm g\, cm^{-3}}}\right) M_7^2 \, , 
\label{eq:bondi}
\end{equation}
where we assumed that the relative velocity between the gas and the
\bh{} is comparable to $c_s$. At this rate of accretion, the gas
residing in the nuclear region of size $\lesssim 0.01$~pc and mass
$1\%$ of the \bh{'s} mass will be accreted in $\sim 10^5$~yr. In order
for the gas flow to persist uninterrupted on longer time scales, gas
needs to be supplied from larger radii, either via an accretion disk
or by accretion from the interstellar medium.

A time scale of $\sim 10^5\,{\rm yr}$ is, in principle, long enough
for the angular momentum of the accreted gas to have an effect on the
orientation of the \bh{} spins\footnote{Note that in the reference
  frame co-rotating with the binary the gas has a rotational velocity
  component, even though its velocity distribution is initially chaotic in the
  frame of the observer.}. Such coupling between the angular momentum
of the accreted gas and the \bh{} rotation can lead to precession of
their spin axes and possibly a partial alignment with the orbital
angular momentum of the binary. This effect, however, is expected to
be more efficient in accretion disks of small and moderate thickness
\citep{np98,na99,king05,lp06,lp07, fragile07, bogdanovic07,
  perego09,dotti09}. Since the time scale over which we follow the
binary and gas in our simulations is much shorter, we capture no spin
alignment effects in our study and the \bbh{} dynamics in all of our
simulations is essentially indistinguishable from the equivalent
vacuum case.

\begin{figure}[ht]
\bigplot{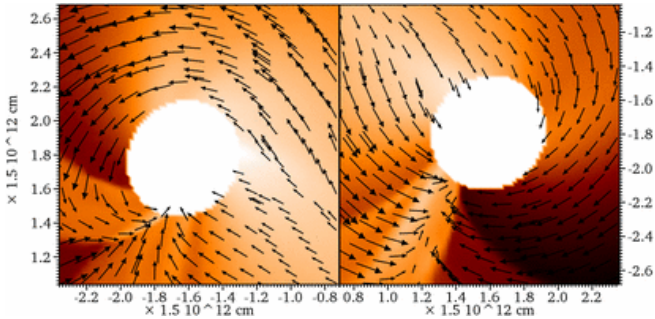}
\caption{Velocity field of the gas plotted in the vicinity of the prograde- 
\bh{} (left) and retrograde-spinning \bh{} (right) 
   $\sim3.3\times10^3\,M_7\,{\rm s}$ before merger, 
   a point at which the accretion rates differ by a factor of $\sim2$.}
\label{fig:2dAccretion}
\end{figure}
\begin{figure}[ht]
\includegraphics[width=0.96\linewidth]{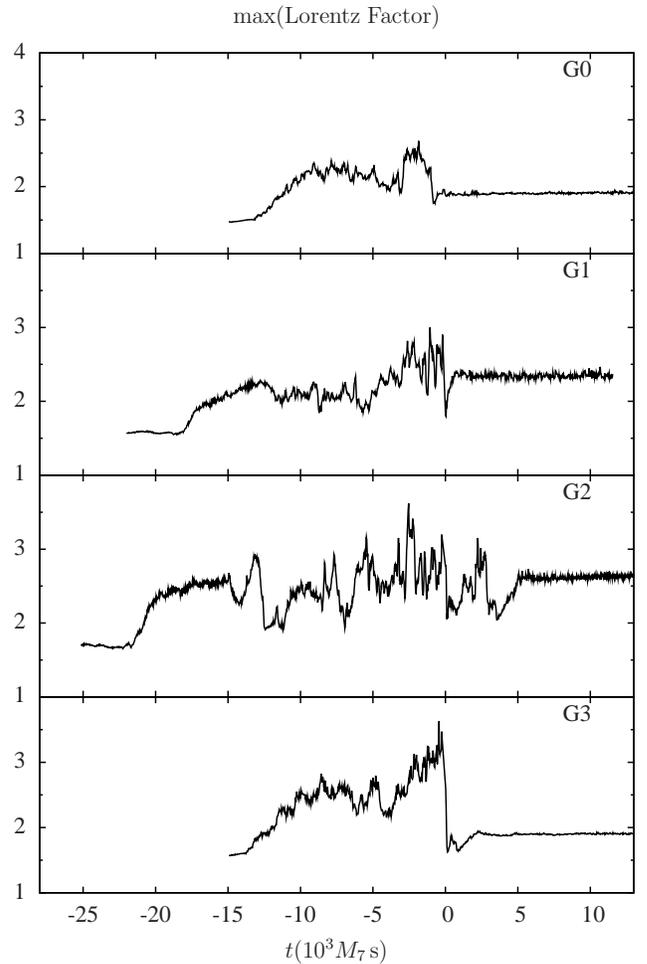}
\caption{Maximum Lorentz factor of the gas outside the \ahz{s}.
  Post-merger values correspond to a steady Bondi-like accretion.
  Merger occurs at $t\sim0$.}
\label{fig:MaxW}
\end{figure}
Since the \bh{s} in our simulations accrete in a Bondi-like fashion,
the expectation is that they draw matter from a region enclosed within
their Bondi radius, $R_{\rm B} = GM/c_s^2$, which approximately
delineates the zone of gravitational influence of a \bh{}.  While the
Bondi radius does not explicitly depend on the spin of a \bh{,} we
find from run G3 that the spin orientation of the merging \bh{s} does
affect their accretion rates. In G3, the rates of accretion of the two
\bh{s} start diverging from each other at approximately $\sim
6.4\times10^3\,M_7\,{\rm s}$ 
before the coalescence and reach a factor of $\sim 2$ difference at
the plunge.  This difference arises as a combination of the frame
dragging and geometry of the density wakes in the vicinity of the
\bh{s}. The panels in Fig.~\ref{fig:2dAccretion} highlight the
differences in the velocity field of the gas around the prograde- and
retrograde-spinning \bh{s} at a time, $\sim3.3\times10^3\,M_7\,{\rm s}$
before merger, when
the difference in accretion rates is highest.  Due to the orbital
motion of the binary, the relative motion of the gas flow in
Fig.~\ref{fig:2dAccretion} appears to be from right to left at the
position of the prograde-spinning \bh{} and from left to right for the
retrograde-spinning \bh{}. A fraction of the gas near each \bh{} tends
to co-rotate with the spin of the \bh{}.  For both \bh{s} the effect
of frame dragging is such that the relative velocity of the gas with
respect to the \bh{} is effectively lower just below the \bh{} and
higher just above it, when compared to the mean velocity of the flow.
This low velocity spot is, according to Eq.~\ref{eq:bondi}, where most
of the accretion is favored to occur across the \ahz{s} of the
\bh{s}. In the case of the prograde-spinning \bh{,} some of the gas
plunges directly into the \bh{} and some rotates around and is
fed to it through the trailing wake. In the case of the
retrograde-spinning \bh{,} the gas is compressed into the front of the
density wake which is then less effective at channeling the gas into
the \bh{}, leading to the lower accretion rate visible in
Fig.~\ref{fig:Accretion}. Nevertheless, given the relatively weak
implicit dependence of the Bondi accretion rate on the \bh{} spin, it
is unlikely that one would be able to infer the spin magnitude or
orientation in this mode of accretion based on the accretion powered
luminosity curve alone.

The discontinuities in the accretion rates observed in
Fig.~\ref{fig:Accretion} at the time of merger occur due to the error
in locating the common, initially highly deformed, \ahz{} of the final
\bh{.} This is the same source of error that we discussed in the
context of the maximum bremsstrahlung luminosity measured at merger
time. Notice that for $t>0$, the exponential decay in the
bremsstrahlung luminosity is also mirrored here as an exponential
decay in the accretion rate of the final \bh{}.  Similarly, the
post-merger variability present in runs G2 and G3
(Fig.~\ref{fig:brems1d}) is also repeated in the accretion rates. As
in the case of the luminosity, this behavior can be explained by the
more turbulent flows of runs G2 and G3. It is also evident that the
accretion rates measured in runs G1 and G2 fall below that in run
G0. Since the orbital hang-ups of runs G1 and G2 result in a longer
inspiral, the two \bh{s} have more time to deplete the surrounding
cloud of gas, leading to a lower density of gas in the vicinity of the
binary and consequently lower accretion rates at later times. This is
also consistent with the behavior of the bremsstrahlung luminosity
curves which are, during the post-merger phase of exponential decay,
lower for the spinning-\bh{} cases which exhibit a hang-up, G1 and G2,
than for the non-spinning case G0.  While details of the exponential
decay phase can be dependent on our choice of initial conditions and
the structure of the cloud, the result that systems with a higher net
spin magnitude may appear dimmer with respect to their non-spinning
counterparts is likely to be real. Without a knowledge of the gas
density in the vicinity of \smbh{} binaries, though, it would be hard
to infer the net value of the spin for a given system based solely on
the inferred accretion curve.

We now briefly discuss our results in the context of
earlier works that considered binary coalescence in non-vacuum,
specifically that of \citet{vanmeter09}. We compare the global maximum
Lorentz factor of the gas calculated from our simulations
(Fig.~\ref{fig:MaxW}) with that in Fig.~2 of \citet{vanmeter09},
corresponding to their case where test particles have a random and
isotropic velocity distribution.  \footnote{Note that we evaluate the 
Lorentz factor given the velocity of a parcel of fluid, while \citet{vanmeter09} 
evaluate the {\it collisional} Lorentz factor from relative velocities of converging 
particles.} Since it is likely that the maximum Lorentz factor is found in the 
immediate vicinity of the \bh{} horizons, in Fig.~\ref{fig:MaxW}, we do not 
include the maximum Lorentz factor during 
$-5\times10^2\,M_7\,{\rm s} \ltsim t \ltsim 0\,{\rm s}$ because of, as
mentioned before, the uncertainty in determining the region outside the
final \ahz{.}

We find that the maximum Lorentz factors measured for the non-spinning 
and spinning binary cases during the inspiral phase appear in good qualitative 
agreement in our respective works. In our simulations, the ramp-up in 
the Lorentz factor during the inspiral and up to coalescence is 
reflected in the broad peak in bremsstrahlung luminosity for all runs, 
thus confirming that the increase in bremsstrahlung luminosity is due to 
shocks that arise in the vicinity of the binary. Moreover, we
confirm that the spinning systems (G1 and G2 cases), on average, have a
higher maximum Lorentz factor than the non-rotating G0 case. This
behavior can be explained as follows: in the case of spinning \bh{s},
the fluid can travel deeper into the potential well of a \bh{} without
being accreted; as a consequence, the fluid can emerge with a higher
kinetic energy compared to the non-spinning case. This interpretation
is correct as long as the fluid moves freely along a geodesic and
before it encounters pressure forces from the surrounding medium. We,
however, do not find that overall higher Lorentz factors in the merger of
rotating \bh{s} result in higher luminosities. This is because the spinning
\bh{s} tend to deplete their surroundings of gas more readily during
the orbital hang-up phase and, in such a way, suppress the luminosity.

\begin{figure*}[tbp]
\bigplot{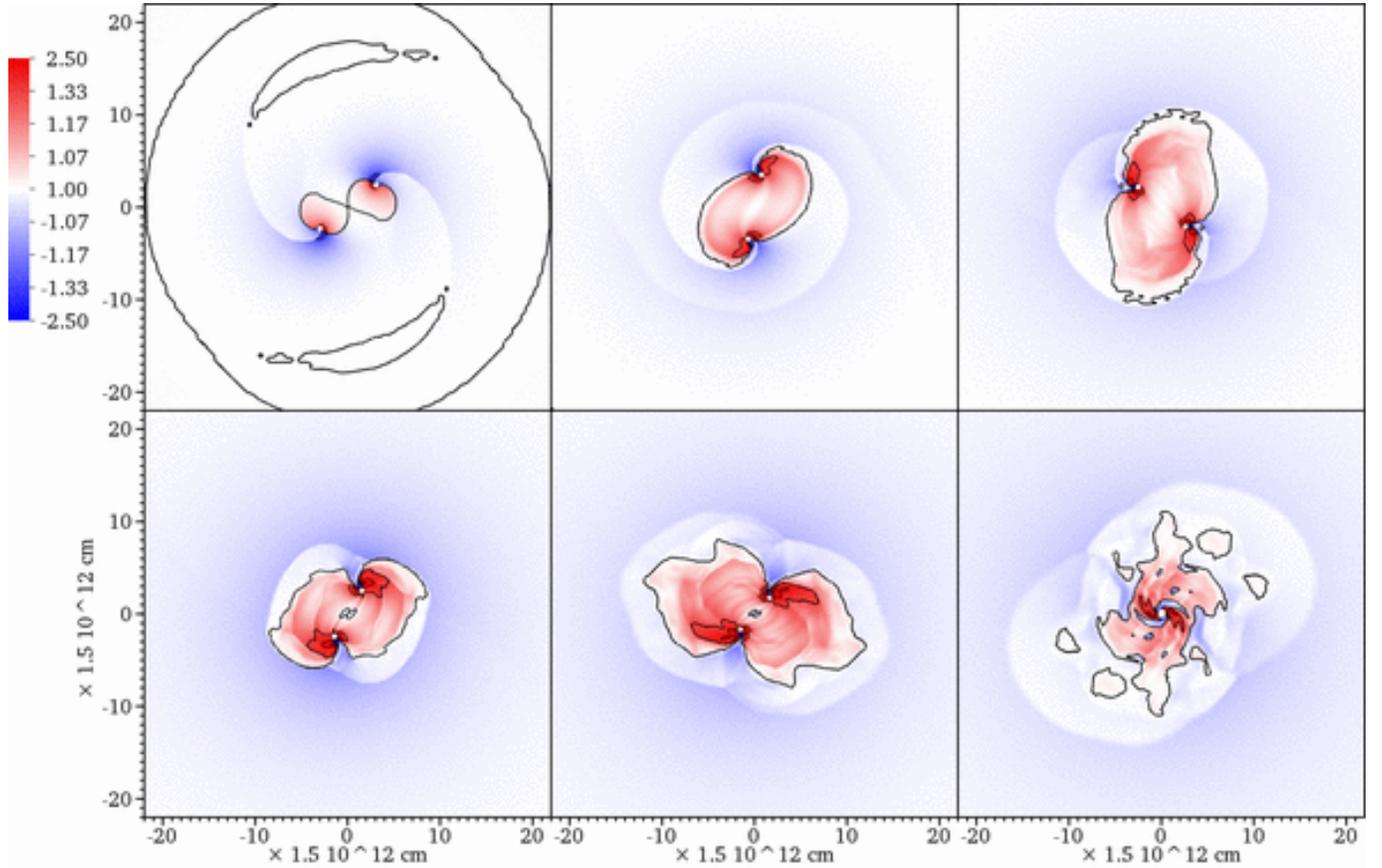}
\caption{Snapshots of the Lorentz factor measured in the plane of the binary in run G2.
Outgoing flows are shown in red and ingoing flows in blue 
(color online only), with the black contour line separating ingoing and
outgoing flows. The gas between the \bh{s} is flowing radially outwards.
The snapshots shown are coincident with those depicted in 
Figures~\ref{fig:2DrhoG2},  \ref{fig:2DMachG2},  and \ref{fig:BremsG2}.}
\label{fig:xyflowsG2}
\end{figure*}

\begin{figure*}[tbp]
\bigplot{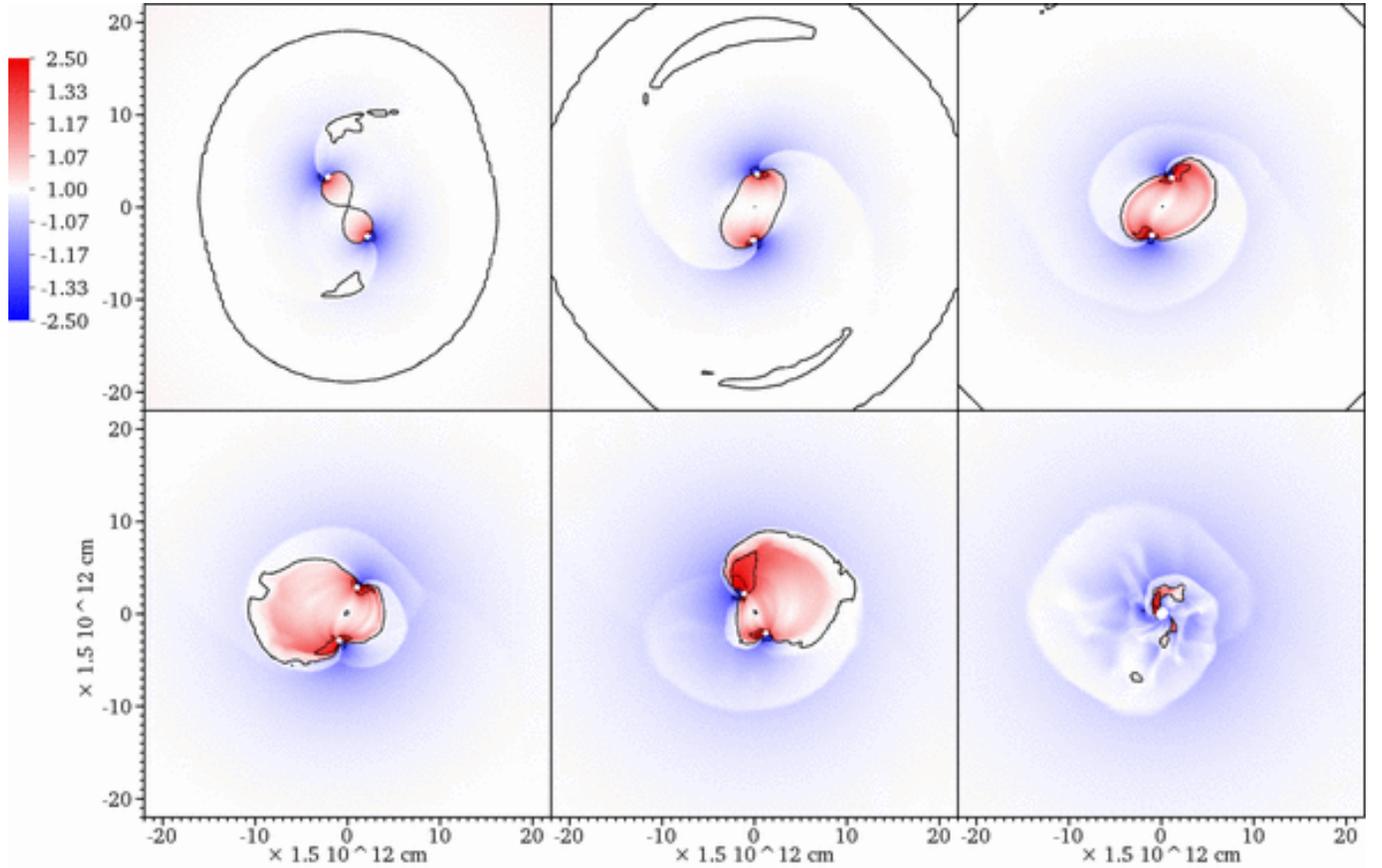}
\caption{Same as in Fig.~\ref{fig:xyflowsG2}  but for the run G3. Snapshots shown are coincident 
with those in Figures~\ref{fig:2DrhoG3},  \ref{fig:2DMachG3}, and \ref{fig:BremsG3}.}
\label{fig:xyflowsG3}
\end{figure*}

Unlike \citet{vanmeter09}, we do not see a stronger pre-merger spike in
the maximum Lorentz factor for the spinning \bh{} cases. We think that
this difference can be explained in the context of gas
pressure forces. Namely, in the test particle treatment considered by
\citet{vanmeter09}, pressure gradients are not taken into account and 
each particle moves along a geodesic with an unaltered velocity.
In our work, pressure forces from the surrounding gas act to
modify the velocity of the fluid and, in the specific case described
above, result in a lower Lorentz factor. This can be seen in 
Figs.~\ref{fig:xyflowsG2} and \ref{fig:xyflowsG3}, which illustrate 
the dynamics of the gas in the immediate vicinity of the binary.
During the inspiral, the motion of the binary leads to an ejection of gas 
which collides with the gas falling inwards, towards the binary. Because the 
ejected gas is unable to proceed further on its outward radial trajectory 
and is still within the gravitational influence of the binary, it eventually falls back
towards the \bh{s}.  In run G2, the gas is being ejected out by the binary on several
occasions, giving rise to a turbulent motion in the gas cloud.  
In run G3 the dynamics of the flow seeded by the 
prograde- and retrograde-spinning \bh{s} results
in a lopsided ejection-infall event.  In both cases this dynamically steered 
turbulence persists after the merger, leading to the post-merger variability 
in the bremsstrahlung luminosity and accretion rates. 

It is also worth noting that we find somewhat higher maximum Lorentz
factors in the post-merger phase compared to the equivalent scenario
in \citet{vanmeter09}, because our Lorentz factors reflect the bulk inflow 
of the gas into the merged BH, after the flow settles into a state of steady 
Bondi-type accretion.

\subsection{Gravitational Waves}\label{S_gw}

\begin{figure}[ht]
\includegraphics[width=0.96\linewidth]{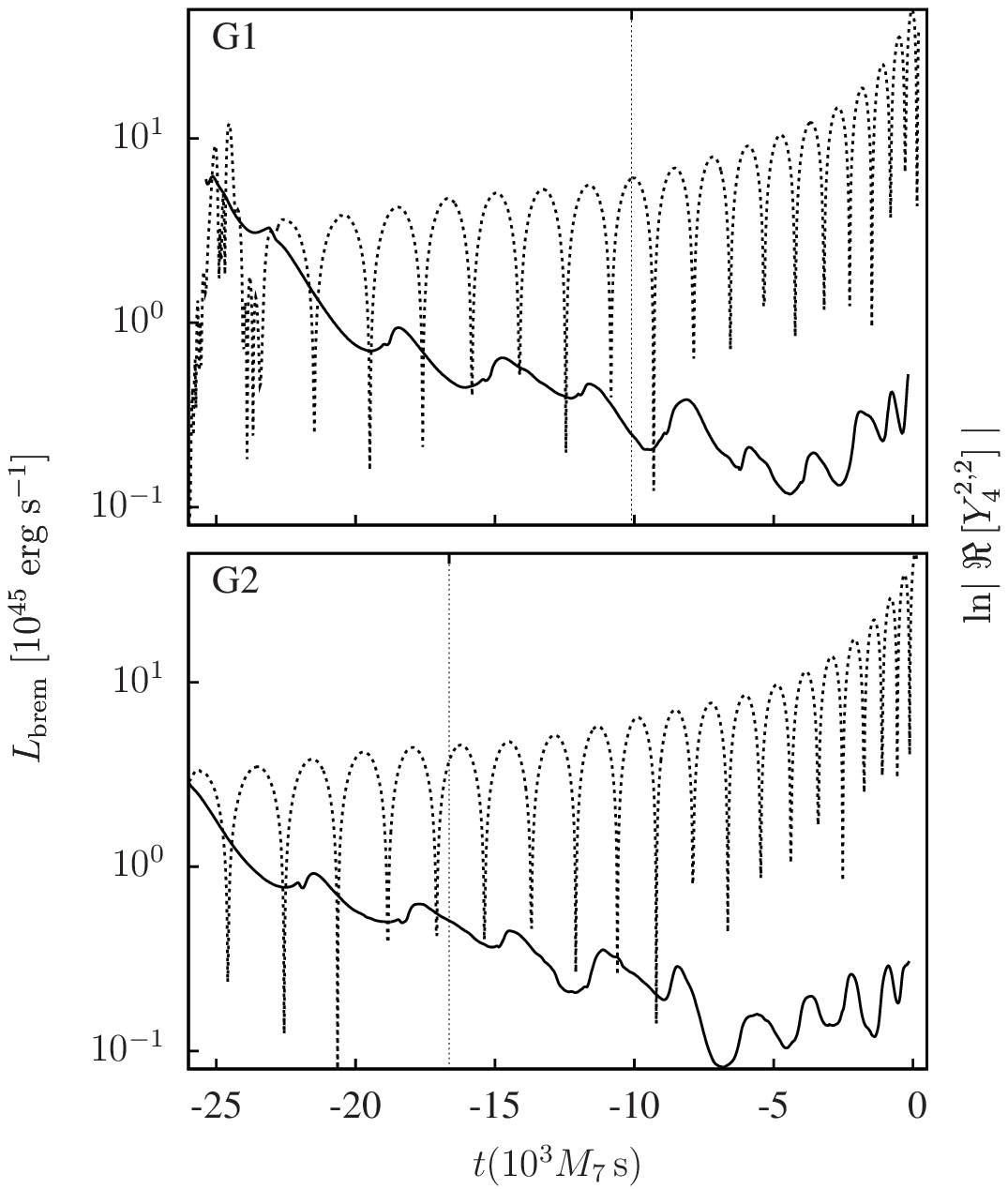}
\caption{Comparison of bremsstrahlung and \gw{} variability for the G1
  (upper panel) and G2 (lower panel) runs.  Real part of the waveform
  $\Re[\Psi_4^{2,2}]$ (dashed line) is superimposed on the top of the
  Doppler-boosted bremsstrahlung light curve (solid line).  Vertical
  lines mark the maxima of the real part of the waveform.}
\label{fig:dbswave}
\end{figure}

We now discuss properties of the \gw{} signatures associated with our
simulated systems as well as correlated \EM{}+\gw{} signatures.  The
gravitational waveforms from our simulations are almost identical to those of
vacuum simulations, with point-wise differences of $< 10^{-2}\,\%$ for
all the cases we considered. This is because the mass of the gas that
is gravitationally bound to our binaries is many orders of magnitude
smaller than the mass of the \bbh{.}

As mentioned before, detection of correlated \EM{}+\gw{} oscillations
from the same object would be a smoking gun for a \bbh{} system on the
way to coalescence and would directly link a detected \gw{} source to
its \EM{} counterpart.  The oscillations observed from the
bremsstrahlung's relativistically beamed light are directly tied to
the orbital dynamics of the binary and thus are also correlated with
the frequency of the \gw{s}.  In order to compare the characteristic
variability of the relativistically beamed bremsstrahlung light curve
(Section~\ref{S_brems}) with that of the \gw{s}, we overlay in
Fig.~\ref{fig:dbswave} the light curves of the beamed bremsstrahlung
emission and the magnitude of the gravitational waveforms for the two
prograde-spinning \bh{} cases (G1 and G2) during the inspiral phase.
Vertical lines have been drawn to mark the peaks of the real part of
the waveform. Clearly the frequency of oscillations in the light curve
closely matches the \gw{} frequency.  No similar variability is
visibly present in the bremsstrahlung light curve in run G3 since the
asymmetry in spins inhibits the formation of the strong, symmetric
density wakes.  As most \bbh{} systems in the universe are expected to be
unequal-mass binaries, and thus have some inherent asymmetry, the
likelihood of observing such correlated oscillations therefore depends
on the extent to which asymmetries can modify the variability as well
as the dominant mechanism that powers the emission.

Another form of EM variability associated with \bh{} coalescences has
been predicted to arise due to the effect of \gw{} mass
loss.  If present, this variability is expected to be most prominent
in geometrically thin circumbinary accretion disks in which transient
shocks ``light up'' the gas disk in response to a changed
gravitational potential of the central \bh{}~\citep{bp07}.  It has
been shown that transient shocks are absent in moderately thick,
hotter disks, where $H/R\gtrsim f$ ($f$ is the fractional mass loss
due to emission of \gw{}). In such flows the velocity of radial
motions that arise in the gas due to perturbations in the
gravitational potential is low with respect to the speed of sound and
thus incapable of producing shocks \citep{oneill09}.  With fractional
mass losses at the level of 3.6\%, 5.0\%, 6.3\%, and 3.6\% in runs
G0-G3, respectively, we find no subsequent EM variability in our
simulated systems due to this effect.  This is consistent with the
expectations based on studies described above since our gas cloud
scenario also falls in the class of hot and geometrically thick
accretion flows.

Shocks and EM variability associated with the gas have also 
been predicted to arise from the \gw{} recoil of the post-coalescence \bh{} 
\citep{2008ApJ...676L...5L,2008ApJ...682..758S, Megevand:2009yx,Corrales:2009nv,rossi09,sk08}. 
From our four cases only the remnant \bh{} in run G3 receives a ``GW kick'' 
as a consequence of an asymmetry in the emission of \gw{s}.
In this case, the up-down configuration of the spin vectors of the two
\bh{s} breaks the symmetry of the system and causes a fraction of
linear momentum carried by GWs to be emitted preferentially in one
direction. As a consequence of the conservation of linear momentum, 
the remnant \bh{} recoils in the opposite direction with respect to the 
center of mass of the pre-coalescence binary with velocity of ${\sim 180 \, {\rm km\,s^{-1}}}$.
No prompt shocks arise in run G3 due to the \gw{} recoil, where the evolution
follows the remnant \bh{} to a total displacement of only $\sim 0.2\,M$ from
the center of mass.  The absence of shocks and \EM{} signatures in our simulations
is again not surprising given the properties of the gas and the short timescales covered by 
our simulations. 
The imprints of the recoil on the gas could, however, be noticeable on much 
longer timescales, in which case a recoiling \bh{} can produce shocks or a 
trail of hot, X-ray emitting gas that extends out of the immediate center of a 
galaxy~\citep{devecchi09}.

\subsection{Observability of coalescences}\label{S_observability}

In this section, we outline some observational strategies for the
detection and monitoring of coalescing binaries based on the
characteristic \EM{} and \gw{} signatures presented in this
work.

Consider a binary of mass $10^7\,\msun$ at a redshift of $z=1$.
With current expected \lisa{} sensitivities, such a binary 
will be detectable by \lisa{} during plunge and coalescence. 
Shortly before merger, the \lisa{} error region could confine the location 
of such a binary to within a few tenths of a square degree on the sky \citep{lh06}. 
The exact size of the error region depends on a number of factors, including 
the location of the object on the sky, \bh{} masses, and spin orientation.
For instance, if the spin axes are misaligned, the orbital precession can 
further reduce the size of the error region to a few hundredths of a 
square degree.  

The \lisa{} error box hence marks a region on the sky relevant to
\EM{} searches for counterparts. Given our assumptions about 
the dominant emission mechanism in the vicinity of the \bh{} binary it 
follows that $L_{\rm bol}\sim L_{\rm brem}$.  The hot plasma cloud 
in our simulations is a natural source of high energy photons 
peaking in the $\gamma$-ray part of the spectrum at energies 
$kT_e \sim 1$~MeV. Because of the higher sensitivity of the X-ray 
over $\gamma$-ray surveys in terms of observed flux, in the remainder of
the discussion we focus on the observational strategy in the context of
X-ray observations.  The X-ray emission from our simulated sources
would comprise a fraction of their bolometric luminosity and we
estimate it by applying an empirical scaling in the form of a
bolometric correction. Assuming a bolometric correction of $C_X =
15.8$, appropriate for low luminosity \agns{}~\citep{ho09}, we infer
the X-ray luminosity associated with our binary, $L_{X} = L_{\rm
  bol}/C_{X}\sim 10^{43}\,{\rm erg\,s^{-1}}$.  A binary at $z=1$ is at
luminosity distance $d_L=6.6\,{\rm Gpc}$ (assuming a flat universe
with matter parameter $\Omega_m=0.27$ and Hubble parameter
$h_0=0.71$). The observed X-ray flux from this system is
$F_{X} \sim 10^{-15}\,{\rm erg\,cm^{-2}\,s^{-1}}$.  According to our
simulations, this level of luminosity will be observable for at least
$t \gtrsim 400\,M = 2\times10^4\, M_7\,{\rm s}$ in the frame of the
binary or, for a source at $z=1$, for $t \gtrsim
4\times10^4\,M_7\,{\rm s}$ in the frame of the observer.

Of the future X-ray observatories that may operate contemporary with
\lisa{}, {\it IXO} and {\it EXIST} have planned sensitivities
sufficient to observe high luminosity obscured \agns{} at $z \sim
0-2.5$ and low luminosity \agns{} at $z <0.5$. Both X-ray instruments
have a planned \FOV{} close to $20'$ in diameter and a flux
sensitivity limit of $10^{-15}\,{\rm erg\,cm^{-2}\,s^{-1}}$ achievable
in $\sim 10^4\,{\rm s}$ of exposure time. Given the above estimate for
the X-ray flux, it follows that some of the brighter candidate \agns{}
could be detected in as little as 1 hour of exposure. X-ray monitoring
with $\sim1$-hour sampling frequency would allow a follow-up of the
quasi-periodic variability close to the coalescence until the point
where the orbital period of the binary becomes $\lesssim 1\,{\rm hr}/(1+z)$.
In order not to be masked by a natural variability intrinsic to most
\agns{,} the magnitude of quasi-periodic oscillations should exceed
few tens of percent of the luminosity of an \agns{.} Given a \lisa{}
error region smaller than $\sim20'$, either X-ray instrument could
cover it in a single exposure, hence allowing continuous monitoring of
the X-ray luminosity curve.

Note that binaries with masses lower than $10^7\,\msun$ will be
detected by \lisa{} during their inspiral phase as well as plunge and
coalescence, thus including more orbital cycles prior to
coalescence. This may facilitate some \EM{} counterpart searches since
they can be triggered earlier, but in some cases earlier detection
will imply a trade-off in strength of \gw{} signal. Because the \gw{}
signal is weaker during the inspiral phase than during the plunge and
coalescence, the \lisa{} error region will be larger, and a day before
the coalescence it may encompass a square degree range \citep{lh06}.
In such a case, because the size of the \lisa{} error region exceeds
the planned \FOV{} of any X-ray instrument, an observational strategy
would require multiple exposure frames until the whole error region is
tiled. In order to detect variability, the procedure would then need
to be repeated and the error region scanned multiple times. Given a
square degree error region and the $20'$ \FOV{} of both future X-ray
instruments, it would take about 9 exposures to cover the error region
once, before a new snapshot of the field can be acquired. Because
lower mass binaries are also expected to be less luminous ($L_{X}\sim
10^{40}\,{\rm erg\,s^{-1}}$ for $10^6\,\msun$ binary), they would fall
into a low luminosity AGN category and could only be monitored in the local
universe out to a distance of $\sim 100\,{\rm Mpc}$. A large initial
error region and low luminosity make detection of the quasi-periodic
variability a challenging prospect for lower mass
systems. Nevertheless, a luminosity rise peaking at coalescence followed
by a sudden drop-off can still be used as robust signatures for the \EM{}
detection of these binary systems. Following a detection, multi-wavelength
coverage of the object on the sky would lead to localization of the
source with an even higher precision, at the level of arcseconds and
higher, and would allow a study of properties of the host galaxy.

While our discussion of observability of \bh{} coalescences focuses 
on ``LISA binaries" with masses of $10^7\,\msun$ or less and the feasibility of 
coincident \EM{+}\gw{} detections, it is worth mentioning that a serendipitous 
\EM{} detection of a \bh{} coalescence may occur even before \lisa{}, in the 
current era of observations. Such a detection would be most likely for more 
massive ($\sim 10^8\,\msun$) and more luminous binary systems that 
could be ``caught" by the current wide field of view observatories 
monitoring the high energy transient sky. 
Other forms of \EM{} variability may also arise from mechanisms that
were not captured by our simulations, such as radio, X-ray and
$\gamma$-ray outbursts due to reconnection and effects of magnetic 
field lines close to the binary. Specifically, if a magnetic field with 
near-equipartition strength is present in the vicinity of a binary the 
synchrotron and inverse Compton radiation may dominate the 
emission in millimeter and high energy bands, respectively. 
Whether these phenomena can give rise to characteristic 
signatures that would uniquely point to a \bbh{}
coalescence event is a question of interest which remains to be
studied in the context of MHD-NR calculations.


\section{Conclusions}\label{sec:conclusions} 

We presented the first fully general relativistic numerical
hydrodynamics simulations of \smbh{} binaries in a gaseous environment
through inspiral, plunge, and coalescence.  The gaseous environment is
a hot and turbulent gas cloud with physical properties reminiscent of
accretion flows in low luminosity \agns{s}. The gas cloud was chosen
as one of the characteristic scenarios representative of conditions in
which pre-coalescence binaries may exist in galactic centers.  Since
the radial inflow speed in a hot gas cloud in our simulations is
sufficiently high to prevent binary torques from evacuating most of
the gas locally, accretion and interaction of the binaries with the
gas continue uninterrupted throughout the merger.  As a sample of the
parameter space, we studied three symmetric, equal-mass \bbh{} with
spins of $a/M={0,0.4,0.6}$ aligned with the orbital angular
momentum. In addition, we considered a fourth case of a binary with
anti-aligned \bh{} spins of magnitude $a/M=0.4$ parallel to the
orbital angular momentum.  The characteristic \EM{} and \gw{}
signatures that arise from such interactions were the focus of this
work. We summarize our most important results as follows:
\begin{itemize}

\item Our simulations show that correlated \EM{}-\gw{} variability
  can occur in merging binary systems immersed in hot gas flows. 
  Specifically, we found \EM{} variability arising due to the effects of
  relativistic beaming and Doppler boosting modulated by the binary
  orbital motion. In these systems the frequency of the EM oscillations 
  is equal to that of the \gw{s} and the maximum amplitude of variations 
  in luminosity is a factor of $\sim2$.
 
\item The variable \EM{} emission in our simulations is powered by
  shocks triggered by the orbiting \bh{s}.  While quasi-periodic variability is
  present in all cases considered from inspiral through the plunge, it is
  not as pronounced in the case of a binary with asymmetric, 
  prograde-retrograde spin configuration.

\item In cases where quasi-periodic variability in luminosity may be
  weak or absent, additional signatures may be sought for in searches
  for \EM{} counterparts. Our models indicate that, in cases where the
  luminosity is dominated by emission from the shocked gas, light
  curves may exhibit a gradual rise, arriving at a peak at the time of
  coalescence, followed by a sudden drop-off. If present, these two
  features are sufficiently robust to allow identification of an \EM{}
  counterpart to a \gw{} source.

\item We estimate that most massive binaries
  detectable in the \lisa{} band may be identified in \EM{} searches
  out to $z\approx1$. However, lower mass binaries and systems with
  gas densities lower than those considered here would fall into a
  class of low luminosity AGNs that could only be identified in the local
  universe.

\end{itemize}
In summary, if coincident variability can be observed in both light
and \gw{s} from the same object, this signature would be convincing
evidence for an impending \bbh{} coalescence. Our results suggest some
encouraging prospects for such detections. Nevertheless, given the
extent of the parameter space involved in coalescing \bbh{s}
interacting with gas, more follow-up work is needed.  In particular,
since most of the \smbh{} binaries in the universe are expected to involve unequal
masses and general spin orientations, it is important to further explore the
parameter space to investigate how common situations with characteristic \EM{}
signatures such as those in the present work may be.  Furthermore, because the
thermodynamic properties of the surrounding gas can significantly influence
the properties of \EM{} signals, in the future we will also consider
other scenarios for gaseous environments around \smbh{s} in galactic
centers, such as a rotating gas cloud and circumbinary accretion disk.


\acknowledgments

We thank Mike Eracleous, Cole Miller, and Sean O'Neill for useful 
comments and suggestions and James Healy for computational infrastructure.  
T. Bogdonavi\'{c} is grateful to Richard Mushotzky, Scott Noble, and 
Tom Maccarone for stimulating discussions.  This
work was supported in part by the NSF grants 0653443, 0855892,
0914553, 0941417, 0903973. Support for Bogdanovi\'c provided by NASA
through Einstein Postdoctoral Fellowship Award PF9-00061 issued by the
Chandra X-ray Observatory Center, which is operated by the Smithsonian
Astrophysical Observatory for and on behalf of the NASA under contract
NAS8-03060.  Computations described in this paper were carried out
under Teragrid allocation TG-MCA08X009.


\bibliography{BBL/phyjabb,BBL/refs,BBL/gas}

\begin{thebibliography}{91}
\expandafter\ifx\csname natexlab\endcsname\relax\def\natexlab#1{#1}\fi
\expandafter\ifx\csname bibnamefont\endcsname\relax
  \def\bibnamefont#1{#1}\fi
\expandafter\ifx\csname bibfnamefont\endcsname\relax
  \def\bibfnamefont#1{#1}\fi
\expandafter\ifx\csname citenamefont\endcsname\relax
  \def\citenamefont#1{#1}\fi
\expandafter\ifx\csname url\endcsname\relax
  \def\url#1{\texttt{#1}}\fi
\expandafter\ifx\csname urlprefix\endcsname\relax\def\urlprefix{URL }\fi
\providecommand{\bibinfo}[2]{#2}
\providecommand{\eprint}[2][]{\url{#2}}

\bibitem[{\citenamefont{{Haehnelt} and {Kauffmann}}(2002)}]{hk02}
\bibinfo{author}{\bibfnamefont{M.~G.} \bibnamefont{{Haehnelt}}}
  \bibnamefont{and}
  \bibinfo{author}{\bibfnamefont{G.}~\bibnamefont{{Kauffmann}}},
  \bibinfo{journal}{\mnras} \textbf{\bibinfo{volume}{336}},
  \bibinfo{pages}{L61} (\bibinfo{year}{2002}), \eprint{arXiv:astro-ph/0208215}.

\bibitem[{\citenamefont{{Volonteri}
  et~al.}(2003{\natexlab{a}})\citenamefont{{Volonteri}, {Haardt}, and
  {Madau}}}]{volonteri03}
\bibinfo{author}{\bibfnamefont{M.}~\bibnamefont{{Volonteri}}},
  \bibinfo{author}{\bibfnamefont{F.}~\bibnamefont{{Haardt}}}, \bibnamefont{and}
  \bibinfo{author}{\bibfnamefont{P.}~\bibnamefont{{Madau}}},
  \bibinfo{journal}{\apj} \textbf{\bibinfo{volume}{582}}, \bibinfo{pages}{559}
  (\bibinfo{year}{2003}{\natexlab{a}}), \eprint{arXiv:astro-ph/0207276}.

\bibitem[{\citenamefont{{Kormendy} and {Richstone}}(1995)}]{kr95}
\bibinfo{author}{\bibfnamefont{J.}~\bibnamefont{{Kormendy}}} \bibnamefont{and}
  \bibinfo{author}{\bibfnamefont{D.}~\bibnamefont{{Richstone}}},
  \bibinfo{journal}{\araa} \textbf{\bibinfo{volume}{33}}, \bibinfo{pages}{581}
  (\bibinfo{year}{1995}).

\bibitem[{\citenamefont{{Richstone} et~al.}(1998)\citenamefont{{Richstone},
  {Ajhar}, {Bender}, {Bower}, {Dressler}, {Faber}, {Filippenko}, {Gebhardt},
  {Green}, {Ho} et~al.}}]{richstone98}
\bibinfo{author}{\bibfnamefont{D.}~\bibnamefont{{Richstone}}},
  \bibinfo{author}{\bibfnamefont{E.~A.} \bibnamefont{{Ajhar}}},
  \bibinfo{author}{\bibfnamefont{R.}~\bibnamefont{{Bender}}},
  \bibinfo{author}{\bibfnamefont{G.}~\bibnamefont{{Bower}}},
  \bibinfo{author}{\bibfnamefont{A.}~\bibnamefont{{Dressler}}},
  \bibinfo{author}{\bibfnamefont{S.~M.} \bibnamefont{{Faber}}},
  \bibinfo{author}{\bibfnamefont{A.~V.} \bibnamefont{{Filippenko}}},
  \bibinfo{author}{\bibfnamefont{K.}~\bibnamefont{{Gebhardt}}},
  \bibinfo{author}{\bibfnamefont{R.}~\bibnamefont{{Green}}},
  \bibinfo{author}{\bibfnamefont{L.~C.} \bibnamefont{{Ho}}},
  \bibnamefont{et~al.}, \bibinfo{journal}{\nat} \textbf{\bibinfo{volume}{395}},
  \bibinfo{pages}{A14+} (\bibinfo{year}{1998}),
  \eprint{arXiv:astro-ph/9810378}.

\bibitem[{\citenamefont{{Peterson} and {Wandel}}(2000)}]{pw00}
\bibinfo{author}{\bibfnamefont{B.~M.} \bibnamefont{{Peterson}}}
  \bibnamefont{and} \bibinfo{author}{\bibfnamefont{A.}~\bibnamefont{{Wandel}}},
  \bibinfo{journal}{\apjl} \textbf{\bibinfo{volume}{540}}, \bibinfo{pages}{L13}
  (\bibinfo{year}{2000}), \eprint{arXiv:astro-ph/0007147}.

\bibitem[{\citenamefont{{Ferrarese} and {Ford}}(2005)}]{ff05}
\bibinfo{author}{\bibfnamefont{L.}~\bibnamefont{{Ferrarese}}} \bibnamefont{and}
  \bibinfo{author}{\bibfnamefont{H.}~\bibnamefont{{Ford}}},
  \bibinfo{journal}{Space Science Reviews} \textbf{\bibinfo{volume}{116}},
  \bibinfo{pages}{523} (\bibinfo{year}{2005}), \eprint{arXiv:astro-ph/0411247}.

\bibitem[{\citenamefont{{Volonteri}
  et~al.}(2003{\natexlab{b}})\citenamefont{{Volonteri}, {Madau}, and
  {Haardt}}}]{volonteri2003}
\bibinfo{author}{\bibfnamefont{M.}~\bibnamefont{{Volonteri}}},
  \bibinfo{author}{\bibfnamefont{P.}~\bibnamefont{{Madau}}}, \bibnamefont{and}
  \bibinfo{author}{\bibfnamefont{F.}~\bibnamefont{{Haardt}}},
  \bibinfo{journal}{Astrophys. J.} \textbf{\bibinfo{volume}{593}},
  \bibinfo{pages}{661} (\bibinfo{year}{2003}{\natexlab{b}}),
  \bibinfo{note}{arXiv:astro-ph/0304389}.

\bibitem[{\citenamefont{{Volonteri} et~al.}(2004)\citenamefont{{Volonteri},
  {Haardt}, {Madau}, and {Sesana}}}]{volonteri2004}
\bibinfo{author}{\bibfnamefont{M.}~\bibnamefont{{Volonteri}}},
  \bibinfo{author}{\bibfnamefont{F.}~\bibnamefont{{Haardt}}},
  \bibinfo{author}{\bibfnamefont{P.}~\bibnamefont{{Madau}}}, \bibnamefont{and}
  \bibinfo{author}{\bibfnamefont{A.}~\bibnamefont{{Sesana}}}, in
  \emph{\bibinfo{booktitle}{Astrophysics and Space Science}}, edited by
  \bibinfo{editor}{\bibfnamefont{M.}~\bibnamefont{{Plionis}}}
  (\bibinfo{year}{2004}), vol. \bibinfo{volume}{301} of
  \emph{\bibinfo{series}{Astrophysics and Space Science Library}}, p.
  \bibinfo{pages}{227}.

\bibitem[{\citenamefont{{Micic} et~al.}(2007)\citenamefont{{Micic},
  {Holley-Bockelmann}, {Sigurdsson}, and {Abel}}}]{micic2007}
\bibinfo{author}{\bibfnamefont{M.}~\bibnamefont{{Micic}}},
  \bibinfo{author}{\bibfnamefont{K.}~\bibnamefont{{Holley-Bockelmann}}},
  \bibinfo{author}{\bibfnamefont{S.}~\bibnamefont{{Sigurdsson}}},
  \bibnamefont{and} \bibinfo{author}{\bibfnamefont{T.}~\bibnamefont{{Abel}}},
  \bibinfo{journal}{Mon. Not. R. Astron. Soc.} \textbf{\bibinfo{volume}{380}},
  \bibinfo{pages}{1533} (\bibinfo{year}{2007}),
  \bibinfo{note}{arXiv:astro-ph/0703540}.

\bibitem[{\citenamefont{{Sesana} et~al.}(2007)\citenamefont{{Sesana},
  {Volonteri}, and {Haardt}}}]{sesana07}
\bibinfo{author}{\bibfnamefont{A.}~\bibnamefont{{Sesana}}},
  \bibinfo{author}{\bibfnamefont{M.}~\bibnamefont{{Volonteri}}},
  \bibnamefont{and} \bibinfo{author}{\bibfnamefont{F.}~\bibnamefont{{Haardt}}},
  \bibinfo{journal}{\mnras} \textbf{\bibinfo{volume}{377}},
  \bibinfo{pages}{1711} (\bibinfo{year}{2007}),
  \eprint{arXiv:astro-ph/0701556}.

\bibitem[{\citenamefont{{Kocsis} et~al.}(2006)\citenamefont{{Kocsis}, {Frei},
  {Haiman}, and {Menou}}}]{kocsis06}
\bibinfo{author}{\bibfnamefont{B.}~\bibnamefont{{Kocsis}}},
  \bibinfo{author}{\bibfnamefont{Z.}~\bibnamefont{{Frei}}},
  \bibinfo{author}{\bibfnamefont{Z.}~\bibnamefont{{Haiman}}}, \bibnamefont{and}
  \bibinfo{author}{\bibfnamefont{K.}~\bibnamefont{{Menou}}},
  \bibinfo{journal}{\apj} \textbf{\bibinfo{volume}{637}}, \bibinfo{pages}{27}
  (\bibinfo{year}{2006}), \eprint{arXiv:astro-ph/0505394}.

\bibitem[{\citenamefont{{Hughes} and {Holz}}(2003)}]{hh03}
\bibinfo{author}{\bibfnamefont{S.~A.} \bibnamefont{{Hughes}}} \bibnamefont{and}
  \bibinfo{author}{\bibfnamefont{D.~E.} \bibnamefont{{Holz}}},
  \bibinfo{journal}{Classical and Quantum Gravity}
  \textbf{\bibinfo{volume}{20}}, \bibinfo{pages}{65} (\bibinfo{year}{2003}),
  \eprint{arXiv:astro-ph/0212218}.

\bibitem[{\citenamefont{{Bogdanovi{\'c}}
  et~al.}(2009)\citenamefont{{Bogdanovi{\'c}}, {Eracleous}, and
  {Sigurdsson}}}]{bogdanovic09}
\bibinfo{author}{\bibfnamefont{T.}~\bibnamefont{{Bogdanovi{\'c}}}},
  \bibinfo{author}{\bibfnamefont{M.}~\bibnamefont{{Eracleous}}},
  \bibnamefont{and}
  \bibinfo{author}{\bibfnamefont{S.}~\bibnamefont{{Sigurdsson}}},
  \bibinfo{journal}{New Astron. Rev.} \textbf{\bibinfo{volume}{53}},
  \bibinfo{pages}{113} (\bibinfo{year}{2009}), \eprint{0909.0516}.

\bibitem[{\citenamefont{{Kazantzidis} et~al.}(2005)\citenamefont{{Kazantzidis},
  {Mayer}, {Colpi}, {Madau}, {Debattista}, {Wadsley}, {Stadel}, {Quinn}, and
  {Moore}}}]{kazantzidis05}
\bibinfo{author}{\bibfnamefont{S.}~\bibnamefont{{Kazantzidis}}},
  \bibinfo{author}{\bibfnamefont{L.}~\bibnamefont{{Mayer}}},
  \bibinfo{author}{\bibfnamefont{M.}~\bibnamefont{{Colpi}}},
  \bibinfo{author}{\bibfnamefont{P.}~\bibnamefont{{Madau}}},
  \bibinfo{author}{\bibfnamefont{V.~P.} \bibnamefont{{Debattista}}},
  \bibinfo{author}{\bibfnamefont{J.}~\bibnamefont{{Wadsley}}},
  \bibinfo{author}{\bibfnamefont{J.}~\bibnamefont{{Stadel}}},
  \bibinfo{author}{\bibfnamefont{T.}~\bibnamefont{{Quinn}}}, \bibnamefont{and}
  \bibinfo{author}{\bibfnamefont{B.}~\bibnamefont{{Moore}}},
  \bibinfo{journal}{\apjl} \textbf{\bibinfo{volume}{623}}, \bibinfo{pages}{L67}
  (\bibinfo{year}{2005}), \eprint{arXiv:astro-ph/0407407}.

\bibitem[{\citenamefont{{Armitage} and {Natarajan}}(2002)}]{an02}
\bibinfo{author}{\bibfnamefont{P.~J.} \bibnamefont{{Armitage}}}
  \bibnamefont{and}
  \bibinfo{author}{\bibfnamefont{P.}~\bibnamefont{{Natarajan}}},
  \bibinfo{journal}{\apjl} \textbf{\bibinfo{volume}{567}}, \bibinfo{pages}{L9}
  (\bibinfo{year}{2002}), \eprint{arXiv:astro-ph/0201318}.

\bibitem[{\citenamefont{{Escala} et~al.}(2004)\citenamefont{{Escala}, {Larson},
  {Coppi}, and {Mardones}}}]{escala04}
\bibinfo{author}{\bibfnamefont{A.}~\bibnamefont{{Escala}}},
  \bibinfo{author}{\bibfnamefont{R.~B.} \bibnamefont{{Larson}}},
  \bibinfo{author}{\bibfnamefont{P.~S.} \bibnamefont{{Coppi}}},
  \bibnamefont{and}
  \bibinfo{author}{\bibfnamefont{D.}~\bibnamefont{{Mardones}}},
  \bibinfo{journal}{\apj} \textbf{\bibinfo{volume}{607}}, \bibinfo{pages}{765}
  (\bibinfo{year}{2004}), \eprint{arXiv:astro-ph/0310851}.

\bibitem[{\citenamefont{{Escala} et~al.}(2005)\citenamefont{{Escala}, {Larson},
  {Coppi}, and {Mardones}}}]{escala05}
\bibinfo{author}{\bibfnamefont{A.}~\bibnamefont{{Escala}}},
  \bibinfo{author}{\bibfnamefont{R.~B.} \bibnamefont{{Larson}}},
  \bibinfo{author}{\bibfnamefont{P.~S.} \bibnamefont{{Coppi}}},
  \bibnamefont{and}
  \bibinfo{author}{\bibfnamefont{D.}~\bibnamefont{{Mardones}}},
  \bibinfo{journal}{\apj} \textbf{\bibinfo{volume}{630}}, \bibinfo{pages}{152}
  (\bibinfo{year}{2005}), \eprint{arXiv:astro-ph/0406304}.

\bibitem[{\citenamefont{{Dotti} et~al.}(2007)\citenamefont{{Dotti}, {Colpi},
  {Haardt}, and {Mayer}}}]{dotti07}
\bibinfo{author}{\bibfnamefont{M.}~\bibnamefont{{Dotti}}},
  \bibinfo{author}{\bibfnamefont{M.}~\bibnamefont{{Colpi}}},
  \bibinfo{author}{\bibfnamefont{F.}~\bibnamefont{{Haardt}}}, \bibnamefont{and}
  \bibinfo{author}{\bibfnamefont{L.}~\bibnamefont{{Mayer}}},
  \bibinfo{journal}{\mnras} \textbf{\bibinfo{volume}{379}},
  \bibinfo{pages}{956} (\bibinfo{year}{2007}), \eprint{arXiv:astro-ph/0612505}.

\bibitem[{\citenamefont{{Mayer} et~al.}(2007)\citenamefont{{Mayer},
  {Kazantzidis}, {Madau}, {Colpi}, {Quinn}, and {Wadsley}}}]{mayer07}
\bibinfo{author}{\bibfnamefont{L.}~\bibnamefont{{Mayer}}},
  \bibinfo{author}{\bibfnamefont{S.}~\bibnamefont{{Kazantzidis}}},
  \bibinfo{author}{\bibfnamefont{P.}~\bibnamefont{{Madau}}},
  \bibinfo{author}{\bibfnamefont{M.}~\bibnamefont{{Colpi}}},
  \bibinfo{author}{\bibfnamefont{T.}~\bibnamefont{{Quinn}}}, \bibnamefont{and}
  \bibinfo{author}{\bibfnamefont{J.}~\bibnamefont{{Wadsley}}},
  \bibinfo{journal}{Science} \textbf{\bibinfo{volume}{316}},
  \bibinfo{pages}{1874} (\bibinfo{year}{2007}), \eprint{0706.1562}.

\bibitem[{\citenamefont{{Colpi} et~al.}(2007)\citenamefont{{Colpi}, {Dotti},
  {Mayer}, and {Kazantzidis}}}]{colpi07}
\bibinfo{author}{\bibfnamefont{M.}~\bibnamefont{{Colpi}}},
  \bibinfo{author}{\bibfnamefont{M.}~\bibnamefont{{Dotti}}},
  \bibinfo{author}{\bibfnamefont{L.}~\bibnamefont{{Mayer}}}, \bibnamefont{and}
  \bibinfo{author}{\bibfnamefont{S.}~\bibnamefont{{Kazantzidis}}},
  \bibinfo{journal}{ArXiv e-prints}  (\bibinfo{year}{2007}),
  \eprint{0710.5207}.

\bibitem[{\citenamefont{{MacFadyen} and {Milosavljevi{\'c}}}(2008)}]{mm08}
\bibinfo{author}{\bibfnamefont{A.~I.} \bibnamefont{{MacFadyen}}}
  \bibnamefont{and}
  \bibinfo{author}{\bibfnamefont{M.}~\bibnamefont{{Milosavljevi{\'c}}}},
  \bibinfo{journal}{\apj} \textbf{\bibinfo{volume}{672}}, \bibinfo{pages}{83}
  (\bibinfo{year}{2008}), \eprint{arXiv:astro-ph/0607467}.

\bibitem[{\citenamefont{{Hayasaki} et~al.}(2008)\citenamefont{{Hayasaki},
  {Mineshige}, and {Ho}}}]{hayasaki08}
\bibinfo{author}{\bibfnamefont{K.}~\bibnamefont{{Hayasaki}}},
  \bibinfo{author}{\bibfnamefont{S.}~\bibnamefont{{Mineshige}}},
  \bibnamefont{and} \bibinfo{author}{\bibfnamefont{L.~C.} \bibnamefont{{Ho}}},
  \bibinfo{journal}{\apj} \textbf{\bibinfo{volume}{682}}, \bibinfo{pages}{1134}
  (\bibinfo{year}{2008}), \eprint{0708.2555}.

\bibitem[{\citenamefont{{Cuadra} et~al.}(2009)\citenamefont{{Cuadra},
  {Armitage}, {Alexander}, and {Begelman}}}]{cuadra09}
\bibinfo{author}{\bibfnamefont{J.}~\bibnamefont{{Cuadra}}},
  \bibinfo{author}{\bibfnamefont{P.~J.} \bibnamefont{{Armitage}}},
  \bibinfo{author}{\bibfnamefont{R.~D.} \bibnamefont{{Alexander}}},
  \bibnamefont{and} \bibinfo{author}{\bibfnamefont{M.~C.}
  \bibnamefont{{Begelman}}}, \bibinfo{journal}{\mnras}
  \textbf{\bibinfo{volume}{393}}, \bibinfo{pages}{1423} (\bibinfo{year}{2009}),
  \eprint{0809.0311}.

\bibitem[{\citenamefont{{Milosavljevi{\'c}} and {Phinney}}(2005)}]{mp05}
\bibinfo{author}{\bibfnamefont{M.}~\bibnamefont{{Milosavljevi{\'c}}}}
  \bibnamefont{and} \bibinfo{author}{\bibfnamefont{E.~S.}
  \bibnamefont{{Phinney}}}, \bibinfo{journal}{\apjl}
  \textbf{\bibinfo{volume}{622}}, \bibinfo{pages}{L93} (\bibinfo{year}{2005}),
  \eprint{arXiv:astro-ph/0410343}.

\bibitem[{\citenamefont{{Quataert}}(1999)}]{quataert99}
\bibinfo{author}{\bibfnamefont{E.~J.~L.} \bibnamefont{{Quataert}}}, Ph.D.
  thesis, \bibinfo{school}{AA(HARVARD UNIVERSITY)} (\bibinfo{year}{1999}).

\bibitem[{\citenamefont{{Ptak} et~al.}(2004)\citenamefont{{Ptak}, {Terashima},
  {Ho}, and {Quataert}}}]{ptak04}
\bibinfo{author}{\bibfnamefont{A.}~\bibnamefont{{Ptak}}},
  \bibinfo{author}{\bibfnamefont{Y.}~\bibnamefont{{Terashima}}},
  \bibinfo{author}{\bibfnamefont{L.~C.} \bibnamefont{{Ho}}}, \bibnamefont{and}
  \bibinfo{author}{\bibfnamefont{E.}~\bibnamefont{{Quataert}}},
  \bibinfo{journal}{\apj} \textbf{\bibinfo{volume}{606}}, \bibinfo{pages}{173}
  (\bibinfo{year}{2004}), \eprint{arXiv:astro-ph/0401524}.

\bibitem[{\citenamefont{{Nemmen} et~al.}(2006)\citenamefont{{Nemmen},
  {Storchi-Bergmann}, {Yuan}, {Eracleous}, {Terashima}, and
  {Wilson}}}]{nemmen06}
\bibinfo{author}{\bibfnamefont{R.~S.} \bibnamefont{{Nemmen}}},
  \bibinfo{author}{\bibfnamefont{T.}~\bibnamefont{{Storchi-Bergmann}}},
  \bibinfo{author}{\bibfnamefont{F.}~\bibnamefont{{Yuan}}},
  \bibinfo{author}{\bibfnamefont{M.}~\bibnamefont{{Eracleous}}},
  \bibinfo{author}{\bibfnamefont{Y.}~\bibnamefont{{Terashima}}},
  \bibnamefont{and} \bibinfo{author}{\bibfnamefont{A.~S.}
  \bibnamefont{{Wilson}}}, \bibinfo{journal}{\apj}
  \textbf{\bibinfo{volume}{643}}, \bibinfo{pages}{652} (\bibinfo{year}{2006}),
  \eprint{arXiv:astro-ph/0512540}.

\bibitem[{\citenamefont{{Elitzur} and {Ho}}(2009)}]{eh09}
\bibinfo{author}{\bibfnamefont{M.}~\bibnamefont{{Elitzur}}} \bibnamefont{and}
  \bibinfo{author}{\bibfnamefont{L.~C.} \bibnamefont{{Ho}}},
  \bibinfo{journal}{\apjl} \textbf{\bibinfo{volume}{701}}, \bibinfo{pages}{L91}
  (\bibinfo{year}{2009}), \eprint{0907.3752}.

\bibitem[{\citenamefont{{Bogdanovi{\'c}}
  et~al.}(2007)\citenamefont{{Bogdanovi{\'c}}, {Reynolds}, and
  {Miller}}}]{bogdanovic07}
\bibinfo{author}{\bibfnamefont{T.}~\bibnamefont{{Bogdanovi{\'c}}}},
  \bibinfo{author}{\bibfnamefont{C.~S.} \bibnamefont{{Reynolds}}},
  \bibnamefont{and} \bibinfo{author}{\bibfnamefont{M.~C.}
  \bibnamefont{{Miller}}}, \bibinfo{journal}{\apjl}
  \textbf{\bibinfo{volume}{661}}, \bibinfo{pages}{L147} (\bibinfo{year}{2007}),
  \eprint{arXiv:astro-ph/0703054}.

\bibitem[{\citenamefont{{Herrmann}
  et~al.}(2007{\natexlab{a}})\citenamefont{{Herrmann}, {Hinder}, {Shoemaker},
  {Laguna}, and {Matzner}}}]{2007PhRvD..76h4032H}
\bibinfo{author}{\bibfnamefont{F.}~\bibnamefont{{Herrmann}}},
  \bibinfo{author}{\bibfnamefont{I.}~\bibnamefont{{Hinder}}},
  \bibinfo{author}{\bibfnamefont{D.~M.} \bibnamefont{{Shoemaker}}},
  \bibinfo{author}{\bibfnamefont{P.}~\bibnamefont{{Laguna}}}, \bibnamefont{and}
  \bibinfo{author}{\bibfnamefont{R.~A.} \bibnamefont{{Matzner}}},
  \bibinfo{journal}{Phys. Rev. D} \textbf{\bibinfo{volume}{76}},
  \bibinfo{pages}{084032} (\bibinfo{year}{2007}{\natexlab{a}}),
  \eprint{arXiv:0706.2541}.

\bibitem[{\citenamefont{{Koppitz} et~al.}(2007)\citenamefont{{Koppitz},
  {Pollney}, {Reisswig}, {Rezzolla}, {Thornburg}, {Diener}, and
  {Schnetter}}}]{2007PhRvL..99d1102K}
\bibinfo{author}{\bibfnamefont{M.}~\bibnamefont{{Koppitz}}},
  \bibinfo{author}{\bibfnamefont{D.}~\bibnamefont{{Pollney}}},
  \bibinfo{author}{\bibfnamefont{C.}~\bibnamefont{{Reisswig}}},
  \bibinfo{author}{\bibfnamefont{L.}~\bibnamefont{{Rezzolla}}},
  \bibinfo{author}{\bibfnamefont{J.}~\bibnamefont{{Thornburg}}},
  \bibinfo{author}{\bibfnamefont{P.}~\bibnamefont{{Diener}}}, \bibnamefont{and}
  \bibinfo{author}{\bibfnamefont{E.}~\bibnamefont{{Schnetter}}},
  \bibinfo{journal}{Physical Review Letters} \textbf{\bibinfo{volume}{99}},
  \bibinfo{pages}{041102} (\bibinfo{year}{2007}), \eprint{arXiv:gr-qc/0701163}.

\bibitem[{\citenamefont{{Campanelli}
  et~al.}(2007{\natexlab{a}})\citenamefont{{Campanelli}, {Lousto}, {Zlochower},
  and {Merritt}}}]{2007ApJ...659L...5C}
\bibinfo{author}{\bibfnamefont{M.}~\bibnamefont{{Campanelli}}},
  \bibinfo{author}{\bibfnamefont{C.}~\bibnamefont{{Lousto}}},
  \bibinfo{author}{\bibfnamefont{Y.}~\bibnamefont{{Zlochower}}},
  \bibnamefont{and}
  \bibinfo{author}{\bibfnamefont{D.}~\bibnamefont{{Merritt}}},
  \bibinfo{journal}{Astrophys. J. Lett.} \textbf{\bibinfo{volume}{659}},
  \bibinfo{pages}{L5} (\bibinfo{year}{2007}{\natexlab{a}}),
  \eprint{arXiv:gr-qc/0701164}.

\bibitem[{\citenamefont{{Campanelli}
  et~al.}(2007{\natexlab{b}})\citenamefont{{Campanelli}, {Lousto}, {Zlochower},
  and {Merritt}}}]{2007PhRvL..98w1102C}
\bibinfo{author}{\bibfnamefont{M.}~\bibnamefont{{Campanelli}}},
  \bibinfo{author}{\bibfnamefont{C.~O.} \bibnamefont{{Lousto}}},
  \bibinfo{author}{\bibfnamefont{Y.}~\bibnamefont{{Zlochower}}},
  \bibnamefont{and}
  \bibinfo{author}{\bibfnamefont{D.}~\bibnamefont{{Merritt}}},
  \bibinfo{journal}{Physical Review Letters} \textbf{\bibinfo{volume}{98}},
  \bibinfo{pages}{231102} (\bibinfo{year}{2007}{\natexlab{b}}),
  \eprint{arXiv:gr-qc/0702133}.

\bibitem[{\citenamefont{{Gonz{\'a}lez}
  et~al.}(2007)\citenamefont{{Gonz{\'a}lez}, {Hannam}, {Sperhake},
  {Br{\"u}gmann}, and {Husa}}}]{2007PhRvL..98w1101G}
\bibinfo{author}{\bibfnamefont{J.~A.} \bibnamefont{{Gonz{\'a}lez}}},
  \bibinfo{author}{\bibfnamefont{M.}~\bibnamefont{{Hannam}}},
  \bibinfo{author}{\bibfnamefont{U.}~\bibnamefont{{Sperhake}}},
  \bibinfo{author}{\bibfnamefont{B.}~\bibnamefont{{Br{\"u}gmann}}},
  \bibnamefont{and} \bibinfo{author}{\bibfnamefont{S.}~\bibnamefont{{Husa}}},
  \bibinfo{journal}{Physical Review Letters} \textbf{\bibinfo{volume}{98}},
  \bibinfo{pages}{231101} (\bibinfo{year}{2007}), \eprint{arXiv:gr-qc/0702052}.

\bibitem[{\citenamefont{{Baker} et~al.}(2007)\citenamefont{{Baker}, {Boggs},
  {Centrella}, {Kelly}, {McWilliams}, {Miller}, and {van
  Meter}}}]{2007ApJ...668.1140B}
\bibinfo{author}{\bibfnamefont{J.~G.} \bibnamefont{{Baker}}},
  \bibinfo{author}{\bibfnamefont{W.~D.} \bibnamefont{{Boggs}}},
  \bibinfo{author}{\bibfnamefont{J.}~\bibnamefont{{Centrella}}},
  \bibinfo{author}{\bibfnamefont{B.~J.} \bibnamefont{{Kelly}}},
  \bibinfo{author}{\bibfnamefont{S.~T.} \bibnamefont{{McWilliams}}},
  \bibinfo{author}{\bibfnamefont{M.~C.} \bibnamefont{{Miller}}},
  \bibnamefont{and} \bibinfo{author}{\bibfnamefont{J.~R.} \bibnamefont{{van
  Meter}}}, \bibinfo{journal}{Astrophys. J.} \textbf{\bibinfo{volume}{668}},
  \bibinfo{pages}{1140} (\bibinfo{year}{2007}),
  \eprint{arXiv:astro-ph/0702390}.

\bibitem[{\citenamefont{{Schnittman} and
  {Buonanno}}(2007)}]{2007ApJ...662L..63S}
\bibinfo{author}{\bibfnamefont{J.~D.} \bibnamefont{{Schnittman}}}
  \bibnamefont{and}
  \bibinfo{author}{\bibfnamefont{A.}~\bibnamefont{{Buonanno}}},
  \bibinfo{journal}{Astrophys. J. Lett.} \textbf{\bibinfo{volume}{662}},
  \bibinfo{pages}{L63} (\bibinfo{year}{2007}), \eprint{arXiv:astro-ph/0702641}.

\bibitem[{\citenamefont{{van Meter} et~al.}(2009)\citenamefont{{van Meter},
  {Wise}, {Miller}, {Reynolds}, {Centrella}, {Baker}, {Boggs}, {Kelly}, and
  {McWilliams}}}]{vanmeter09}
\bibinfo{author}{\bibfnamefont{J.~R.} \bibnamefont{{van Meter}}},
  \bibinfo{author}{\bibfnamefont{J.~H.} \bibnamefont{{Wise}}},
  \bibinfo{author}{\bibfnamefont{M.~C.} \bibnamefont{{Miller}}},
  \bibinfo{author}{\bibfnamefont{C.~S.} \bibnamefont{{Reynolds}}},
  \bibinfo{author}{\bibfnamefont{J.~M.} \bibnamefont{{Centrella}}},
  \bibinfo{author}{\bibfnamefont{J.~G.} \bibnamefont{{Baker}}},
  \bibinfo{author}{\bibfnamefont{W.~D.} \bibnamefont{{Boggs}}},
  \bibinfo{author}{\bibfnamefont{B.~J.} \bibnamefont{{Kelly}}},
  \bibnamefont{and} \bibinfo{author}{\bibfnamefont{S.~T.}
  \bibnamefont{{McWilliams}}}, \bibinfo{journal}{ArXiv e-prints}
  (\bibinfo{year}{2009}), \eprint{0908.0023}.

\bibitem[{\citenamefont{{Palenzuela}
  et~al.}(2009{\natexlab{a}})\citenamefont{{Palenzuela}, {Anderson}, {Lehner},
  {Liebling}, and {Neilsen}}}]{palenzuela09}
\bibinfo{author}{\bibfnamefont{C.}~\bibnamefont{{Palenzuela}}},
  \bibinfo{author}{\bibfnamefont{M.}~\bibnamefont{{Anderson}}},
  \bibinfo{author}{\bibfnamefont{L.}~\bibnamefont{{Lehner}}},
  \bibinfo{author}{\bibfnamefont{S.~L.} \bibnamefont{{Liebling}}},
  \bibnamefont{and}
  \bibinfo{author}{\bibfnamefont{D.}~\bibnamefont{{Neilsen}}},
  \bibinfo{journal}{Physical Review Letters} \textbf{\bibinfo{volume}{103}},
  \bibinfo{pages}{081101} (\bibinfo{year}{2009}{\natexlab{a}}),
  \eprint{0905.1121}.

\bibitem[{\citenamefont{{Palenzuela}
  et~al.}(2009{\natexlab{b}})\citenamefont{{Palenzuela}, {Lehner}, and
  {Yoshida}}}]{ply09}
\bibinfo{author}{\bibfnamefont{C.}~\bibnamefont{{Palenzuela}}},
  \bibinfo{author}{\bibfnamefont{L.}~\bibnamefont{{Lehner}}}, \bibnamefont{and}
  \bibinfo{author}{\bibfnamefont{S.}~\bibnamefont{{Yoshida}}},
  \bibinfo{journal}{ArXiv e-prints}  (\bibinfo{year}{2009}{\natexlab{b}}),
  \eprint{0911.3889}.

\bibitem[{\citenamefont{{Washik} et~al.}(2008)\citenamefont{{Washik}, {Healy},
  {Herrmann}, {Hinder}, {Shoemaker}, {Laguna}, and
  {Matzner}}}]{2008PhRvL.101f1102W}
\bibinfo{author}{\bibfnamefont{M.~C.} \bibnamefont{{Washik}}},
  \bibinfo{author}{\bibfnamefont{J.}~\bibnamefont{{Healy}}},
  \bibinfo{author}{\bibfnamefont{F.}~\bibnamefont{{Herrmann}}},
  \bibinfo{author}{\bibfnamefont{I.}~\bibnamefont{{Hinder}}},
  \bibinfo{author}{\bibfnamefont{D.~M.} \bibnamefont{{Shoemaker}}},
  \bibinfo{author}{\bibfnamefont{P.}~\bibnamefont{{Laguna}}}, \bibnamefont{and}
  \bibinfo{author}{\bibfnamefont{R.~A.} \bibnamefont{{Matzner}}},
  \bibinfo{journal}{Physical Review Letters} \textbf{\bibinfo{volume}{101}},
  \bibinfo{pages}{061102} (\bibinfo{year}{2008}), \eprint{arXiv:0802.2520}.

\bibitem[{\citenamefont{{Hinder} et~al.}(2008)\citenamefont{{Hinder},
  {Vaishnav}, {Herrmann}, {Shoemaker}, and {Laguna}}}]{2008PhRvD..77h1502H}
\bibinfo{author}{\bibfnamefont{I.}~\bibnamefont{{Hinder}}},
  \bibinfo{author}{\bibfnamefont{B.}~\bibnamefont{{Vaishnav}}},
  \bibinfo{author}{\bibfnamefont{F.}~\bibnamefont{{Herrmann}}},
  \bibinfo{author}{\bibfnamefont{D.~M.} \bibnamefont{{Shoemaker}}},
  \bibnamefont{and} \bibinfo{author}{\bibfnamefont{P.}~\bibnamefont{{Laguna}}},
  \bibinfo{journal}{Phys. Rev. D} \textbf{\bibinfo{volume}{77}},
  \bibinfo{pages}{081502} (\bibinfo{year}{2008}), \eprint{arXiv:0710.5167}.

\bibitem[{\citenamefont{{Herrmann}
  et~al.}(2007{\natexlab{b}})\citenamefont{{Herrmann}, {Hinder}, {Shoemaker},
  {Laguna}, and {Matzner}}}]{2007ApJ...661..430H}
\bibinfo{author}{\bibfnamefont{F.}~\bibnamefont{{Herrmann}}},
  \bibinfo{author}{\bibfnamefont{I.}~\bibnamefont{{Hinder}}},
  \bibinfo{author}{\bibfnamefont{D.}~\bibnamefont{{Shoemaker}}},
  \bibinfo{author}{\bibfnamefont{P.}~\bibnamefont{{Laguna}}}, \bibnamefont{and}
  \bibinfo{author}{\bibfnamefont{R.~A.} \bibnamefont{{Matzner}}},
  \bibinfo{journal}{Astrophys. J.} \textbf{\bibinfo{volume}{661}},
  \bibinfo{pages}{430} (\bibinfo{year}{2007}{\natexlab{b}}),
  \eprint{arXiv:gr-qc/0701143}.

\bibitem[{\citenamefont{{Herrmann}
  et~al.}(2007{\natexlab{c}})\citenamefont{{Herrmann}, {Hinder}, {Shoemaker},
  and {Laguna}}}]{2007CQGra..24...33H}
\bibinfo{author}{\bibfnamefont{F.}~\bibnamefont{{Herrmann}}},
  \bibinfo{author}{\bibfnamefont{I.}~\bibnamefont{{Hinder}}},
  \bibinfo{author}{\bibfnamefont{D.}~\bibnamefont{{Shoemaker}}},
  \bibnamefont{and} \bibinfo{author}{\bibfnamefont{P.}~\bibnamefont{{Laguna}}},
  \bibinfo{journal}{Classical and Quantum Gravity}
  \textbf{\bibinfo{volume}{24}}, \bibinfo{pages}{33}
  (\bibinfo{year}{2007}{\natexlab{c}}), \eprint{arXiv:gr-qc/0601026}.

\bibitem[{\citenamefont{{Bode} et~al.}(2008)\citenamefont{{Bode}, {Shoemaker},
  {Herrmann}, and {Hinder}}}]{2008PhRvD..77d4027B}
\bibinfo{author}{\bibfnamefont{T.}~\bibnamefont{{Bode}}},
  \bibinfo{author}{\bibfnamefont{D.}~\bibnamefont{{Shoemaker}}},
  \bibinfo{author}{\bibfnamefont{F.}~\bibnamefont{{Herrmann}}},
  \bibnamefont{and} \bibinfo{author}{\bibfnamefont{I.}~\bibnamefont{{Hinder}}},
  \bibinfo{journal}{Phys. Rev. D} \textbf{\bibinfo{volume}{77}},
  \bibinfo{pages}{044027} (\bibinfo{year}{2008}), \eprint{arXiv:0711.0669}.

\bibitem[{\citenamefont{Bode et~al.}(2009)}]{Bode:2009fq}
\bibinfo{author}{\bibfnamefont{T.}~\bibnamefont{Bode}} \bibnamefont{et~al.},
  \bibinfo{journal}{Phys. Rev.} \textbf{\bibinfo{volume}{D80}},
  \bibinfo{pages}{024008} (\bibinfo{year}{2009}), \eprint{0902.1127}.

\bibitem[{\citenamefont{{Campanelli}
  et~al.}(2006{\natexlab{a}})\citenamefont{{Campanelli}, {Lousto},
  {Marronetti}, and {Zlochower}}}]{2006PhRvL..96k1101C}
\bibinfo{author}{\bibfnamefont{M.}~\bibnamefont{{Campanelli}}},
  \bibinfo{author}{\bibfnamefont{C.~O.} \bibnamefont{{Lousto}}},
  \bibinfo{author}{\bibfnamefont{P.}~\bibnamefont{{Marronetti}}},
  \bibnamefont{and}
  \bibinfo{author}{\bibfnamefont{Y.}~\bibnamefont{{Zlochower}}},
  \bibinfo{journal}{Physical Review Letters} \textbf{\bibinfo{volume}{96}},
  \bibinfo{pages}{111101} (\bibinfo{year}{2006}{\natexlab{a}}),
  \eprint{arXiv:gr-qc/0511048}.

\bibitem[{\citenamefont{{Baker} et~al.}(2006)\citenamefont{{Baker},
  {Centrella}, {Choi}, {Koppitz}, and {van Meter}}}]{2006PhRvL..96k1102B}
\bibinfo{author}{\bibfnamefont{J.~G.} \bibnamefont{{Baker}}},
  \bibinfo{author}{\bibfnamefont{J.}~\bibnamefont{{Centrella}}},
  \bibinfo{author}{\bibfnamefont{D.-I.} \bibnamefont{{Choi}}},
  \bibinfo{author}{\bibfnamefont{M.}~\bibnamefont{{Koppitz}}},
  \bibnamefont{and} \bibinfo{author}{\bibfnamefont{J.}~\bibnamefont{{van
  Meter}}}, \bibinfo{journal}{Physical Review Letters}
  \textbf{\bibinfo{volume}{96}}, \bibinfo{pages}{111102}
  (\bibinfo{year}{2006}), \eprint{arXiv:gr-qc/0511103}.

\bibitem[{\citenamefont{Husa et~al.}(2006)\citenamefont{Husa, Hinder, and
  Lechner}}]{Husa:2004ip}
\bibinfo{author}{\bibfnamefont{S.}~\bibnamefont{Husa}},
  \bibinfo{author}{\bibfnamefont{I.}~\bibnamefont{Hinder}}, \bibnamefont{and}
  \bibinfo{author}{\bibfnamefont{C.}~\bibnamefont{Lechner}},
  \bibinfo{journal}{Computer Physics Communications}
  \textbf{\bibinfo{volume}{174}}, \bibinfo{pages}{983} (\bibinfo{year}{2006}),
  \eprint{gr-qc/0404023}.

\bibitem[{\citenamefont{Allen et~al.}(1999)\citenamefont{Allen, Goodale, and
  Seidel}}]{Allen99a}
\bibinfo{author}{\bibfnamefont{G.}~\bibnamefont{Allen}},
  \bibinfo{author}{\bibfnamefont{T.}~\bibnamefont{Goodale}}, \bibnamefont{and}
  \bibinfo{author}{\bibfnamefont{E.}~\bibnamefont{Seidel}}, in
  \emph{\bibinfo{booktitle}{7th Symposium on the Frontiers of Massively
  Parallel Computation-Frontiers 99}} (\bibinfo{publisher}{IEEE},
  \bibinfo{address}{New York}, \bibinfo{year}{1999}).

\bibitem[{\citenamefont{Schnetter et~al.}(2004)\citenamefont{Schnetter, Hawley,
  and Hawke}}]{Schnetter-etal-03b}
\bibinfo{author}{\bibfnamefont{E.}~\bibnamefont{Schnetter}},
  \bibinfo{author}{\bibfnamefont{S.~H.} \bibnamefont{Hawley}},
  \bibnamefont{and} \bibinfo{author}{\bibfnamefont{I.}~\bibnamefont{Hawke}},
  \bibinfo{journal}{Class. Quant. Grav.} \textbf{\bibinfo{volume}{21}},
  \bibinfo{pages}{1465} (\bibinfo{year}{2004}).

\bibitem[{Whisky-web()}]{whisky-web}
Whisky-web, \bibinfo{note}{{Whisky}, EU Network GR Hydrodynamics Code:\\{\tt
  http://www.whiskycode.org}}.

\bibitem[{\citenamefont{Baiotti et~al.}(2003)\citenamefont{Baiotti, Hawke,
  Montero, and Rezzolla}}]{Baiotti03a}
\bibinfo{author}{\bibfnamefont{L.}~\bibnamefont{Baiotti}},
  \bibinfo{author}{\bibfnamefont{I.}~\bibnamefont{Hawke}},
  \bibinfo{author}{\bibfnamefont{P.}~\bibnamefont{Montero}}, \bibnamefont{and}
  \bibinfo{author}{\bibfnamefont{L.}~\bibnamefont{Rezzolla}}, in
  \emph{\bibinfo{booktitle}{Computational Astrophysics in Italy: Methods and
  Tools}}, edited by
  \bibinfo{editor}{\bibfnamefont{R.}~\bibnamefont{Capuzzo-Dolcetta}}
  (\bibinfo{publisher}{Mem. Soc. Astron. It. Suppl.},
  \bibinfo{address}{Trieste}, \bibinfo{year}{2003}), vol.~\bibinfo{volume}{1},
  p. \bibinfo{pages}{327}.

\bibitem[{\citenamefont{Mart\'{\i} et~al.}(1991)\citenamefont{Mart\'{\i},
  Ib{\'a}{\~nez}, and Miralles}}]{Marti91}
\bibinfo{author}{\bibfnamefont{J.~M.} \bibnamefont{Mart\'{\i}}},
  \bibinfo{author}{\bibfnamefont{J.~M.} \bibnamefont{Ib{\'a}{\~nez}}},
  \bibnamefont{and} \bibinfo{author}{\bibfnamefont{J.~M.}
  \bibnamefont{Miralles}}, \bibinfo{journal}{Phys. Rev. D}
  \textbf{\bibinfo{volume}{43}}, \bibinfo{pages}{3794} (\bibinfo{year}{1991}).

\bibitem[{\citenamefont{Banyuls et~al.}(1997)\citenamefont{Banyuls, Font,
  Ib{\'a}{\~nez}, Mart\'{\i}, and Miralles}}]{Banyuls97}
\bibinfo{author}{\bibfnamefont{F.}~\bibnamefont{Banyuls}},
  \bibinfo{author}{\bibfnamefont{J.~A.} \bibnamefont{Font}},
  \bibinfo{author}{\bibfnamefont{J.~M.} \bibnamefont{Ib{\'a}{\~nez}}},
  \bibinfo{author}{\bibfnamefont{J.~M.} \bibnamefont{Mart\'{\i}}},
  \bibnamefont{and} \bibinfo{author}{\bibfnamefont{J.~A.}
  \bibnamefont{Miralles}}, \bibinfo{journal}{Astrophys. J.}
  \textbf{\bibinfo{volume}{476}}, \bibinfo{pages}{221} (\bibinfo{year}{1997}).

\bibitem[{\citenamefont{Ib{\'a}{\~nez}
  et~al.}(2001)\citenamefont{Ib{\'a}{\~nez}, Aloy, Font, Mart\'{\i}, Miralles,
  and Pons}}]{Ibanez01}
\bibinfo{author}{\bibfnamefont{J.}~\bibnamefont{Ib{\'a}{\~nez}}},
  \bibinfo{author}{\bibfnamefont{M.}~\bibnamefont{Aloy}},
  \bibinfo{author}{\bibfnamefont{J.}~\bibnamefont{Font}},
  \bibinfo{author}{\bibfnamefont{J.}~\bibnamefont{Mart\'{\i}}},
  \bibinfo{author}{\bibfnamefont{J.}~\bibnamefont{Miralles}}, \bibnamefont{and}
  \bibinfo{author}{\bibfnamefont{J.}~\bibnamefont{Pons}}, in
  \emph{\bibinfo{booktitle}{Godunov methods: theory and applications}}, edited
  by \bibinfo{editor}{\bibfnamefont{E.}~\bibnamefont{Toro}}
  (\bibinfo{publisher}{Kluwer Academic/Plenum Publishers},
  \bibinfo{address}{New York}, \bibinfo{year}{2001}).

\bibitem[{\citenamefont{{Faber} et~al.}(2007)\citenamefont{{Faber},
  {Baumgarte}, {Etienne}, {Shapiro}, and {Taniguchi}}}]{2007PhRvD..76j4021F}
\bibinfo{author}{\bibfnamefont{J.~A.} \bibnamefont{{Faber}}},
  \bibinfo{author}{\bibfnamefont{T.~W.} \bibnamefont{{Baumgarte}}},
  \bibinfo{author}{\bibfnamefont{Z.~B.} \bibnamefont{{Etienne}}},
  \bibinfo{author}{\bibfnamefont{S.~L.} \bibnamefont{{Shapiro}}},
  \bibnamefont{and}
  \bibinfo{author}{\bibfnamefont{K.}~\bibnamefont{{Taniguchi}}},
  \bibinfo{journal}{Phys. Rev. D} \textbf{\bibinfo{volume}{76}},
  \bibinfo{pages}{104021} (\bibinfo{year}{2007}), \eprint{arXiv:0708.2436}.

\bibitem[{\citenamefont{Ansorg et~al.}(2004)\citenamefont{Ansorg, Br\"ugmann,
  and Tichy}}]{Ansorg:2004ds}
\bibinfo{author}{\bibfnamefont{M.}~\bibnamefont{Ansorg}},
  \bibinfo{author}{\bibfnamefont{B.}~\bibnamefont{Br\"ugmann}},
  \bibnamefont{and} \bibinfo{author}{\bibfnamefont{W.}~\bibnamefont{Tichy}},
  \bibinfo{journal}{Phys. Rev. D} \textbf{\bibinfo{volume}{70}},
  \bibinfo{pages}{064011} (\bibinfo{year}{2004}).

\bibitem[{\citenamefont{{Bowen} and {York}}(1980)}]{1980PhRvD..21.2047B}
\bibinfo{author}{\bibfnamefont{J.~M.} \bibnamefont{{Bowen}}} \bibnamefont{and}
  \bibinfo{author}{\bibfnamefont{J.~W.} \bibnamefont{{York}},
  \bibfnamefont{Jr.}}, \bibinfo{journal}{\prd} \textbf{\bibinfo{volume}{21}},
  \bibinfo{pages}{2047} (\bibinfo{year}{1980}).

\bibitem[{\citenamefont{York}(1979)}]{York79}
\bibinfo{author}{\bibfnamefont{J.~W.} \bibnamefont{York}}, in
  \emph{\bibinfo{booktitle}{Sources of Gravitational Radiation}}, edited by
  \bibinfo{editor}{\bibfnamefont{L.~L.} \bibnamefont{Smarr}}
  (\bibinfo{publisher}{Cambridge University Press},
  \bibinfo{address}{Cambridge, UK}, \bibinfo{year}{1979}), pp.
  \bibinfo{pages}{83--126}, ISBN \bibinfo{isbn}{0-521-22778-X}.

\bibitem[{\citenamefont{Tolman}(1939)}]{Tolman39}
\bibinfo{author}{\bibfnamefont{R.~C.} \bibnamefont{Tolman}},
  \bibinfo{journal}{Phys. Rev.} \textbf{\bibinfo{volume}{55}},
  \bibinfo{pages}{364} (\bibinfo{year}{1939}).

\bibitem[{\citenamefont{Oppenheimer and Volkoff}(1939)}]{Oppenheimer39b}
\bibinfo{author}{\bibfnamefont{J.~R.} \bibnamefont{Oppenheimer}}
  \bibnamefont{and} \bibinfo{author}{\bibfnamefont{G.}~\bibnamefont{Volkoff}},
  \bibinfo{journal}{Phys. Rev.} \textbf{\bibinfo{volume}{55}},
  \bibinfo{pages}{374} (\bibinfo{year}{1939}).

\bibitem[{\citenamefont{Misner et~al.}(1973)\citenamefont{Misner, Thorne, and
  Wheeler}}]{Misner73}
\bibinfo{author}{\bibfnamefont{C.~W.} \bibnamefont{Misner}},
  \bibinfo{author}{\bibfnamefont{K.~S.} \bibnamefont{Thorne}},
  \bibnamefont{and} \bibinfo{author}{\bibfnamefont{J.~A.}
  \bibnamefont{Wheeler}}, \emph{\bibinfo{title}{Gravitation}}
  (\bibinfo{publisher}{W. H. Freeman}, \bibinfo{address}{San Francisco},
  \bibinfo{year}{1973}).

\bibitem[{\citenamefont{{Hayasaki} et~al.}(2007)\citenamefont{{Hayasaki},
  {Mineshige}, and {Sudou}}}]{hayasaki07}
\bibinfo{author}{\bibfnamefont{K.}~\bibnamefont{{Hayasaki}}},
  \bibinfo{author}{\bibfnamefont{S.}~\bibnamefont{{Mineshige}}},
  \bibnamefont{and} \bibinfo{author}{\bibfnamefont{H.}~\bibnamefont{{Sudou}}},
  \bibinfo{journal}{\pasj} \textbf{\bibinfo{volume}{59}}, \bibinfo{pages}{427}
  (\bibinfo{year}{2007}), \eprint{arXiv:astro-ph/0609144}.

\bibitem[{\citenamefont{Buonanno et~al.}(2006)\citenamefont{Buonanno, Chen, and
  Damour}}]{Buonanno:104005}
\bibinfo{author}{\bibfnamefont{A.}~\bibnamefont{Buonanno}},
  \bibinfo{author}{\bibfnamefont{Y.}~\bibnamefont{Chen}}, \bibnamefont{and}
  \bibinfo{author}{\bibfnamefont{T.}~\bibnamefont{Damour}},
  \bibinfo{journal}{Physical Review D (Particles, Fields, Gravitation, and
  Cosmology)} \textbf{\bibinfo{volume}{74}}, \bibinfo{eid}{104005}
  (pages~\bibinfo{numpages}{26}) (\bibinfo{year}{2006}), \eprint{0508067},
  \urlprefix\url{http://link.aps.org/abstract/PRD/v74/e104005}.

\bibitem[{\citenamefont{Kidder}(1995)}]{PhysRevD.52.821}
\bibinfo{author}{\bibfnamefont{L.~E.} \bibnamefont{Kidder}},
  \bibinfo{journal}{Phys. Rev. D} \textbf{\bibinfo{volume}{52}},
  \bibinfo{pages}{821} (\bibinfo{year}{1995}).

\bibitem[{\citenamefont{{Campanelli}
  et~al.}(2006{\natexlab{b}})\citenamefont{{Campanelli}, {Lousto}, and
  {Zlochower}}}]{2006PhRvD..74d1501C}
\bibinfo{author}{\bibfnamefont{M.}~\bibnamefont{{Campanelli}}},
  \bibinfo{author}{\bibfnamefont{C.~O.} \bibnamefont{{Lousto}}},
  \bibnamefont{and}
  \bibinfo{author}{\bibfnamefont{Y.}~\bibnamefont{{Zlochower}}},
  \bibinfo{journal}{Phys. Rev. D} \textbf{\bibinfo{volume}{74}},
  \bibinfo{pages}{041501} (\bibinfo{year}{2006}{\natexlab{b}}),
  \eprint{arXiv:gr-qc/0604012}.

\bibitem[{\citenamefont{{Ichimaru}}(1977)}]{ichimaru77}
\bibinfo{author}{\bibfnamefont{S.}~\bibnamefont{{Ichimaru}}},
  \bibinfo{journal}{\apj} \textbf{\bibinfo{volume}{214}}, \bibinfo{pages}{840}
  (\bibinfo{year}{1977}).

\bibitem[{\citenamefont{{Rees} et~al.}(1982)\citenamefont{{Rees}, {Begelman},
  {Blandford}, and {Phinney}}}]{rees82}
\bibinfo{author}{\bibfnamefont{M.~J.} \bibnamefont{{Rees}}},
  \bibinfo{author}{\bibfnamefont{M.~C.} \bibnamefont{{Begelman}}},
  \bibinfo{author}{\bibfnamefont{R.~D.} \bibnamefont{{Blandford}}},
  \bibnamefont{and} \bibinfo{author}{\bibfnamefont{E.~S.}
  \bibnamefont{{Phinney}}}, \bibinfo{journal}{\nat}
  \textbf{\bibinfo{volume}{295}}, \bibinfo{pages}{17} (\bibinfo{year}{1982}).

\bibitem[{\citenamefont{{Narayan} and {Yi}}(1994)}]{ny94}
\bibinfo{author}{\bibfnamefont{R.}~\bibnamefont{{Narayan}}} \bibnamefont{and}
  \bibinfo{author}{\bibfnamefont{I.}~\bibnamefont{{Yi}}},
  \bibinfo{journal}{\apjl} \textbf{\bibinfo{volume}{428}}, \bibinfo{pages}{L13}
  (\bibinfo{year}{1994}), \eprint{arXiv:astro-ph/9403052}.

\bibitem[{\citenamefont{{Rybicki} and {Lightman}}(1986)}]{rl86}
\bibinfo{author}{\bibfnamefont{G.~B.} \bibnamefont{{Rybicki}}}
  \bibnamefont{and} \bibinfo{author}{\bibfnamefont{A.~P.}
  \bibnamefont{{Lightman}}}, \emph{\bibinfo{title}{{Radiative Processes in
  Astrophysics}}} (\bibinfo{year}{1986}).

\bibitem[{\citenamefont{{Quataert}}(2003)}]{quataert03}
\bibinfo{author}{\bibfnamefont{E.}~\bibnamefont{{Quataert}}},
  \bibinfo{journal}{Astronomische Nachrichten Supplement}
  \textbf{\bibinfo{volume}{324}}, \bibinfo{pages}{435} (\bibinfo{year}{2003}),
  \eprint{arXiv:astro-ph/0304099}.

\bibitem[{\citenamefont{Farris et~al.}(2009)\citenamefont{Farris, Liu, and
  Shapiro}}]{Farris:2009mt}
\bibinfo{author}{\bibfnamefont{B.~D.} \bibnamefont{Farris}},
  \bibinfo{author}{\bibfnamefont{Y.~T.} \bibnamefont{Liu}}, \bibnamefont{and}
  \bibinfo{author}{\bibfnamefont{S.~L.} \bibnamefont{Shapiro}}
  (\bibinfo{year}{2009}), \eprint{0912.2096}.

\bibitem[{\citenamefont{{Natarajan} and {Pringle}}(1998)}]{np98}
\bibinfo{author}{\bibfnamefont{P.}~\bibnamefont{{Natarajan}}} \bibnamefont{and}
  \bibinfo{author}{\bibfnamefont{J.~E.} \bibnamefont{{Pringle}}},
  \bibinfo{journal}{\apjl} \textbf{\bibinfo{volume}{506}}, \bibinfo{pages}{L97}
  (\bibinfo{year}{1998}), \eprint{arXiv:astro-ph/9808187}.

\bibitem[{\citenamefont{{Natarajan} and {Armitage}}(1999)}]{na99}
\bibinfo{author}{\bibfnamefont{P.}~\bibnamefont{{Natarajan}}} \bibnamefont{and}
  \bibinfo{author}{\bibfnamefont{P.~J.} \bibnamefont{{Armitage}}},
  \bibinfo{journal}{\mnras} \textbf{\bibinfo{volume}{309}},
  \bibinfo{pages}{961} (\bibinfo{year}{1999}), \eprint{arXiv:astro-ph/9812001}.

\bibitem[{\citenamefont{{King} et~al.}(2005)\citenamefont{{King}, {Lubow},
  {Ogilvie}, and {Pringle}}}]{king05}
\bibinfo{author}{\bibfnamefont{A.~R.} \bibnamefont{{King}}},
  \bibinfo{author}{\bibfnamefont{S.~H.} \bibnamefont{{Lubow}}},
  \bibinfo{author}{\bibfnamefont{G.~I.} \bibnamefont{{Ogilvie}}},
  \bibnamefont{and} \bibinfo{author}{\bibfnamefont{J.~E.}
  \bibnamefont{{Pringle}}}, \bibinfo{journal}{\mnras}
  \textbf{\bibinfo{volume}{363}}, \bibinfo{pages}{49} (\bibinfo{year}{2005}),
  \eprint{arXiv:astro-ph/0507098}.

\bibitem[{\citenamefont{{Lodato} and {Pringle}}(2006)}]{lp06}
\bibinfo{author}{\bibfnamefont{G.}~\bibnamefont{{Lodato}}} \bibnamefont{and}
  \bibinfo{author}{\bibfnamefont{J.~E.} \bibnamefont{{Pringle}}},
  \bibinfo{journal}{\mnras} \textbf{\bibinfo{volume}{368}},
  \bibinfo{pages}{1196} (\bibinfo{year}{2006}),
  \eprint{arXiv:astro-ph/0602306}.

\bibitem[{\citenamefont{{Lodato} and {Pringle}}(2007)}]{lp07}
\bibinfo{author}{\bibfnamefont{G.}~\bibnamefont{{Lodato}}} \bibnamefont{and}
  \bibinfo{author}{\bibfnamefont{J.~E.} \bibnamefont{{Pringle}}},
  \bibinfo{journal}{\mnras} \textbf{\bibinfo{volume}{381}},
  \bibinfo{pages}{1287} (\bibinfo{year}{2007}), \eprint{0708.1124}.

\bibitem[{\citenamefont{{Fragile} et~al.}(2007)\citenamefont{{Fragile},
  {Blaes}, {Anninos}, and {Salmonson}}}]{fragile07}
\bibinfo{author}{\bibfnamefont{P.~C.} \bibnamefont{{Fragile}}},
  \bibinfo{author}{\bibfnamefont{O.~M.} \bibnamefont{{Blaes}}},
  \bibinfo{author}{\bibfnamefont{P.}~\bibnamefont{{Anninos}}},
  \bibnamefont{and} \bibinfo{author}{\bibfnamefont{J.~D.}
  \bibnamefont{{Salmonson}}}, \bibinfo{journal}{\apj}
  \textbf{\bibinfo{volume}{668}}, \bibinfo{pages}{417} (\bibinfo{year}{2007}),
  \eprint{0706.4303}.

\bibitem[{\citenamefont{{Perego} et~al.}(2009)\citenamefont{{Perego}, {Dotti},
  {Colpi}, and {Volonteri}}}]{perego09}
\bibinfo{author}{\bibfnamefont{A.}~\bibnamefont{{Perego}}},
  \bibinfo{author}{\bibfnamefont{M.}~\bibnamefont{{Dotti}}},
  \bibinfo{author}{\bibfnamefont{M.}~\bibnamefont{{Colpi}}}, \bibnamefont{and}
  \bibinfo{author}{\bibfnamefont{M.}~\bibnamefont{{Volonteri}}},
  \bibinfo{journal}{ArXiv e-prints}  (\bibinfo{year}{2009}),
  \eprint{0907.3742}.

\bibitem[{\citenamefont{{Dotti} et~al.}(2009)\citenamefont{{Dotti},
  {Volonteri}, {Perego}, {Colpi}, {Ruszkowski}, and {Haardt}}}]{dotti09}
\bibinfo{author}{\bibfnamefont{M.}~\bibnamefont{{Dotti}}},
  \bibinfo{author}{\bibfnamefont{M.}~\bibnamefont{{Volonteri}}},
  \bibinfo{author}{\bibfnamefont{A.}~\bibnamefont{{Perego}}},
  \bibinfo{author}{\bibfnamefont{M.}~\bibnamefont{{Colpi}}},
  \bibinfo{author}{\bibfnamefont{M.}~\bibnamefont{{Ruszkowski}}},
  \bibnamefont{and} \bibinfo{author}{\bibfnamefont{F.}~\bibnamefont{{Haardt}}},
  \bibinfo{journal}{ArXiv e-prints}  (\bibinfo{year}{2009}),
  \eprint{0910.5729}.

\bibitem[{\citenamefont{{Bode} and {Phinney}}(2007)}]{bp07}
\bibinfo{author}{\bibfnamefont{N.}~\bibnamefont{{Bode}}} \bibnamefont{and}
  \bibinfo{author}{\bibfnamefont{S.}~\bibnamefont{{Phinney}}},
  \bibinfo{journal}{APS Meeting Abstracts} pp. \bibinfo{pages}{1010--+}
  (\bibinfo{year}{2007}).

\bibitem[{\citenamefont{{O'Neill} et~al.}(2009)\citenamefont{{O'Neill},
  {Miller}, {Bogdanovi{\'c}}, {Reynolds}, and {Schnittman}}}]{oneill09}
\bibinfo{author}{\bibfnamefont{S.~M.} \bibnamefont{{O'Neill}}},
  \bibinfo{author}{\bibfnamefont{M.~C.} \bibnamefont{{Miller}}},
  \bibinfo{author}{\bibfnamefont{T.}~\bibnamefont{{Bogdanovi{\'c}}}},
  \bibinfo{author}{\bibfnamefont{C.~S.} \bibnamefont{{Reynolds}}},
  \bibnamefont{and} \bibinfo{author}{\bibfnamefont{J.~D.}
  \bibnamefont{{Schnittman}}}, \bibinfo{journal}{\apj}
  \textbf{\bibinfo{volume}{700}}, \bibinfo{pages}{859} (\bibinfo{year}{2009}),
  \eprint{0812.4874}.

\bibitem[{\citenamefont{{Lippai} et~al.}(2008)\citenamefont{{Lippai}, {Frei},
  and {Haiman}}}]{2008ApJ...676L...5L}
\bibinfo{author}{\bibfnamefont{Z.}~\bibnamefont{{Lippai}}},
  \bibinfo{author}{\bibfnamefont{Z.}~\bibnamefont{{Frei}}}, \bibnamefont{and}
  \bibinfo{author}{\bibfnamefont{Z.}~\bibnamefont{{Haiman}}},
  \bibinfo{journal}{Ap. J. Lett.} \textbf{\bibinfo{volume}{676}},
  \bibinfo{pages}{L5} (\bibinfo{year}{2008}), \eprint{arXiv:0801.0739}.

\bibitem[{\citenamefont{{Shields} and {Bonning}}(2008)}]{2008ApJ...682..758S}
\bibinfo{author}{\bibfnamefont{G.~A.} \bibnamefont{{Shields}}}
  \bibnamefont{and} \bibinfo{author}{\bibfnamefont{E.~W.}
  \bibnamefont{{Bonning}}}, \bibinfo{journal}{Astrophys. J.}
  \textbf{\bibinfo{volume}{682}}, \bibinfo{pages}{758} (\bibinfo{year}{2008}),
  \eprint{arXiv:0802.3873}.

\bibitem[{\citenamefont{Megevand et~al.}(2009)}]{Megevand:2009yx}
\bibinfo{author}{\bibfnamefont{M.}~\bibnamefont{Megevand}}
  \bibnamefont{et~al.}, \bibinfo{journal}{Phys. Rev.}
  \textbf{\bibinfo{volume}{D80}}, \bibinfo{pages}{024012}
  (\bibinfo{year}{2009}), \eprint{0905.3390}.

\bibitem[{\citenamefont{Corrales et~al.}(2009)\citenamefont{Corrales, Haiman,
  and MacFadyen}}]{Corrales:2009nv}
\bibinfo{author}{\bibfnamefont{L.~R.} \bibnamefont{Corrales}},
  \bibinfo{author}{\bibfnamefont{Z.}~\bibnamefont{Haiman}}, \bibnamefont{and}
  \bibinfo{author}{\bibfnamefont{A.}~\bibnamefont{MacFadyen}}
  (\bibinfo{year}{2009}), \eprint{0910.0014}.

\bibitem[{\citenamefont{{Rossi} et~al.}(2009)\citenamefont{{Rossi}, {Lodato},
  {Armitage}, {Pringle}, and {King}}}]{rossi09}
\bibinfo{author}{\bibfnamefont{E.~M.} \bibnamefont{{Rossi}}},
  \bibinfo{author}{\bibfnamefont{G.}~\bibnamefont{{Lodato}}},
  \bibinfo{author}{\bibfnamefont{P.~J.} \bibnamefont{{Armitage}}},
  \bibinfo{author}{\bibfnamefont{J.~E.} \bibnamefont{{Pringle}}},
  \bibnamefont{and} \bibinfo{author}{\bibfnamefont{A.~R.}
  \bibnamefont{{King}}}, \bibinfo{journal}{\mnras} pp. \bibinfo{pages}{1726--+}
  (\bibinfo{year}{2009}), \eprint{0910.0002}.

\bibitem[{\citenamefont{{Schnittman} and {Krolik}}(2008)}]{sk08}
\bibinfo{author}{\bibfnamefont{J.~D.} \bibnamefont{{Schnittman}}}
  \bibnamefont{and} \bibinfo{author}{\bibfnamefont{J.~H.}
  \bibnamefont{{Krolik}}}, \bibinfo{journal}{\apj}
  \textbf{\bibinfo{volume}{684}}, \bibinfo{pages}{835} (\bibinfo{year}{2008}),
  \eprint{0802.3556}.

\bibitem[{\citenamefont{{Devecchi} et~al.}(2009)\citenamefont{{Devecchi},
  {Rasia}, {Dotti}, {Volonteri}, and {Colpi}}}]{devecchi09}
\bibinfo{author}{\bibfnamefont{B.}~\bibnamefont{{Devecchi}}},
  \bibinfo{author}{\bibfnamefont{E.}~\bibnamefont{{Rasia}}},
  \bibinfo{author}{\bibfnamefont{M.}~\bibnamefont{{Dotti}}},
  \bibinfo{author}{\bibfnamefont{M.}~\bibnamefont{{Volonteri}}},
  \bibnamefont{and} \bibinfo{author}{\bibfnamefont{M.}~\bibnamefont{{Colpi}}},
  \bibinfo{journal}{\mnras} \textbf{\bibinfo{volume}{394}},
  \bibinfo{pages}{633} (\bibinfo{year}{2009}), \eprint{0805.2609}.

\bibitem[{\citenamefont{{Lang} and {Hughes}}(2006)}]{lh06}
\bibinfo{author}{\bibfnamefont{R.~N.} \bibnamefont{{Lang}}} \bibnamefont{and}
  \bibinfo{author}{\bibfnamefont{S.~A.} \bibnamefont{{Hughes}}},
  \bibinfo{journal}{\prd} \textbf{\bibinfo{volume}{74}},
  \bibinfo{pages}{122001} (\bibinfo{year}{2006}), \eprint{arXiv:gr-qc/0608062}.

\bibitem[{\citenamefont{{Ho}}(2009)}]{ho09}
\bibinfo{author}{\bibfnamefont{L.~C.} \bibnamefont{{Ho}}},
  \bibinfo{journal}{\apj} \textbf{\bibinfo{volume}{699}}, \bibinfo{pages}{626}
  (\bibinfo{year}{2009}), \eprint{0906.4104}.

\end{thebibliography}

\end{document}